\documentclass[usenatbib]{mn2e}
\usepackage{multirow}
\usepackage{rotating}
\usepackage{colortbl}
\usepackage{color}
\usepackage[fleqn]{amsmath}
\usepackage{amssymb}
\usepackage{amsfonts}
\usepackage{verbatim}
\usepackage{scalefnt}
\usepackage{lscape}
\usepackage{booktabs}

\usepackage[bookmarks=False,bookmarksnumbered,colorlinks=true, citecolor=blue, linkcolor=cyan]{hyperref}

\newcommand{\apjl}{ApJ}
\newcommand{\apjs}{ApJS}
\newcommand{\ao}{Appl.~Opt.}
\newcommand{\aap}{A\&A}

\newcommand{\beq}{\begin{equation}}
\newcommand{\eeq}{\end{equation}}

\definecolor{grey}{rgb}{0.5,0.6,0.7}
\def \simlt { \lower .75ex \hbox{$\sim$} \llap{\raise .27ex \hbox{$<$}} }
\definecolor{purple}{rgb}{0.65,0.15,0.9}
\definecolor{darkorange}{rgb}{0.8,0.3,0}
\definecolor{olive}{rgb}{0.4,0.6,0.25}
\definecolor{darkgreen}{rgb}{0,0.7,0}
\definecolor{darkred}{rgb}{0.5,0,0}

\title[On the population of BHs]{Supermassive black holes in cosmological simulations I: $M_{\rm BH}-M_{\star}$ relation and black hole mass function}
\author[Habouzit et al.]{M\'{e}lanie Habouzit$^{1,2}$\thanks{E-mail: habouzit@mpia.de}, Yuan Li$^{3}$, Rachel S. Somerville$^{4,5}$, Shy Genel$^{4,6}$,
\newauthor
Annalisa Pillepich$^1$, Marta Volonteri$^7$, Romeel Dav\'e$^{8}$, Yetli Rosas-Guevara$^{9}$,
\newauthor
Stuart McAlpine$^{10}$, S\'ebastien Peirani $^{11}$, Lars Hernquist$^{12}$, Daniel Angl\'es-Alc\'azar$^{13,4}$, 
\newauthor 
Amy Reines$^{14}$, Richard Bower$^{15}$, Yohan Dubois$^7$, Dylan Nelson$^{16}$, 
\newauthor
Christophe Pichon$^7$, Mark Vogelsberger$^{17}$\\
 $^1$ Max-Planck-Institut f\"ur Astronomie, K\"onigstuhl 17, D-69117 Heidelberg, Germany\\
  $^2$ Zentrum für Astronomie der Universit\"at Heidelberg,
 ITA, Albert-Ueberle-Str. 2, D-69120 Heidelberg, Germany\\
 $^3$ Astronomy Department, University of California, Berkeley, CA 94720, USA\\
 $^4$ Center for Computational Astrophysics, Flatiron Institute, New York, NY 10010, USA\\
 $^{5}$ Department of Physics and Astronomy, Rutgers University, 136 Frelinghuysen Rd, Piscataway, NY 08854, USA\\
 $^6$ Columbia Astrophysics Laboratory, Columbia University, 550 West 120th Street, New York, NY 10027, USA\\
 $^7$ Institut d'Astrophysique de Paris, Sorbonne Universit\'es, CNRS, UMR 7095, 98 bis bd Arago, 75014 Paris, France\\
 $^{8}$ Institute for Astronomy, Royal Observatory, University of Edinburgh, Edinburgh EH9 3HJ, UK\\ 
 $^{9}$  Donostia International Physics Centre (DIPC), Paseo Manuel de Lardizabal 4, 20018 Donostia-San Sebastian, Spain\\
  $^{10}$ Department of Physics, University of Helsinki, Gustaf H\"{a}llstr\"{o}min katu 2a P.O. Box 64, FI-00014 University of Helsinki, Finland\\
 $^{11}$ Université Côte d'Azur, Observatoire de la Côte d'Azur, Laboratoire Lagrange, Bd de l'Observatoire, CS 34229, Nice, France\\
 $^{12}$ Harvard-Smithsonian Center for Astrophysics, 60 Garden Street, Cambridge, MA 02138, USA\\
 $^{13}$ Department of Physics, University of Connecticut, 196 Auditorium Road, U-3046, Storrs, CT 06269-3046, USA\\
 $^{14}$ eXtreme Gravity Institute, Department of Physics, Montana State University, Bozeman, MT 59717, USA\\
 $^{15}$ Institute for Computational Cosmology, Department of Physics, University of Durham, South Road, Durham DH1 3LE, UK\\
 $^{16}$ Max-Planck-Institut f\"ur Astrophysik, 
 Karl-Schwarzschild-Strasse 1, D-85740 Garching bei M\"unchen, Germany\\
 $^{17}$ Department of Physics, Kavli Institute for Astrophysics and Space Research, MIT, Cambridge, MA 02139, USA
}

\date{2020}                  

\begin{document}
\maketitle

\begin{abstract}
The past decade has seen significant progress in understanding galaxy formation and evolution using large-scale cosmological simulations. While these simulations produce galaxies in overall good agreement with observations, they employ different sub-grid models for galaxies and supermassive black holes (BHs).
We investigate the impact of the sub-grid models on the BH mass properties of the Illustris, TNG100, TNG300, Horizon-AGN, EAGLE, and SIMBA simulations, focusing on the $M_{\rm BH}-M_{\star}$ relation and the BH mass function.
All simulations predict tight $M_{\rm BH}-M_{\star}$ relations, and struggle to produce BHs of $M_{\rm BH}\leqslant 10^{7.5}\, \rm M_{\odot}$ in galaxies of $M_{\star}\sim 10^{10.5}-10^{11.5}\, \rm M_{\odot}$. 
While the time evolution of the mean $M_{\rm BH}-M_{\star}$ relation is mild ($\rm \Delta M_{\rm BH}\leqslant 1\, dex$ for $0\leqslant z \leqslant 5$) for all the simulations, its linearity (shape) and normalization varies from simulation to simulation.
The strength of SN feedback has a large impact on the linearity and time evolution for $M_{\star}\leqslant 10^{10.5}\, \rm M_{\odot}$. We find that the low-mass end is a good discriminant of the simulation models, and highlights the need for new observational constraints.
At the high-mass end, strong AGN feedback can suppress the time evolution of the relation normalization.
Compared with observations of the local Universe, we find an excess of BHs with $M_{\rm BH}\geqslant 10^{9}\, \rm M_{\odot}$ in most of the simulations. 
The BH mass function is dominated by efficiently accreting BHs ($\log_{10}\,f_{\rm Edd}\geqslant -2$) at high redshifts, and transitions progressively from the high-mass to the low-mass end to be governed by inactive BHs. The transition time and the contribution of active BHs are different among the simulations, and can be used to evaluate models against observations.

\end{abstract}

\section{Introduction}

Supermassive black holes (BHs) of millions of solar masses and greater reside in the center of most galaxies in the local Universe \citep{Magorrian1998,2009ApJ...698..198G,2011Natur.480..215M,2012NatCo...3E1304G,2013ARA&A..51..511K}. 
These objects were likely already in place early on in the history of our Universe. Evidence for the presence of these massive objects in the early Universe ($z>5$) include observations of extremely bright quasars powered by $10^{9-10}\, \rm M_{\odot}$ BHs when the Universe was about 1 Gyr \citep{Mortlock2011,2018ApJ...856L..25B}, but also relativistic jets and centrally driven winds at various redshifts \citep[down to $z=0$,][]{2014ApJ...796....7G,2014ApJ...787...38F,2016Natur.533..504C}. 
Accreting BHs, known as active galactic nuclei (AGN) and observed in a broad range of redshifts, can release large amounts of energy in their host galaxies, via AGN feedback.
One of the major outstanding issues in modern astrophysics is how BHs form and evolve from high redshift until today, and how they interact with their host galaxies.

Over the last decade, we have been able to numerically address galaxy formation in a cosmological context.
Large-scale cosmological simulations with a box length of $\sim 100\, \rm cMpc$ and a spatial resolution of $\sim 1\, \rm kpc$ have successfully demonstrated that it was possible to achieve reasonable agreements with current observational constraints, in terms of, e.g., galaxy clustering, stellar mass content of galaxies/halos, galaxy star formation rates, morphologies, sizes, color bimodality \citep[e.g.,][]{Dimatteo2008,2008MNRAS.390.1326O,2014MNRAS.444.1453D,2014MNRAS.442.2304H,2014MNRAS.444.1518V,2014MNRAS.445..175G,2014Natur.509..177V,2015MNRAS.446..521S,2016MNRAS.462.3265D,2017arXiv170302970P,2019MNRAS.486.2827D}.

Some discrepancies among galaxy properties have been found as well, and they help to improve the simulation sub-grid physics models. 
For example, the galaxy sizes in Illustris were too large \citep{2017arXiv170302970P,2015A&C....13...12N} compared to observational constraints.
The galaxy mass function of Illustris showed an excess of low-mass galaxies with $M_{\star}\leqslant 10^{10}\, \rm M_{\odot}$ \citep{2017arXiv170302970P}.
Horizon-AGN also produced an excess of galaxies with $M_{\star}\leqslant 10^{10}\, \rm M_{\odot}$ at $z=0$, before the knee of the galaxy mass function.
The gas fractions in the Illustris galaxies were elevated compared to observations, while the halo gas fractions were too low. This was a consequence of the {\it hot bubble} mode of the AGN feedback of Illustris, whose injection of thermal energy was displaced from the galaxies, and impacting more the host halos. As a result, the stellar to halo mass ratios of Illustris were on the upper limit of observational constraints, and too high for massive halos of $M_{\rm h}\geqslant \rm a \, few\, 10^{12}\, M_{\odot}$ \citep{2014MNRAS.444.1518V}. 
Most cosmological simulations (Horizon-AGN, EAGLE, MUFASA, Illustris) present lower fractions of quenched galaxies (i.e., galaxies with suppressed star formation rates) for $M_{\star}\geqslant 10^{10}\, \rm M_{\odot}$ compared to SDSS constraints \citep[e.g.,][]{2019ApJ...872..160H,2019MNRAS.485.4817D}.
Many papers address comparisons between these simulations and observations, but they often use different observational constraints and definitions of the studied quantities; drawing global conclusions from these analyses is difficult.

The resolution of large-scale cosmological simulations is not sufficient to resolve processes across the entire dynamical range needed, from BH accretion disks to large-scale filaments. Processes related to BH formation, growth, and feedback, as well as any other baryonic processes taking place at the galactic scale, are modeled with sub-grid physics. Simulations all use different sub-grid models, e.g. different location and mass for BH seeding, different models to compute the accretion onto BHs, different efficiencies and models to inject energy released by AGN, some models assuming that AGN feedback channels explicitly depend on BH mass, some others assuming a uniform feedback.   
Therefore, it is crucial to understand whether the BH populations produced by these simulations are in good agreement with observational constraints, and how the discrepancies could help us to improve sub-grid models.

We can generally divide the population of BHs in three categories. The first one is the population of high-redshift quasars, i.e., the most extremely bright and massive BHs that we observed at high redshift, in optical and near-IR surveys. These monsters probe the most extreme end of the BH mass spectrum, and are a challenge for our understanding of BH formation/accretion: they would require an almost continuous growth at the Eddington limit for the entire life of these objects. 
On the other end of the BH mass spectrum, we find the tiny BHs of $M_{\rm BH}\geqslant 10^{4}\, \rm M_{\odot}$ found in dwarf galaxies of $M_{\star}\leqslant 10^{9}\, \rm M_{\odot}$ \citep{2016PASA...33...54R,2019arXiv191109678G}, and theoretically expected to be the relics of BH seeds formed at high redshift.
These two first regimes are unfortunately not covered by the large-scale cosmological simulations studied here. Dwarf galaxies are indeed barely resolved and BH formation sub-grid models are too simplistic. These simulations are also limited by their volumes to form and evolve the rare quasars \citep[but see][for a larger simulation of $400 h^{-1} \, \rm Mpc$ run down to $z=7$]{2017MNRAS.467.4243D}.
Between these two BH mass ranges, cosmological simulations fully cover the regime of all the other BHs with mass $M_{\rm BH}=10^{5}-10^{10}\, \rm M_{\odot}$ observed at $z=4-0$. 
When accreting matter, these BHs power the AGN that we see e.g., in X-ray surveys (e.g., XMM-Newton, Chandra, NuSTAR) with X-ray luminosities of $L_{\rm x}=10^{42}-10^{46}\, \rm erg/s$ \citep[][and references therein]{2014ApJ...786..104U,2015MNRAS.451.1892A,2015ApJ...802...89B,2015ApJ...804..104M}.
This is the category of BHs that we investigate in this series of papers.

With time, we have accumulated evidence for correlations between the mass of BHs and the properties of their host galaxies, such as the total stellar mass, bulge mass, velocity dispersion, S{\'e}rsic index, and infrared luminosity \citep{MarconiHunt2003,Haring2004,Merloni2004,Shankar2004,2009ApJ...690...20S,Gultekin2009,2012AdAst2012E...7K,2013ARA&A..51..511K,2013ApJ...764..184M,2014ApJ...780...70L,2015ApJ...798...54G,2015arXiv150806274R}. These scaling relations are mostly observed in the local Universe since it is extremely challenging to measure these quantities at higher redshifts. For this reason, we only study the $M_{\rm BH}-M_{\star}$ relation in this paper ($M_{\star}$ being easier to measure), even if BH mass scales more tightly with the galaxy velocity dispersion locally \citep{2016ApJ...818...47S,2016MNRAS.460.3119S,2019MNRAS.490..600D,2020FrP.....8...61M}.
The scaling relations potentially imply the co-evolution of BHs and their host galaxies. 
However, several works have also emphasized the possibility that scaling relations may not require co-evolution processes according to the central-limit theorem \citep{2010MNRAS.407.1016H,2011ApJ...734...92J}. In these studies, scaling relations can be explained by the hierarchical assembly of BH and bulge growth by galaxy mergers. A large scatter in the high-redshift Universe will be reduced with time. 
Assuming that BHs only grow by mergers, \citet{2010MNRAS.407.1016H} show that the scatter of the $M_{\rm BH}-M_{\rm bulge}$ relation decreases with an increasing number of mergers, and so with time.

The time evolution of such scaling relations is difficult to investigate in high-redshift observations. Studies are often based on broad-line AGN targets, where BH masses are estimated from the virial method. The underlying assumption is that broad-line AGN behave the same way as any other galaxies, and obey the same scaling relations, which may not be the case \citep{2015arXiv150806274R}.
Most observational studies (including sometimes fainter AGN as well) have found mild or null evolution of the scaling relations \citep{2003ApJ...583..124S,2009ApJ...706L.215J,2013ApJ...764...80S,2013ApJ...767...13S,2015ApJ...805...96S}.
For instance, \citet{2013ApJ...764...80S} find no time evolution of the $M_{\rm BH}-\sigma_{\star}$ relation of quasars for $0.1<z<1.2$ and for $10^{7.5}<M_{\rm BH}/\rm M_{\odot}<10^{9}$; the relation evolves for lower and more massive BHs, but this can be attributed to observational uncertainties, as well as the large intrinsic scatter in the relation.
Similar results are obtained in \citet{2011ApJ...741L..11C} for 32 broad-line AGN with $10^{7.2}<M_{\rm BH}/M_{\star}<10^{8.7}$ in the redshift range $0.3<z<0.9$, and in \citet{2013ApJ...767...13S} with 18 X-ray selected broad-line AGN of the CDFS in the range $0.5<z<1.2$.
\citet{2009ApJ...706L.215J} find no evolution of the $M_{\rm BH}-M_{\star}$ relation while comparing $1<z<2$ objects with the local $M_{\rm BH}-M_{\rm bulge}$ relation.

The evolution of the scaling relation is often studied by comparing the high-redshift $M_{\rm BH}-M_{\star}$ to the same relation scaled from the local bulge mass relation. Going from the $M_{\rm BH}-M_{\star}$ plane to the $M_{\rm BH}-M_{\rm bulge}$ plane implies a redistribution of the stellar mass from the galactic disk to form a bulge, by mergers or secular processes. The studies above find that a large fraction of the high-redshift galaxies have, indeed, a disk component. The mild evolution found in observations tends to indicate that no addition of stellar mass would be required to build the bulges, and the $M_{\rm BH}-M_{\rm bulge}$ relation, from the high-redshift galaxies. The presence of disks in the high-redshift galaxies also means that their bulges can be considered as under-massive, which implies that the time evolution of the $M_{\rm BH}-M_{\rm bulge}$ relation may be stronger than the evolution of the $M_{\rm BH}-M_{\star}$ relation. It also favors the idea that the assembly of BHs takes place before the assembly of their host galaxy bulges.
More recently, \citet{2015ApJ...802...14S} (70 Herschel-detected broad-line AGN) found no evidence for redshift evolution of the $M_{\rm BH}-M_{\star}$ relation since $z=2$. \citet{2012ApJ...753L..30M} reached the same conclusion based on an X-ray stacking study ($z\leqslant 2$), as well as \citet{2019arXiv191202824S} with a sample of 100 X-ray selected moderately-luminous AGN at $z\leqslant 2.5$, combined with brighter AGN from the literature.
However, recently \citet{2019arXiv191011875D} used 32 X-ray selected broad-line AGN within $1.2<z<1.7$ and showed that the ratio $M_{\rm BH}/M_{\star}$ and $M_{\rm BH}/M_{\rm bulge}$ were larger than in the local Universe ($\sim 0.43\, \rm dex$), meaning that the relations are slightly offset toward more massive BHs or lower stellar mass/bulge mass compared to the relation at $z=0$.
Finally, \citet{2006MNRAS.368.1395M} find that the ratio $M_{\rm BH}/M_{\star}$\footnote{\citet{2006MNRAS.368.1395M} actually use the spheroid mass of the galaxies, which can be approximated by $\sim M_{\star}$ for massive galaxies.} of radio-loud AGN galaxies increases with redshift. Since their sample is based on the most massive early-type galaxies in the redshift range $0<z<2$, we cannot rule out that the results are biased. Nevertheless, it may also indicate that BH growth in these massive galaxies is completed first, and then galaxies catch up. In this paper, we will see that there is no consensus on the time evolution of the $M_{\rm BH}-M_{\star}$ mean relation in cosmological simulations.

Several observational studies have pointed out that the $M_{\rm BH}-M_{\star}$ relation may not be linear over the full range of stellar masses. The shape of the relation could depend e.g., on BH mass, stellar mass, galaxy types, star-forming activity. The correlation between the quiescence of the host galaxies and the shape of the $M_{\rm BH}-M_{\star}$ relation has been discussed in e.g., \citet[][]{2015ApJ...798...54G}. Different slopes of the relation have been found for samples of early-type/core-Sersic and late-type/Sersic galaxies in observations \citep[][]{2018ApJ...869..113D,2019ApJ...887...10S}, but also for different types of galaxies and activity of the BHs \citep[i.e., AGN or quiescent BHs;][]{2015arXiv150806274R}. Recently, a nonlinear shape of the $M_{\rm BH}-M_{\star}$ relation was also found in cosmological simulations \citep[e.g.,][]{2017MNRAS.468.3935H,2017MNRAS.465...32B,2017MNRAS.472L.109A} and semi-analytical galaxy formation models \citep[e.g.,][]{2015MNRAS.453.4112F}. In the following, we compare the shape and normalization of the $M_{\rm BH}-M_{\star}$ relation in several cosmological simulations to understand how this relation is impacted by the simulations subgrid physics.

In addition to scaling relations between BH mass and galaxy properties, the BH mass function describes the evolution of the mass distribution of BHs \citep[][for a review]{2012AdAst2012E...7K}, and is crucial to understand BH mass growth and the effects of BH accretion and feedback. 
The comparison with observational constraints can be tricky as these constraints often rely on some physical modeling/assumptions. The empirical BH mass functions have been used to constrain physical models of BH evolution. However, due to the uncertainties, it is actually challenging to rule out a significant number of models.
Estimating the BH mass function suffers from incompleteness and mass estimations. BH mass measurements are only possible for a small subset of local BHs. Therefore, these analyses rely on scaling relations (with bulge mass, bulge luminosity, width of the broad emission lines) to estimate BH mass for a large number of BHs, and not direct measurements of BH mass. 
In this case, the local BH mass function is estimated by combining one of the scaling relations with the local number density of galaxies (as a function of their velocity dispersion or bulge luminosity). This method is employed, for example, in \citet{Merloni08,2009ApJ...690...20S}. Several works have also investigated the evolution of the BH mass function with redshift, by combining the local BH mass function to the AGN luminosity function \citep[e.g.][]{2009ApJ...690...20S,2010ApJ...725..388C}.

Empirical BH mass functions \citep{Merloni08,2009ApJ...690...20S,2010ApJ...725..388C} and theoretical models based on semi-analytical models \citep{2009ApJ...704...89S,Fanidakis11,VB2010} or hydrodynamical simulations \citep{2008ApJS..175..356H,Dimatteo2008} agree well at $z=0$, but show large discrepancies at larger redshifts \citep[e.g., $z=2$][]{2012AdAst2012E...7K}. These differences particularly emerge at the low-mass end of the BH mass function for $M_{\rm BH}\leqslant 10^{8.5}\, \rm M_{\odot}$. Some observational constraints predict a turnover for lower-mass BHs \citep{Merloni08} when some others do not \citep{2009ApJ...690...20S,2010ApJ...725..388C}. The contribution of active and inactive BHs to the BH mass function is also crucial to understand the build-up of the BH population with time. In this paper, we compare the BH mass functions of the different simulations, and analyze how they evolve with time, and the relative contribution of active and inactive BHs.

Over the last decade, we have pushed the development of large-scale cosmological simulations to understand galaxy formation in a cosmological context. These simulations were mainly meant to reproduce galaxy properties measured in observations, and not so much the BH properties. 
Observational constraints become more and more accurate and numerous, so today is an excellent time to review the BH populations in cosmological simulations and how they compare to observations. 
In this first paper of a series, we aim at drawing conclusions on the BH populations from the six large-scale cosmological simulations Illustris, TNG100, TNG300, Horizon-AGN, EAGLE, and SIMBA. 
We compare these simulations to one another in a coherent way, i.e. applying the same post processing analyses to all the simulations.
We also compare their results with observational constraints, when possible. In this paper, we do not make close apple-to-apple comparisons of observations with simulated mock observations; our comparisons with observations are mostly to guide the eye of the readers. 
We point out the different properties that are successfully reproduced and the properties that are not. Our main goal is to understand how the different sub-grid models of these simulations can lead to differences in the BH populations, and to evaluate the range of possible values produced by the simulations for a given quantity. 
This will help us to understand what we are missing in our modeling of galaxy and BH formation/evolution in large-scale cosmological simulations, but also what we need to focus on and improve in the future.
In this first paper, we address the mass properties of the simulated BH populations, i.e., the $M_{\rm BH}-M_{\star}$ scaling relation and the BH mass function. 

In section 2, we describe the six Illustris, TNG100, TNG300, Horizon-AGN, EAGLE, and SIMBA large-scale simulations. We present the $M_{\rm BH}-M_{\star}$ diagram in Section 3, and analyze the scaling relations of the simulations in Section 4. In Section 5, we focus on the Illustris and TNG100 simulations, which share the same initial conditions and are, therefore, ideal to understand how the different sub-grid models create different features in the $M_{\rm BH}-M_{\star}$ relation. Finally, in Section 6 we present the BH mass function for all the simulations. We discuss our results and conclude in Section 7 and Section 8.

\section{Simulation models}
In this section, we introduce the six large-scale cosmological simulations. Only BH-related subgrid models and the main aspects of SN feedback modeling are described.
The analysis of the BH population of the Illustris simulation has been presented in \citet[][]{2015MNRAS.452..575S}, of the TNG simulations in \citet{2017MNRAS.465.3291W,2018MNRAS.479.4056W,2019MNRAS.484.4413H} and references therein.
The analysis of the Horizon-AGN BH population has been conducted in \citet{2016MNRAS.460.2979V}, and the analysis of EAGLE in \citet{2015MNRAS.446..521S,2016MNRAS.462..190R,2017MNRAS.468.3395M,2018MNRAS.481.3118M}. Finally, the BH population of SIMBA has been studied in \citet{2019MNRAS.486.2827D,2019MNRAS.487.5764T}.
In the following, we apply the same cut in the total stellar mass of galaxies, i.e, we only consider galaxies with $M_{\star}\geqslant 5\times10^{8}\, \rm M_{\odot}$, which is a common minimum galaxy stellar mass resolved in all simulations.

\subsection{Illustris}
The Illustris simulation consists of a volume of $(106\, \rm cMpc)^{3}$, and was run with the moving Voronoi mesh code {\sc Arepo} \citep{2010MNRAS.401..791S}. The gravitational softening for DM particles is 1.4 ckpc (comoving kpc), and the collisionless baryonic particle softening is 1.4 ckpc for $z\geqslant 1$, and 0.7 pkpc otherwise.

The BH seeding of the simulation is based on dark matter halo mass. Every halo reaching a mass of $M_{\rm h, thres}=7.10 \times 10^{10}\, \rm M_{\odot}$ is seeded with a BH of initial mass $M_{\rm seed}=1.42\times 10^{5}\, \rm M_{\odot}$.
The accretion onto BHs follows the Bondi-Hoyle-Lyttleton formalism and is capped at the Eddington limit:
\begin{eqnarray}
\dot{M}_{\rm BH}&=&\min (\dot{M}_{\rm Bondi},\dot{M}_{\rm Edd})\\
&=&\min (\alpha \, 4\pi G^{2}M_{\rm BH}^{2}\bar{\rho}/\bar{c_{\rm s}}^{3}, 4\pi G M_{\rm BH} m_{p}/\epsilon_{\rm r}\sigma_{\rm T}c),
\end{eqnarray}
with $\bar{c_{\rm s}}$ the kernel weighted ambient sound speed, $\epsilon_{\rm r}$ the radiative efficiency, and $\alpha$ the boost factor. The boost factor is introduced to compensate for the fact that the ISM multi-phase structure is not resolved \citep{Springel2005,Booth2009}. As simulations tend to underestimate the density around the BHs, the boost factor allows to increase the accretion rates onto the BHs.

Illustris employs a two-mode AGN feedback, both with release of thermal energy. 
AGN are able to deposit thermal energy in their surroundings with a coupling efficiency of 0.05 in the high-accretion mode, i.e. for BHs with high Eddington ratios of $f_{\rm Edd}=\dot{M}_{\rm BH}/\dot{M}_{\rm Edd}>0.05$. For BHs with $f_{\rm Edd}< 0.05$, thermal energy is released in hot bubbles within a radius of $\sim 100$ kpc around the BHs and couple to the gas with an efficiency of 0.35.

Illustris employs a kinetic wind model for the SN feedback, which is characterised by a mass loading factor and an initial wind velocity. 
The initial wind velocity can be written as:
\begin{eqnarray}
v_{\rm wind}=k_{\rm wind}\,  \sigma_{\rm DM},
\end{eqnarray}
 with $k_{\rm wind}=3.7$ a dimensionless efficiency factor, and $\sigma_{\rm DM}$ the one-dimensional dark matter velocity dispersion. 
\noindent The mass loading factor is defined by:
\begin{eqnarray}
\eta_{\rm wind}&=&\frac{2}{v^{2}_{\rm wind}} e_{\rm wind}\\
 &=&\frac{2}{v^{2}_{\rm wind}} \times 1.89\times 10^{-2} \times 10^{51}\rm erg/M_{\odot},
\end{eqnarray}
with the available $\rm SN_{\rm II}$ energy per formed stellar mass $e_{\rm wind}=1.89 \times 10^{-2} \times 10^{51}=1.09\times 1.73\times 10^{-2} \times {\rm E_{\rm SN_{II}}}$, with $1.73\times 10^{-2}$ the number of $\rm SN_{\rm II}$ per stellar mass formed, and $E_{\rm SN_{II}}=10^{51}\, \rm erg/M_{\odot}$ the available energy available for each $\rm SN_{\rm II}$.

\subsection{IllustrisTNG}
\label{sec:SN_feedback}
The simulations IllustrisTNG \citep[hereafter TNG100 and TNG300,][]{2017arXiv170703397S,2017arXiv170703406P,2018MNRAS.475..624N,2018MNRAS.480.5113M,2018MNRAS.477.1206N} have a volume of $(111\, \rm cMpc)^{3}$ and $(302\, \rm cMpc)^{3}$, respectively. TNG100 shares its initial conditions with the previous Illustris simulation.
The softening of the stellar particles and of DM particles is the same ($\epsilon_{\rm DM}=1.48, 2.96 \, \rm ckpc$ for TNG100 and TNG300, respectively) up to $z=1$, and fixed at their $z=1$ proper values for later redshifts  ($\epsilon_{\rm DM, z=0}=0.74, 1.48 \, \rm kpc$ for TNG100 and TNG300, respectively). The minimum softening of the gas is $\epsilon_{\rm gas,min}=0.19, 0.37 \, \rm ckpc$ for TNG100 and TNG300.

Dark matter halos with a mass exceeding $M_{\rm h, thres}=7.38\times 10^{10}\, \rm M_{\odot}$ are seeded in their center with BHs. The initial BH mass is set to $M_{\rm seed}= 1.18\times 10^{6}\, \rm M_{\odot}$, one order of magnitude higher than in the previous Illustris simulation.

The accretion onto BHs also follows the Bondi-Hoyle-Lyttleton model, but no boost factor is added. The TNG100 and TNG300 are magneto-hydrodynamical simulations; therefore the kernel weighted ambient sound speed is now written as $\bar{c_{\rm s}}=(c^{2}_{\rm s, therm} + (B^{2}/4\pi \rho))^{1/2}$, and includes a term for the magnetic signal propagation speed. The addition of the magnetic fields can impact the relationship between the BHs and the properties of their host galaxies. In TNG, the normalization of the $M_{\rm BH}-M_{\star}$ mean relation is higher with magnetic fields \citep{2017arXiv170302970P}.
Illustris and TNG have two different technical implementations of the Bondi model: Illustris only uses the parent gas cell of the BHs to compute the accretion rates onto the BHs, whereas TNG100 employs a kernel-weighted accretion rate over about 256 neighboring cells. As a consequence, the accretion rates onto the TNG BHs correlate with the gas properties of the central region of the galaxies, while the rates in the Illustris depend on the gas properties at the location of the BHs and could be more stochastic.

TNG includes a two mode AGN feedback: injection of thermal energy in the surroundings of the BHs accreting at high rates, and directional injection of momentum, with each event oriented in a random direction\footnote{While the injection of momentum is directional, the random direction of injection for each event produces, in practise, an overall isotropic feedback after a few events.} for low accretion rate BHs \citep{2017MNRAS.465.3291W,2018MNRAS.479.4056W}. The transition between modes is set by the threshold:
\begin{eqnarray}
f_{\rm Edd}=\min\left( 2\times 10^{-3}\times \left(\frac{M_{\rm BH}}{10^{8}\, \rm M_{\odot}}\right)^{2}, 0.1 \right).
\end{eqnarray}
Only the TNG and SIMBA simulations include a dependence on BH mass for the transition between AGN feedback modes.

The stellar and gas properties in the TNG simulations have been shown to be in generally better agreement with many observations than the Illustris simulation \citep{2017arXiv170302970P,2017arXiv170703406P}. In particular, the gas fraction was too high in galaxies and too small in the circumgalactic medium in massive halos. These discrepancies can be attributed to the low-accretion AGN feedback mode of Illustris, for which thermal energy was injected in bubbles displaced from the center of galaxies \citep{2015MNRAS.452..575S}.
Galaxies were also too large in Illustris by a factor of a few for galaxies of $M_{\star}\leqslant 10^{10.7}\, \rm M_{\odot}$, while the mass-size relation in TNG100 is now in better agreement \citep{2018MNRAS.474.3976G}.

In the following, we describe the modeling of SN feedback in TNG (both with the velocity of the winds and the wind mass loading factor), its evolution with redshift and halo mass, and how it compares to the previous Illustris SN feedback model.
Our description below only discusses the strength of the SN feedback at time of injection of the winds, and do not mean that the final impact of the winds will follow the same trends\footnote{We refer the reader to \citet{2017arXiv170302970P} for a complete description of the SN feedback model at injection (see their section 3.2.1, and their Fig. 6 and Fig. 7), and comparison with the Illustris model, and to \citet{2019MNRAS.490.3234N} for an analysis of the outflow properties in TNG50 compared to their properties at injection.}. The feedback of the SNe will have an important impact on the evolution of the BH populations, as demonstrated later in this paper.

The velocity of the winds in TNG100 is written as:
\begin{eqnarray}
v_{\rm wind}=\max \left(k_{\rm wind}\,  \sigma_{\rm DM} \left(\frac{H_{\rm 0}}{H(z)}\right)^{1/3}, v_{\rm wind, min} \right),
\end{eqnarray}
with $k_{\rm wind}=7.4$ ($k_{\rm wind}=3.7,$ for Illustris), $\sigma_{\rm DM}$ the one-dimensional dark matter velocity dispersion, $H_{\rm 0}, H(z)$ the Hubble constants, and $v_{\rm wind, min}=350\, \rm km/s$ the wind velocity floor imposed in TNG. 
Compared to the Illustris model, TNG100 has a new dependence on the Hubble constant, as well as the addition of a velocity floor \citep{2017arXiv170302970P}. 
This overall parametrization implies that at injection, the TNG winds are faster (higher wind velocity), for all halo masses and at all redshifts.
The addition of $v_{\rm w, min}$ prevents low-mass galaxies to have a unphysically low velocity of the winds, and makes the feedback in these galaxies more impactful than in the Illustris galaxies.
In conclusion, SN feedback in TNG is globally more impactful in low-mass galaxies at all times.

In TNG, a given fraction $\tau_{\rm wind}$ of the energy is released as thermal energy, and the mass loading factor is defined by:
\begin{eqnarray}
\eta_{\rm wind}&=&\frac{2}{v^{2}_{\rm wind}} e_{\rm wind} \left(1-\tau_{\rm wind}\right)\\
&=&\frac{2}{v^{2}_{\rm wind}} \, \times [1.4-5.6]\times 10^{-2}\times 10^{51}\rm erg/M_{\odot},
\end{eqnarray}
with [1.4-5.6] an efficiency factor which depends on the metallicity of the gas cell (1.4 for gas cell metallicity of 0.02):
The metallicity dependence makes the SN wind mass loading factor (and the available wind energy) smaller in metal-enriched environments. The overall effect is a more effective SN feedback at injection in low-mass galaxies in TNG than in Illustris.
The mass loading factor at injection is smaller in TNG than in Illustris at $z=0$ for all halo masses. The mass loading factor always increases with decreasing redshift in Illustris, and is constant (for the smallest halos of $M_{\rm h}\leqslant 10^{10}\, \rm M_{\odot}$) or decreasing (for more massive halos) in TNG. Consequently, the wind mass loading factor of TNG is higher than in Illustris only at high redshift ($z\geqslant 2-3$, depending on halo mass) for halos of $M_{\rm h}\geqslant 10^{11}\, \rm M_{\odot}$.

We discussed here the strength of the feedback at injection. 
The final effect seen in the TNG simulation is an overall stronger SN feedback in low-mass galaxies. This is shown by the lower normalization of the galaxy stellar mass function in TNG compared to Illustris (Fig.~\ref{fig:trends}) in the low-mass regime  \citep[see also][]{2017arXiv170302970P}. In this paper, we will analyze the effect of the modeling of SN feedback on the growth of BHs.

\subsection{Horizon-AGN}
Horizon-AGN \citep{2014MNRAS.444.1453D,2016MNRAS.463.3948D} has a volume of $(142\, \rm cMpc)^{3}$, and was run with the adaptive mesh refinement (AMR) code {\sc Ramses} \citep{2002A&A...385..337T}.
The seeding of Horizon-AGN is different from the precedent simulations, which all use a fixed threshold in the dark matter halo mass. In Horizon-AGN, BHs form in dense gas cells (i.e., in cells for which the gas density exceeds the threshold for star formation, here $n_{0} = 0.1 \, \rm H/cm^{3}$) with stellar velocity dispersion greater than $100\, \rm km/s$. BH seeds are not formed closer than 50 ckpc of an existing BH.
Finally, BHs are prevented from forming after $z=1.5$. At that time, all the progenitors of the $M_{\star}\geqslant 10^{10}\, \rm M_{\odot}$ galaxies at $z=0$ are formed and seeded with BHs \citep{2016MNRAS.460.2979V}.

The accretion rate onto BHs is computed following the Bondi-Hoyle-Lyttleton formalism with a boost factor $\alpha=(\rho/\rho_0)^{2}$ when the density $\rho$ is higher than the resolution-dependent threshold $\rho_{0}$ and $\alpha=1$ otherwise \citep{Booth2009}.

Horizon-AGN includes a two mode AGN feedback. In the high-accretion mode ($f_{\rm Edd}>0.01$), thermal energy is isotropically released within a sphere of radius a few resolution elements. The energy deposition rate is $\dot{E}_{\rm AGN}=\epsilon_f \epsilon r \dot{M}_{\rm BH} c^2= 0.015  \dot{M}_{\rm BH} c^2$. 
In the low-accretion mode, energy is injected into a bipolar outflow with a velocity of $10^4 \, \rm km/s$, to mimic the formation of a jet. The energy rate in this mode is $\dot{E}_{\rm AGN}= 0.1  \dot{M}_{\rm BH} c^2$.
The technical details of BH formation, growth and feedback modeling of Horizon-AGN can be found in \citet{2012MNRAS.420.2662D}.

Horizon-AGN employs a kinetic SN feedback, including momentum, mechanical energy and metals from type II, Type Ia SNe, and stellar winds \citep[details in][]{2017MNRAS.tmp..224K}. The feedback is modelled as kinetic release of energy on timescale below 50 Myr, and a thermal energy after 50 Myr. The feedback is also pulsed: energy is accumulated until sufficient to propagate the blast wave to at least two cells. The energy released depends on the SSP modeled asssumed and the metallicity of the gas, and is about $e_{\rm SN}\sim10^{49}\, \rm erg/M_{\odot}$.

\subsection{EAGLE}
The EAGLE simulation has a volume of $(100\, \rm cMpc)^{3}$ \citep[][]{2015MNRAS.446..521S,2015MNRAS.450.1937C,2016A&C....15...72M}, and was run with the code {\sc ANARCHY} \citep[Dalla Vecchia et al., in prep,][]{2015MNRAS.454.2277S}, which is based on the Smoothed Particle Hydrodynamics (SPH) method and is a modified version of the code GADGET3 \citep{2005MNRAS.364.1105S}. The DM mass resolution is $M_{\rm DM, res}=9.7\times 10^{6} \, \rm M_{\odot}$, and the baryonic mass resolution $M_{\rm baryon,res}=1.81\times 10^6 \, \rm M_{\odot}$. 
The gravitational interaction between particles is determined by a Plummer potential with a softening length of 2.66 ckpc and limited to a maximum physical length of 0.70 pkpc

All DM halos more massive than $M_{\rm DM}=1.48\times 10^{10}\, \rm M_{\odot}$ are seeded with a BH of $M_{\rm seed}=1.48\times 10^{5}\, \rm M_{\odot}$.

Accretion rates onto BHs are computed using a modified Bondi-Hoyle-Lyttleton model \citep{2015MNRAS.454.1038R,2016MNRAS.462..190R}:
\begin{eqnarray}
\dot{M}_{\rm BH}=\min (
\dot{M}_{\rm Bondi}\times \min \left((c_{\rm s}/V_{\rm \Phi})^{3}/C_{\rm visc},1\right),
\dot{M}_{\rm Edd}),
\end{eqnarray}
with $c_{\rm s}$ the sound speed of the surrounding gas, $V_{\Phi}$ the SPH-average circular speed of the gas around BHs, $C_{\rm visc}$ is a free 
parameter referring to the viscosity of the sub-grid accretion disk. 
In the default Bondi model, i.e., in the absence of angular momentum of the gas, the gas within the Bondi radius ($=GM_{\rm BH}/c^{2}_{s}$) directly falls onto the BHs. In the presence of angular momentum the gas will instead form an accretion disk.
Taking into account the angular momentum of the infalling material reduces BH accretion rates in small galaxies compared to the default Bondi model.
No additional boost factor is present in the accretion rate model.

The simulation employs a single-mode AGN feedback \citep{Booth2009}. 
A fraction of the accreted gas is stochastically injected as thermal energy with a net efficiency of $\epsilon_{r}\epsilon_f=0.1\times 0.15=0.015$. 
To prevent AGN feedback from becoming inefficient due to numerical loses and overcooling, the injection of thermal energy only occurs when the BH has accreted a sufficient amount of mass whose equivalent energy can raise the temperature of a gas particle by $\Delta T = 10^{8.5} \, \rm K$.

Feedback from SNe is injected stochastically as thermal energy \citep{2012MNRAS.426..140D}. 
SN energy of $10^{51}\rm \, erg/M_{\odot}$ is released 30 Myr after the stellar particle is born. The available energy injected depends on the local gas metallicity and density, and therefore, the feedback is stronger at higher redshift.
The metallicity dependence is physically motivated: the efficiency is reduced for metallicities $Z>0.1 \rm Z_{\odot}$ where metal-line cooling becomes important. The efficiency increases with higher densities to compensate for numerical overcooling \citep[the galaxies would be too compact without the density dependence,][]{2015MNRAS.450.1937C}.

\subsection{SIMBA}
The simulation SIMBA \citep{2019MNRAS.486.2827D} was run with the code GIZMO \citep{2015MNRAS.450...53H,2017arXiv171201294H}, with a box side length of $147\, \rm cMpc$. The DM mass resolution is $M_{\rm DM, res}=9.6\times 10^{7} \, \rm M_{\odot}$, and the baryonic mass resolution $M_{\rm baryon,DM}=1.8\times 10^{7} \, \rm M_{\odot}$. 
The softening is adaptive for all particle types with minimum value of $\epsilon=0.74\, \rm pkpc$.

A friends-of-friends algorithm identifies galaxies on the fly, and seed them with a BH of $M_{\rm seed}=1.4\times 10^{4}\, \rm M_{\odot}$ 
when their mass exceed the limit $M_{\star}\geqslant 10^{9.5}\, \rm M_{\odot}$.
This relatively high stellar mass threshold is motivated by higher resolution simulations showing that stellar feedback strongly suppresses black hole growth in lower mass galaxies \citep{2017MNRAS.468.3935H,2017MNRAS.472L.109A}.

While all the simulations presented above use a single model to compute the accretion onto the BHs, SIMBA uses a two mode model: torque-limited accretion model for cold gas ($T<10^{5}\, \rm K$) and the Bondi-Hoyle-Lyttleton model for hot gas ($T>10^{5}\, \rm K$). The accretion rate is written as:
\begin{eqnarray}
\dot{M}_{\rm BH}=(1-\epsilon_{\rm r})\times \notag \\
\left[ \min(\dot{M}_{\rm Bondi},\dot{M}_{\rm Edd}) 
+ \min(\dot{M}_{\rm Torque},3\, \dot{M}_{\rm Edd})
\right],
\end{eqnarray}
where $\epsilon_{\rm r}=0.1$, and $\dot{M}_{\rm Torque}$ is the gas inflow rate driven by gravitational instabilities from the scale of the galaxy to the accretion disk of the BH, i.e. here within $R_{0}=2h^{-1}\, \rm kpc$ \citep{2011MNRAS.415.1027H,2015ApJ...800..127A,2017MNRAS.464.2840A}. 
The gas inflow rate $\dot{M}_{\rm Torque}$ is defined as:
\begin{eqnarray}
\dot{M}_{\rm Torque}= \epsilon_{\rm T} \, f^{5/2}_{\rm d} 
\times \left( \frac{M_{\rm BH}}{10^{8}\, \rm M_{\odot}}\right)^{1/6} 
\times \left( \frac{M_{\rm enc} (R_{0})}{10^{9}\, \rm M_{\odot}}\right) \notag\\
\times \left( \frac{R_{0}}{100\, \rm pc}\right)^{-3/2} \times \left(1+\frac{f_{0}}{f_{\rm gas}} \right)^{-1}\, \rm M_{\odot}/yr,
\end{eqnarray}
with $\epsilon_{\rm T}$ a normalization parameter, $f_{\rm d}$ the mass fraction (gas + stellar content) of the disk, $f_{\rm gas}$ the gas fraction of the disk, $M_{\rm enc} (R_{0})$ the mass of the gas and stellar content.
A normalization parameter $\alpha=0.1$ is included in the Bondi mode to represent the efficiency of gas transport from the accretion disk down to the black hole (for consistency with the gravitational torque rate).

The simulation employs a modified version of the kinetic AGN feedback model of \citet{2017MNRAS.464.2840A}. In the high accretion rate mode ($f_{\rm Edd}\geqslant 0.2$, radiative mode), high mass loading outflows are employed, and the velocity of the outflows increases with $\log_{10} M_{\rm BH}$. In this mode the outflow velocoties are typically $<1000\, \rm km/s$.
In the low accretion rate mode of AGN feedback ($f_{\rm Edd}<0.2$, jet mode), lower mass loading but faster outflows are used. In this mode the velocity of the jets increases for decreasing $f_{\rm Edd}$. Full speed jets with $v_{\rm jet}\sim 8000\, \rm km/s$ are only achieved at $f_{\rm Edd}< 0.02$.
Only BHs of $M_{\rm BH}\geqslant 10^{7.5}\, \rm M_{\odot}$ are allowed to transition into this jet mode. 
Finally, X-ray feedback \citep[following][]{2012ApJ...754..125C} is included for galaxies with low gas fraction ($f_{\rm gas}<0.2$) and velocity jets of $v_{\rm jet}\geqslant 7000\, \rm km/s$ \citep{2019MNRAS.486.2827D}.
Finally, the feedback kinetic efficiencies $\epsilon=0.03$ (radiative feedback mode) and $\epsilon=0.3$ (jet), listed in Table~\ref{table:table_params}, depend on the outflow velocity. The numbers quoted here are representative of BHs with $M_{\rm BH}=10^{9} \, \rm M_{\odot}$.

Stellar feedback in SIMBA is modeled via hydrodynamically-decoupled winds as in Mufasa \citep{2016MNRAS.462.3265D}. The mass loading factor depends on the stellar mass of the galaxy following scalings from the FIRE simulations \citep{2017MNRAS.470.4698A}.  Similar to TNG, the mass loading factor is flat in galaxies with $M_{\star}\leqslant 3\times 10^{8}\, \rm M_{\odot}$ and is reduced at high redshift to avoid excessive feedback in these galaxies, and allows them to grow when poorly resolved.
The typical wind velocity are substantially lower than TNG100 ($\sim 1.6\times v_{\rm circ}$), and $30\%$ of the winds are ejected hot.
SIMBA also includes Type Ia SNe and AGB wind heating \citep{2019MNRAS.486.2827D}.\\

In the following, we generally refer to the mass of the BHs as the mass of the most massive BHs in the galaxies. For Illustris and TNG, BH mass refers to the sum of the BHs in the galaxies, but since BHs are automatically merged when their galaxies merge it does not make any difference (in practise only one or two galaxies per snapshots host several BHs).

In most of the simulations BHs (not in Horizon-AGN) are re-positioned every timestep to the local gravitational potential minimum of the galaxies or within smaller regions, which favors the accretion of gas onto the BHs. 
For example in EAGLE, BHs are moved to a particle with a lower potential but only within their kernel, and only if the relative velocity of the particle is less than $0.25\,c_{\rm s}$ \citep{2015MNRAS.446..521S}, and if its distance is within three gravitational softening lengths. 
Finally, BHs merge if they are close enough to each other. Some simulations have additional criteria, such as Horizon-AGN, EAGLE, SIMBA, and TNG. 
For instance, in the EAGLE simulation BHs are merged if their separation distance is smaller than both the SPH smoothing kernel of the BH and three gravitational softening lengths and if the relative velocities of the BHs at separations of the SPH kernel are less than the circular velocity at this distance \citep{2016MNRAS.463..870S}. In TNG, BHs merge if within the accretion/feedback region of another BH.

\subsection{Calibration of the simulations}
\subsubsection{Calibration of the BH sub-grid models}
Sub-grid models usually involve some {\it efficiency} parameters that can be tuned in order to produce a population of objects, here BHs, matching a given observational constraint. These models include the modeling of the accretion onto the BHs and of AGN feedback. The parameters related to seeding, e.g., the initial mass of the BH particles, are also often chosen to produce a BH population matching one of the $z=0$ empirical scaling relations.
We show some of the empirical relations commonly used in the literature in Fig.~\ref{fig:all_scaling_relations}, i.e., the relation $M_{\rm BH}-M_{\rm bulge}$ of \citet{Haring2004,2013ARA&A..51..511K,2013ApJ...764..184M}, and the relation $M_{\rm BH}-M_{\star}$ of \citet{2015arXiv150806274R,2019MNRAS.487.3404B}.

All the simulations were calibrated to one of the empirical $M_{\rm BH}-M_{\rm bulge}$ relations, with some variations in the computation of $M_{\rm bulge}$.
The Illustris simulation was calibrated to the $M_{\rm BH}-M_{\rm bulge}$ relation of \citet{2013ARA&A..51..511K}, assuming that the total stellar mass within the stellar half-mass radius of the simulated galaxies was a proxy for $M_{\rm bulge}$. 
Horizon-AGN was calibrated on the local scaling $M_{\rm BH}-M_{\rm bulge}$ relation of \citet{Haring2004} \citep{2012MNRAS.420.2662D} (using $M_{\rm bulge}$ for the simulated galaxies as well), to determine what fraction of rest-mass accreted energy should be released as AGN feedback. 
The simulation EAGLE was calibrated to the \citet{2013ApJ...764..184M} relation between the mass of BHs and the bulge mass of their host galaxies, assuming that the stellar mass of the observed galaxies was dominated by their bulge. The simulation was therefore calibrated assuming that $M_{\rm bulge}=M_{\star}$, which may be a correct for the highest-mass galaxies. 
The SIMBA simulation was calibrated on the amplitude of the $M_{\rm BH}-M_{\rm bulge}$ relation of \citet{2013ARA&A..51..511K} (assuming that $M_{\rm bulge}=M_{\star}$) at $\log_{10} M_{\rm BH}/\rm M_{\odot} \sim 8-9$ through the efficiency parameter of the accretion model/AGN model \citep[see][for more details]{2017MNRAS.464.2840A}. 
All simulations based on Bondi accretion (Illustris, TNG, Horizon-AGN, Eagle) match the normalization of the $M_{\rm BH}-M_{\rm bulge}$ relation by tunning a feedback efficiency parameter, while SIMBA (based on gravitational torque accretion) calibrates the accretion efficiency to match the $M_{\rm BH}-M_{\rm bulge}$ relation. This difference is primarily driven by the BH mass dependence of the accretion parameterization \citep{2015ApJ...800..127A,2017MNRAS.464.2840A}.

\subsubsection{General calibration of the simulations}
More generally, the simulations are calibrated with observational constraints of galaxies, either by directly using the fits of these constraints (EAGLE), or by simple agreement with the constraints (Illustris, IllustrisTNG).
IllustrisTNG was not directly calibrated by eye with any new observational constraints, but instead the sub-grid models of the previous Illustris were adapted to better match the Illustris galaxy properties to observations used for the calibration of the Illustris simulation. 
The EAGLE simulation was calibrated to fits of observational data representing the galaxy mass function at $z=0$, galaxy sizes at a function of galaxy mass at $z=0.1$, and the local $M_{\rm BH}-M_{\star}$ relation \citep{2015MNRAS.446..521S,2015MNRAS.450.1937C}. The Illustris simulation was calibrated based on cosmic star formation rate density, galaxy mass function at $z=0$, stellar to halo mass ratios at $z=0$, as well the gas metallicity mass relation at $z=0$, and the local $M_{\rm BH}-M_{\star}$ relation \citep{2013MNRAS.436.3031V,2014MNRAS.438.1985T}. 
Similarly, IllustrisTNG was calibrated based on the cosmic star formation rate density, the galaxy mass function, the stellar to halo ratios, and the $M_{\rm BH}-M_{\star}$ as well, by comparison to the results of the Illustris simulation. Calibrations to the gas fraction at $z=0$, and to the galaxy sizes as a function of galaxy masses at $z=0$ were added.
The SIMBA simulation was calibrated on the galaxy stellar mass function and its evolution with time. No calibration on galaxy sizes or galaxy/halo gas content was made. 
Finally, the Horizon-AGN simulation was simply calibrated with the $M_{\rm BH}-M_{\star}$ relation to finalize the AGN feedback model, and the rest of the subgrid physics (star formation, SN feedback) was just the result of the underlying model (no calibration on the galaxy stellar mass function for instance).

\subsection{BH luminosity}
We compute in post-processing the luminosity of the BHs with the model of \citet{Churazov2005}, i.e. explicitly distinguishing radiatively efficient and radiatively inefficient AGN. The bolometric luminosity of radiatively efficient BHs, i.e. with an Eddington ratio of $f_{\rm Edd}>0.1$, is defined as:
\begin{eqnarray}
L_{\rm bol}=\frac{\epsilon_{\rm r}}{1-\epsilon_{\rm r}} \dot{M}_{\rm BH} c^{2}.
\end{eqnarray}
BHs with small Eddington ratio of $f_{\rm Edd}\leqslant 0.1$ are considered radiatively inefficient and their bolometric luminosities are computed as:
\begin{eqnarray}
L_{\rm bol}=0.1 L_{\rm Edd} (10 f_{\rm Edd})^{2}=(10 f_{\rm Edd}) \epsilon_{\rm r} \dot{M}_{\rm BH} c^{2}.
\end{eqnarray}
The distinction between radiatively efficient and inefficient AGN is often not made when computing the luminosity of the AGN in simulations. Therefore, our post-processing computation of the luminosities differs from the intrinsic luminosity that goes into AGN feedback in the simulations.
We use the radiative efficiency parameter which were used to derive the accretion rate self-consistently in the simulations, i.e., $\epsilon_{\rm r}=0.2$ for Illustris, TNG100, and TNG300, and  $\epsilon_{\rm r}=0.1$ for Horizon-AGN, EAGLE and SIMBA.

\section{$M_{\rm BH}-M_{\star}$ diagrams of the cosmological simulations}
We focus our investigation on the scaling relation between BH mass and the total stellar mass of their host galaxies, a quantity that can be measured beyond the local Universe. 
Empirical scaling relations have been derived with other galaxy properties (e.g., stellar velocity dispersion, bulge luminosity, bulge mass), and we discuss these quantities in Appendix \ref{other_scaling_relations}. Given the differences found in the $M_{\rm BH}-\sigma$ relation of Illustris and TNG100 and observations \citep{2019arXiv191000017L}, we prefer to not investigate this relation here.
In this section, we present several versions of the $M_{\rm BH}-M_{\star}$ diagrams of the simulations. While we do not intend to broadly discuss accretion properties of the BHs and the correlation with star-forming activity of their host galaxies, we present here some insights into these correlations.

\subsection{$M_{\rm BH}-M_{\star}$ diagram in observations}
\begin{figure*}
\centering
\includegraphics[scale=0.45]{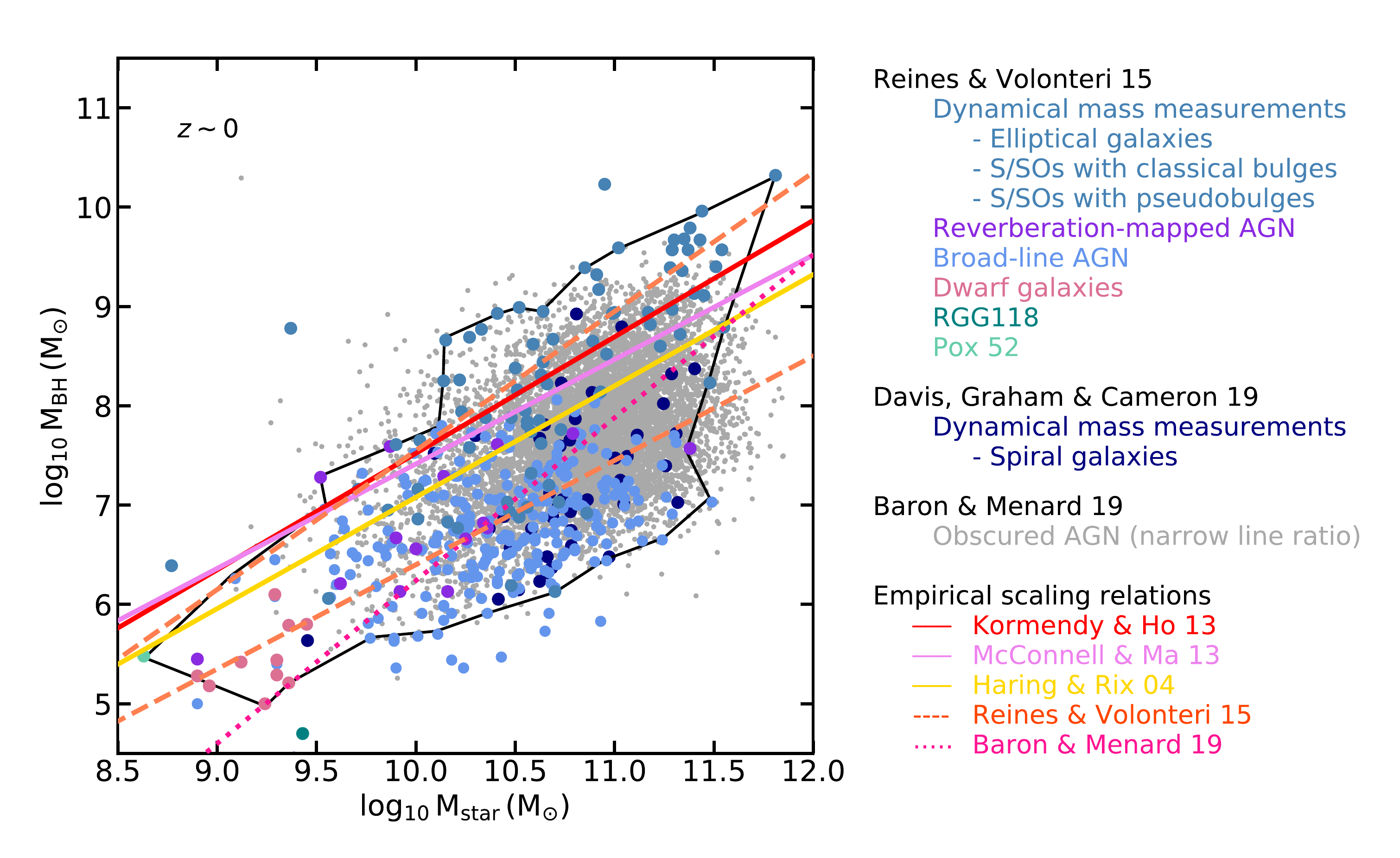}
\caption{Diagram showing BH mass as a function of total stellar mass of BH host galaxies. We show some empirical scaling relations from the literature, both for $M_{\rm BH}-M_{\rm bulge}$ and $M_{\rm BH}-M_{\star}$.  
We show the relations $M_{\rm BH}-M_{\rm bulge}$ derived in \citet{2013ARA&A..51..511K} as a solid red line, in \citet{2013ApJ...764..184M} as a solid purple line, and the relation of \citet{Haring2004} as a solid yellow line. We also show the two $M_{\rm BH}-M_{\star}$ relations of \citet{2015arXiv150806274R} in dashed orange lines: the line on the top for elliptical and spiral/S0 galaxies with classical bulges, and the bottom line for the broad-line AGN. The full sample of \citet{2015arXiv150806274R} is shown as well with colored dots. 
We also show the sample of spiral galaxies of \citet{2018ApJ...869..113D} with dynamical BH mass measurements. 
Finally, we also reproduce here the sample of AGN of \citet{2019MNRAS.487.3404B}, which includes obscured AGN, and their scaling $M_{\rm BH}-M_{\star}$ scaling relation as a dotted pink line.
We draw here a {\it cartoon} black region including a large fraction of the observations that we use in the following.}
\label{fig:all_scaling_relations}
\end{figure*}

Before investigating the $M_{\rm BH}-M_{\star}$ diagram in cosmological simulations, we show in Fig.~\ref{fig:all_scaling_relations} several observational samples. The sample of  \citet{2015arXiv150806274R} includes a small sample of dwarf galaxies (shown in pink), 262 broad-line AGN (light blue), reverberation-mapped AGN (in purple), and 79 galaxies with dynamical BH mass measurements (dark blue). 
The BHs present in RGG118 \citep{2015ApJ...809L..14B} and Pox 52 \citep{Barth2004} are shown in green. 
We add the sample of \citet{2018ApJ...869..113D} with dynamical mass measurements in dark blue. Finally, we show the recent observational sample of \citet{2019MNRAS.487.3404B}, which include obscured AGN. In the following, we compare the simulations to the sample of \citet{2015arXiv150806274R}, which covers the same $M_{\rm BH}-M_{\star}$ regions as most other observations.
In Fig.~\ref{fig:all_scaling_relations} we add several empirical scaling relations that are commonly used in the literature, both $M_{\rm BH}-M_{\rm bulge}$ \citep{Haring2004,2013ARA&A..51..511K,2013ApJ...764..184M} and $M_{\rm BH}-M_{\star}$ relations \citep{2015arXiv150806274R,2019MNRAS.487.3404B}. 
The empirical scaling relations and even the samples themselves can be biased, in many different ways. 
That is why we do not aim at an apple-to-apple comparison with observations, which would require to take into account all the selection effects and biases (e.g., galaxy selection, BH mass measurements).
Moreover, most of the samples are biased towards luminous unobscured AGN \citep[but see][]{2019MNRAS.487.3404B}, or massive BHs whose sphere of influence can be resolved for mass dynamical measurements. For example, selection bias in dynamically-measured BH sample, whose host galaxies often appear to have higher velocity dispersion $\sigma$, could artificially enhance BH masses \citep{Bernardi2007}. Since the empirical $M_{\rm BH}-\sigma$ relations are employed to estimate BH masses of AGN, this could impact more broadly the observational samples by a factor of a few \citep{2016MNRAS.460.3119S}

\begin{landscape}
\begin{table}
\caption{Parameters and models of BH and galaxy formation/evolution in the simulation Illustris, TNG100, TNG300, Horizon-AGN, EAGLE, and SIMBA. We include the quantities related to the volume and resolution of the simulations, the seeding prescriptions (i.e. minimum halo mass seeded, minimum cell density, velocity dispersion), and the BH mass of the seeds, the parameters of the BH accretion models (models and boost factors), the SN feedback models (models, efficiencies, and energy released per core collapse SN), and finally the parameters related to AGN feedback (number of modes, models, efficiencies, transition between modes).}
\begin{center}
\begin{tabular}{lcccccc}
\hline

\hline
& Illustris & TNG100 & TNG300 & Horizon-AGN & EAGLE & SIMBA\\
\hline

\hline
{\bf Cosmology} &&&&&\\
$\Omega_{\rm \Lambda}$ &0.7274&0.6911&0.6911&0.728&0.693&0.7\\
$\Omega_{\rm m}$ &0.2726&0.3089&0.3089&0.272&0.307&0.3\\
$\Omega_{\rm b}$ &0.0456&0.0486&0.0486&0.045&0.0483&0.048\\
$\sigma_{\rm 8}$ &0.809&0.8159&0.8159&0.81&0.8288&0.82\\
$n_{\rm s}$ &0.963&0.9667&0.9667&0.967&0.9611&0.97\\
$H_{\rm 0} \, \rm (km\, s^{-1}\, Mpc^{-1})$ &70.4&67.74&67.74&70.4&67.77&68\\
\hline
{\bf Resolution} &&&&&\\
Box side length ($\rm cMpc$)                                & 106.5 & 110.7 & 302.6 & 142.0 & 100.0& 147.1\\
Dark matter mass reso. ($\rm M_{\odot}$)     & $6.26\times 10^{6}$ & $7.5 \times 10^{6}$ & $5.9 \times 10^{7}$ & $8\times 10^{7}$ & $9.7\times 10^{6}$ & $9.6\times 10^{7}$\\
Baryonic mass reso. ($\rm M_{\odot}$)          & $1.26 \times 10^{6}$ & $1.4 \times 10^{6}$ & $1.1 \times 10^{7}$ & $2\times 10^{6}$  & $1.81\times 10^{6}$ & $1.82\times 10^{7}$\\
Spatial resolution ($\rm pkpc$)				   & 0.71 & 0.74 & 1.48 & 1.0 & 0.7 & 0.74\\
Gravitational softening ($\rm ckpc$)   & 1.4 	& 1.48 ($z\geqslant1$) & 2.96 ($z\geqslant1$)&	& 2.66 ($z\geqslant2.8$)	& 0.74\\
& & /0.74 pkpc & /1.48 pkpc	& & / max 0.7 pkpc &\\
Baryonic softening ($\rm ckpc$)       		& 1.4 ckpc ($z\geqslant 1$)	& 1.48 ($z\geqslant1$)	&2.96 ($z\geqslant1$)	&	& 2.66 ($z\geqslant2.8$) & 0.74	\\
& /0.7 pkpc 	&	/0.74 pkpc &	/1.48 pkpc &	&  / max 0.7 pkpc&	\\
\hline

{\bf Seeding} &&&&&\\
BH seed mass ($\rm M_{\odot}$)			  & $1.42\times 10^{5}$ & $1.18\times 10^{6}$ & $1.18\times  10^{6}$ & $10^{5}$ & $1.48\times 10^{5}$ & $1.43\times 10^{4}$\\
Seeding prescriptions                                           &  $M_{\rm h}/\rm M_{\odot}\geqslant$ & $M_{\rm h}/\rm M_{\odot}\geqslant$& $M_{\rm h}/\rm M_{\odot}\geqslant$ & $n\geqslant 0.1\, \rm H/cm^{3}$ & $M_{\rm h}/\rm M_{\odot}\geqslant$ & $M_{\star}/\rm M_{\odot}>$\\
&   									$7.1\times 10^{10}$ & $7.4\times 10^{10}$ & $7.4\times 10^{10}$ & $\sigma \geqslant 100\, \rm km/s$ & $1.48\times 10^{10}$ & $10^{9.5}$\\
\hline
Radiative efficiency $\epsilon_{\rm r}$                  & 0.2 & 0.2 & 0.2 & 0.1 & 0.1 & 0.1\\

\hline
{\bf Accretion} &&&&&\\
Model & Bondi  & Bondi + mag. field & Bondi + mag. field & Bondi  & Bondi + visc. & Bondi + torques\\
Boost factor & $\alpha=100$ & - & - & density-dependent  & -  & $\alpha=0.1$ \\
\hline

{\bf SN feedback}&&&&&\\
Model & kinetic & kinetic & kinetic & kinetic/thermal  & thermal & kinetic \\
\hline
{\bf AGN feedback} &&&&&\\
Single or 2 modes &2 modes &2 modes&2 modes&2 modes& single mode & 2 modes\\
High acc rate model & isotropic thermal  & isotropic thermal & isotropic thermal & isotropic thermal & isotropic thermal & kinetic \\
Feedback efficiency & $0.05\times0.2=0.01$ & $0.1\times0.2=0.02$ & $0.1\times0.2=0.02$ &$0.15\times0.1=0.015$& $0.1\times 0.15=0.015$ & $0.03\times0.1=0.003$\\

Low acc rate model & thermal hot bubble & pure kinetic winds & pure kinetic winds & kinetic bicanonical winds & -  & kinetic/ X-ray\\
Feedback efficiency & $0.35\times 0.2=0.07$ & $\leqslant0.2\times0.2= 0.04$ & $\leqslant0.2\times0.2= 0.04$ & $1\times0.1=0.1$ & - & $0.3\times0.1=0.03$ \\

Transition btw. modes  & $f_{\rm Edd}=0.05$ & $\min(0.002\left(\frac{M_{\rm BH}}{10^{8}\, \rm M_{\odot}}\right)^{2}, 0.1)$ & $\min(0.002\left(\frac{M_{\rm BH}}{10^{8}\, \rm M_{\odot}}\right)^{2}, 0.1)$ & $0.01$ & - &$0.2$ \\
\hline

\hline
\end {tabular}
\end{center}
\label{table:table_params}
\end{table}
\end{landscape}

\begin{figure*}
\centering
\includegraphics[scale=0.445]{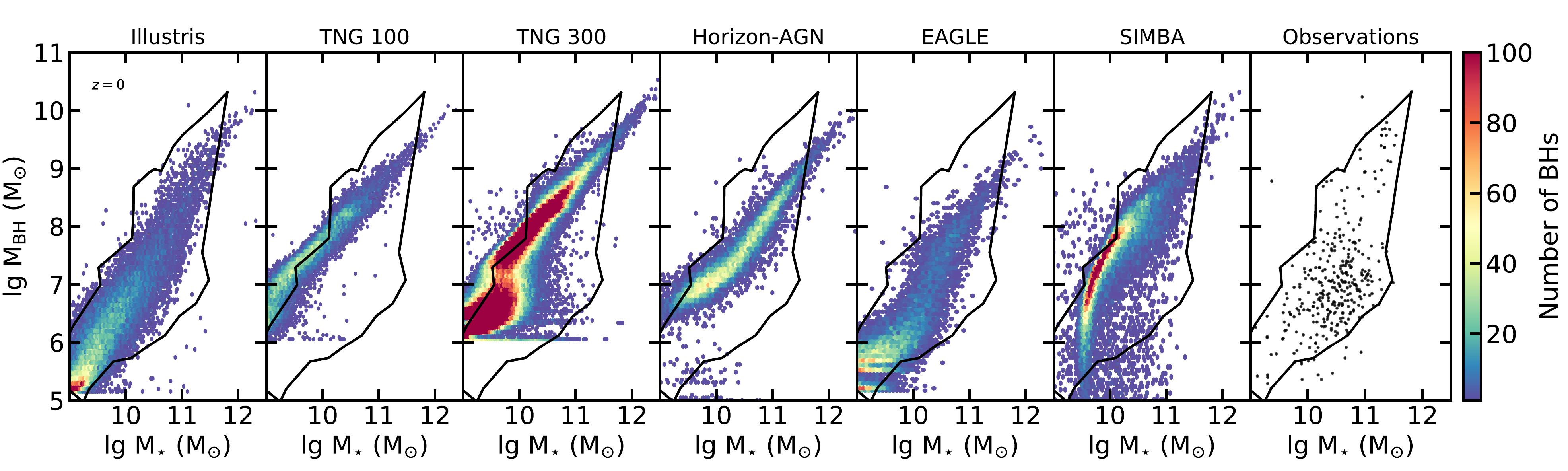}
\caption{BH populations in the $M_{\rm BH}-M_{\star}$ diagram of the simulations at $z=0$. Hexabins are color coded by the number of BHs in each bin. 
We show the observational sample of \citet{2015arXiv150806274R} for the local Universe ($z\sim0$) in black dots in the last panel on the right (uncertainties on BH masses are $\sim 0.5\, \rm dex$, and $\sim 0.3\, \rm dex$ for the stellar masses). To guide the eye, we define and duplicate in each panel a {\it cartoon} region of the diagram including most of the observations. Two main discrepancies emerge from this figure: some simulations do not produce the most massive BHs observed in galaxies with $M_{\star}\sim 10^{11}\,\rm M_{\odot}$ and the broad-line AGN of $M_{\rm BH}\sim 10^{6-7}\, \rm M_{\odot}$ observed in galaxies of $M_{\star}\sim 10^{10.5-11}\,\rm M_{\odot}$.}
\label{fig:scaling_relation}
\end{figure*}

\subsection{$M_{\rm BH}-M_{\star}$ diagrams at $z=0$ in simulations}
We present the $M_{\rm BH}-M_{\star}$ diagrams of the simulations at $z=0$ in Fig.~\ref{fig:scaling_relation} in hexabins color-coded by the number of BHs in each bin.
For comparison, we show the observational sample of \citet{2015arXiv150806274R} in the right panel of the figure. Their AGN sample from SDSS goes out to $z\leqslant 0.055$, corresponding to a distance of $\sim 230 \, \rm Mpc$ (similar to the simulations length) but does not cover the entire sky. 
To guide the eye, we report these observations with a black region in all the panels. This black region is not rigorously defined, but is drawn by eye to include all the observations except a few isolated data points. The region purposely looks like a cartoon to show that we do not make any analysis of the observations here. 

The simulated populations of BHs/galaxies form a tight correlation between BH mass and the stellar mass of the host galaxies in the diagrams, and in most cases, the scatter of the observed BH population is not fully reproduced in the simulations. Here, we list some of the differences between simulated and observed samples of BHs.
\begin{itemize}
\item The broad-line AGN located in the region enclosed within $M_{\rm BH}=10^{6-8}\rm \, M_{\odot}$ and $M_{\star}=10^{10.5-11}\, \rm M_{\odot}$ are not very well reproduced by the simulations, in particular by the TNG100 and the Horizon-AGN simulations.

\item The BHs of $M_{\rm BH}=10^{7}\,\rm M_{\odot}$ in galaxies of $M_{\star}\geqslant 10^{11}\, \rm M_{\odot}$ are not produced by any of the simulations, with the exception of SIMBA and TNG300.

\item Similarly, some of the simulations, such as TNG100, Horizon-AGN, and especially EAGLE, also have a hard time producing the most massive BHs that we observe in galaxies with total stellar mass of a few $10^{10}\, \rm M_{\odot}$ and $10^{11}\, \rm M_{\odot}$. These BHs are the most massive BHs at fixed stellar mass, and their number in observations is low. Therefore, their assembly in the different simulations could be limited by the small simulated volumes ($\sim (100\,\rm cMpc)^{3}$).

\item All the simulations predict BHs of $M_{\rm BH}\sim10^{10}\, \rm M_{\odot}$ in galaxies of $M_{\star}\sim 10^{12}\, \rm M_{\odot}$ at $z=0$ (colored dots located beyond the left side of the black line region, which is constrained by this observational sample). The mass of BHs in massive galaxies is often determined by dynamical measurements, which is only possible for close systems. In observations, these rare massive galaxies are often found on the center of clusters.

\item All the simulations (with the exception of TNG) predict BHs of $M_{\rm BH}\leqslant 10^{6}\, \rm M_{\odot}$ in galaxies of $M_{\star}= 10^{10-11}\, \rm M_{\odot}$ at $z=0$, although with different number densities. SIMBA produces quite a lot of these BHs compared to the other simulations. This region of the diagram is only slightly populated in the observed plane. In the observation panel on the left of Fig.~\ref{fig:scaling_relation}, these BHs appear below the black line region. This BH mass range is not probed by the TNG model that employs a higher seeding mass.

\item Only very few BHs have been observed with $M_{\rm BH}> 10^{8}\, \rm M_{\odot}$ in galaxies of $M_{\star}<10^{10}\, \rm M_{\odot}$; only one BH is present in the sample of \citet{2015arXiv150806274R} (black dot above the black shape in this galaxy mass range). These BHs are rare in simulations. TNG300 is the simulation forming the highest number of these BHs, due to its larger volume. Additionally, we also note that the observations shown here only include a small number of BHs with $M_{\rm BH}> 10^{6.5}\, \rm M_{\odot}$ in galaxies of $M_{\star}<10^{10}\, \rm M_{\odot}$ (BHs above the black lines), while most of the simulations predict a non-negligible number of them. EAGLE is the only simulation is good agreement in this stellar and BH mass range.

\end{itemize}

Here, we compared the simulated BH populations with the observational sample of \citet{2015arXiv150806274R}, which was made to only include dynamical measurements or estimates based on single-epoch virial masses for broad-line AGN. As shown in Fig.~\ref{fig:all_scaling_relations}, this sample includes massive BHs in massive inactive galaxies, which are elliptical and spiral/S0 galaxies with classical bulges, and lower-mass BHs found in spiral and pseudo-bulge galaxies. 
Recently, \citet{2019MNRAS.487.3404B} has discussed the possibility of using only narrow emission lines to estimate BH masses (while broad lines are generally used), therefore allowing BH mass determination for obscured AGN (type II) in addition to the non-obscured AGN (type I) used in \citet{2015arXiv150806274R}. 
The sample of \citet{2019MNRAS.487.3404B} covered the region between the lower-mass broad-line AGN of \citet{2015arXiv150806274R} and its dynamical mass measurement BHs, i.e. the region defined by $M_{\rm BH}=10^{7}-10^{9}\, \rm M_{\odot}$ BHs in $M_{\star}\geqslant 10^{10}\, \rm M_{\odot}$ galaxies.
The sample of \citet{2019MNRAS.487.3404B} confirms that the simulations are generally missing some of the most massive BHs at fixed stellar mass below $M_{\star}=10^{11}\, \rm M_{\odot}$, and confirms that compared to the observations the simulations do not produce enough BHs in the bottom right side of the $M_{\rm BH}-M_{\star}$ diagrams, i.e. BHs of $M_{\rm BH}=10^{7}-10^{8.5}\, \rm M_{\odot}$ in galaxies of $M_{\star}\geqslant 10^{11}\, \rm M_{\odot}$.
\citet{2015arXiv150806274R} establish two distinct scaling relations for massive BHs in quiescent elliptical galaxies and lower-mass BHs observed as AGN, showing that these populations may be distinct populations of BHs. They also discuss the possibility that they could also be sub-populations of the global BH-galaxy population, and that e.g., quiescent low-mass BHs could overlap with the AGN but would simply not be detectable. 
The work of \citet{2019MNRAS.487.3404B} reinforces this idea, and shows that AGN (obscured AGN) in massive galaxies can overlap with quiescent BHs in quiescent elliptical galaxies.
Whether these populations are part of the same population will become clearer in time as we increase our ability to accurately measure BH masses in the local Universe.

\subsection{Sub-grid modeling features in the $M_{\rm BH}-M_{\star}$ diagrams}
In Fig.~\ref{fig:scaling_relation_Lbol}, we show the $M_{\rm BH}-M_{\star}$ diagrams color-coded by BH bolometric luminosity. While we do not intend to analyze in detail the accretion properties of the AGN in this paper (this is the focus of the second paper of our series), the accretion rates and luminosities are important to understand the different processes involved in the evolution of the BH populations. Below, we describe some specific aspects of the BH sub-grid physics that can be identified on Fig.~\ref{fig:scaling_relation_Lbol}.

\subsubsection{On the seeding of BHs}
\indent The high seeding mass of the TNG simulations can be easily seen on Fig.~\ref{fig:scaling_relation_Lbol}: no BHs form with masses lower than $10^{6}\, \rm M_{\odot}$. As indicated in the last bottom left panel of Fig.~\ref{fig:scaling_relation}, lower-mass BHs have been observed in the local Universe. These BHs of $M_{\rm BH}\sim 10^{4-5}\, \rm M_{\odot}$ are detected as AGN in dwarf galaxies of $M_{\star}\sim 10^{9}\, \rm M_{\odot}$ \citep{2013ApJ...775..116R,2015ApJ...809L..14B,2015arXiv150806274R,2016ApJ...817...20M}; probably only the most massive/luminous BHs are detected. 
The presence of BHs in dwarf galaxies is also not covered in the simulation SIMBA, which starts forming BHs in galaxies with $M_{\star}\geqslant 10^{9.5}\, \rm M_{\odot}$.
While modeling BH formation in low-mass galaxies is crucial to understand BH formation in the high-redshift Universe, as well as to understand the current populations of BHs in local dwarf galaxies, the regime of low-mass galaxies is barely resolved in such large-scale simulations of $\geqslant 100\, \rm cMpc$ on a side \citep[but see][and references therein for BH formation in low-mass galaxies from high redshift to low redshift]{2017MNRAS.468.3935H}. 
All the other simulations employ lower seeding masses than the TNG simulations. In Horizon-AGN fewer BHs of $M_{\rm BH}\leqslant 10^{6}\, \rm M_{\odot}$ are present at $z=0$ than at higher redshift. This is because BH formation stops at $z=1.5$ by design in this simulation.

In simulations, the initial mass of BHs was shown to affect the low-mass end of the $M_{\rm BH}-M_{\star}$ diagram, and the overall normalization of the scaling relation \citep[e.g.,][]{2017MNRAS.465...32B}. 
For higher stellar masses, the relations for different seeding masses converge to the same relation (whose normalization depends on the BH accretion efficiency parameter) as a result of self-regulation.
The seed mass is important for the simulations using the Bondi accretion model (e.g., Illustris, TNG, Horizon-AGN, EAGLE), for which there is a degeneracy between the seeding mass and the boost factor to produce the same normalization of the $M_{\rm BH}-M_{\star}$ mean relation. 
In simulations using a torque accretion model (SIMBA), the accretion rate onto the BHs does not strongly depend on their masses ($\dot{M}_{\rm BH}\propto M_{\rm BH}^{1/6}$, contrary to the Bondi model $\dot{M}_{\rm BH}\propto M_{\rm BH}^{2}$). Therefore, the low-mass BHs can grow more efficiently than in the Bondi model, and they converge into the $M_{\rm BH}-M_{\star}$ mean relation regardless of their seed mass \citep{2013ApJ...770....5A,2015ApJ...800..127A}. We point here again that the seeding in large-scale cosmological simulations is often not physically motivated (e.g., fixed BH mass, threshold in galaxy or halo mass to form a BH), but instead parameters of the seeding models are chosen to reproduce the local scaling relation.

\subsubsection{On the growth of BHs}
\indent Getting the BHs on the main scaling relation depends both on the seeding BH mass and BH accretion rate. The parametrizations of these models are degenerate. Therefore, different choices allow the BH population to get on to the same empirical scaling relation. For example, the TNG model seeds BHs with a higher initial BH mass than the Illustris model (by one order of magnitude) but also does not include any boost factor in the accretion model (while Illustris does include a boost factor of 100).

The color code of Fig.~\ref{fig:scaling_relation_Lbol} provides information on the accretion rates onto the BHs.
At $z\sim2$, it is clear that the accretion properties are very different among the different simulations. For example, BHs of $M_{\rm BH}\sim 10^{6}\, \rm M_{\odot}$ embedded in galaxies of $M_{\star}\sim 10^{10}\, \rm M_{\odot}$ have high bolometric luminosities of $L_{\rm bol}\geqslant 10^{44}\, \rm erg/s$ (limit commonly employed to define an AGN) in TNG, Horizon-AGN and SIMBA, while a large fraction of these BHs in Illustris and EAGLE have much lower luminosities.
Interestingly, we see that the initial growth of BHs in SIMBA is quite efficient, and it allows them to get rapidly on the track of the main scaling relation even if only galaxies of $M_{\star}\geqslant 10^{9.5}\, \rm M_{\odot}$ get seeded. 
The accretion model of SIMBA based on gravitational torques is more efficient in the regime of these low-mass BHs than the Bondi accretion (which scales as $M_{\rm BH} ^{2}$) \citep{2015ApJ...800..127A,2017MNRAS.464.2840A}.
In SIMBA, the torque model and the Bondi model always co-exist, but in practise the torque model dominates at early times in gas rich galaxies: cold gas dominates the gas reservoir in the center of galaxies, and there is only little hot gas to be accreted through the Bondi model (Angles-Alcazar, in prep).

\subsubsection{On AGN feedback}
\begin{figure*}
\centering
\includegraphics[scale=0.548]{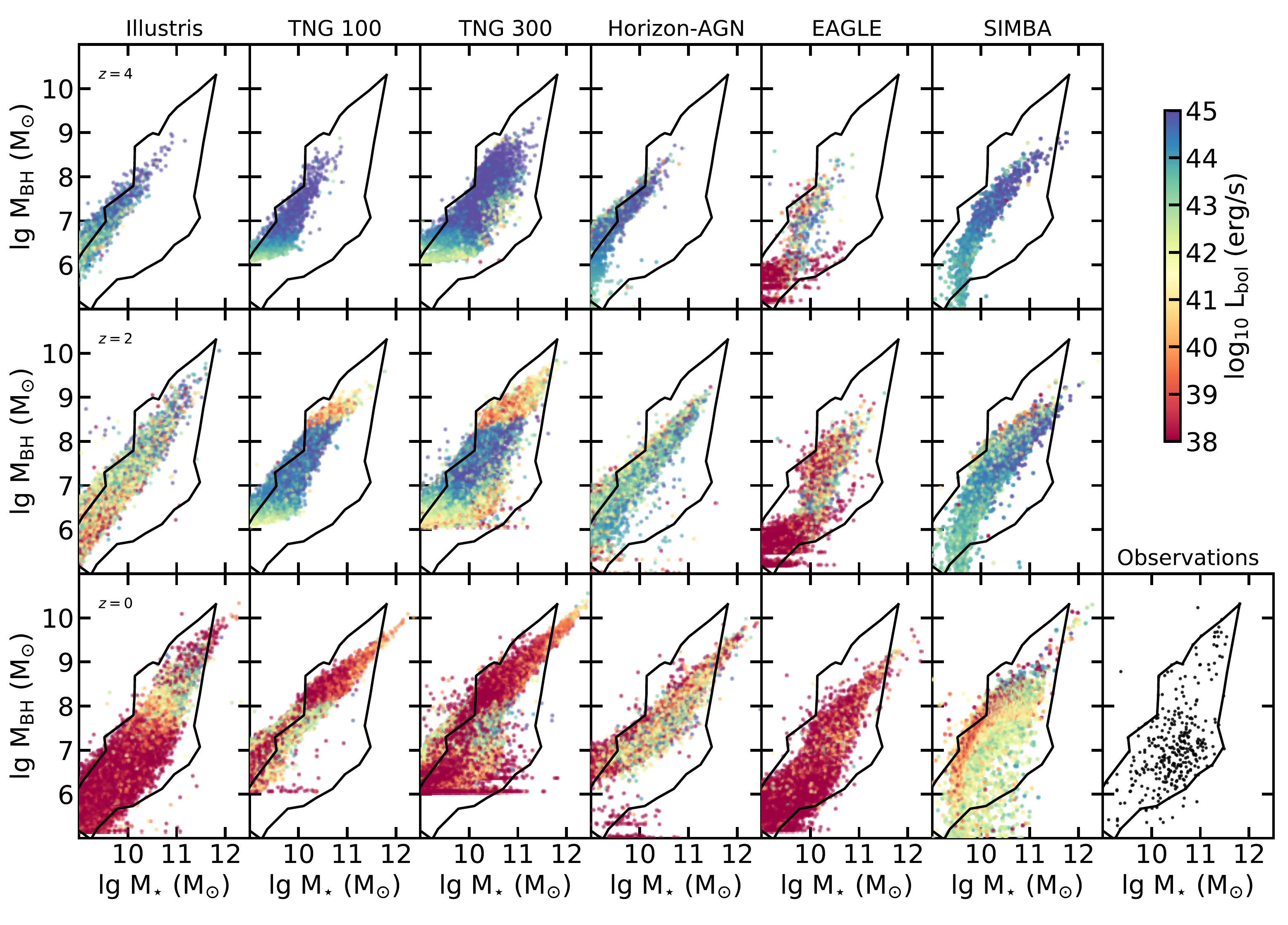}
\caption{Time evolution of the BH population in the $M_{\rm BH}-M_{\star}$ diagram. From the top panels to the bottoms panels, we show the populations at $z=4,2,0$ in hexabins color-coded by the bolometric luminosity of the BHs (in erg/s).
For reference, we show the black cartoon region representing the observations at $z=0$ in each panel.
The simulations have different time evolution of the accretion properties of their BHs. In the TNG panels, we can identify a strong signature of the BH mass dependence of the AGN feedback modeling: BHs with mass lower than $M_{\rm BH}\sim 10^{8}\, \rm M_{\odot}$ are under the thermal mode of AGN feedback and still able to accrete gas, while the accretion of more massive BHs is strongly stunted by the effective kinetic AGN feedback mode.}
\label{fig:scaling_relation_Lbol}
\end{figure*}

\indent In Fig.~\ref{fig:scaling_relation_Lbol}, we can identify a clear cut at fixed BH mass in the bolometric luminosity of BHs in the TNG100 and TNG300 simulations (at $M_{\rm BH}\sim 10^{8}\, \rm M_{\odot}$). Above this characteristic BH mass the accretion rate onto the BHs and consequently their luminosity is strongly reduced. 
In TNG, most of the $M_{\rm BH}\sim 10^{8}\, \rm M_{\odot}$ BHs with low accretion rates transition from the thermal {\it high accretion} feedback mode to the kinetic wind {\it low accretion} feedback mode, which is the mode responsible for efficiently quenching massive galaxies \citep{2018MNRAS.479.4056W,2019MNRAS.484.4413H}.

In EAGLE, we note the presence of faint AGN or inactive BHs with $L_{\rm bol}\leqslant 10^{38}\, \rm erg/s$ and $M_{\rm BH}\geqslant 10^{7}\, \rm M_{\odot}$ in galaxies of $M_{\star}\sim 10^{10}\, \rm M_{\odot}$ (mostly visible at $z=4,2$). This is due to AGN feedback.
BHs in the EAGLE simulation are first regulated by the efficient SN feedback in low-mass galaxies of $M_{\star}\leqslant 10^{10}\, \rm M_{\odot}$, and then have a period of efficient growth, before being regulated by AGN feedback in galaxies of $M_{\star}\geqslant 10^{10}\, \rm M_{\odot}$ \citep{2017MNRAS.468.3395M,2017MNRAS.465...32B,2018MNRAS.481.3118M}.

In the other simulations, the different AGN feedback models do not produce any strong signature in the $M_{\rm BH}-M_{\star}$ diagrams.
We note a small signature in SIMBA at $z=0$, i.e. a weak horizontal line of demarcation at $M_{\rm BH}\geqslant 10^{7.5}\, \rm M_{\odot}$ where BHs seem to have lower luminosities on average.
In SIMBA, only BHs of $M_{\rm BH}\geqslant 10^{7.5}\, \rm M_{\odot}$ and with accretion rates such that $f_{\rm Edd}\leqslant 0.2$ are allowed to enter the efficient kinetic mode of AGN feedback. Once they enter this mode, BHs regulate themselves, as well as the star formation in their host galaxies \citep{2019MNRAS.487.5764T} similarly to TNG.

\subsection{$M_{\rm BH}-M_{\star}$ diagrams for star-forming and quiescent galaxies}
\begin{figure*}
\centering
\includegraphics[scale=0.44]{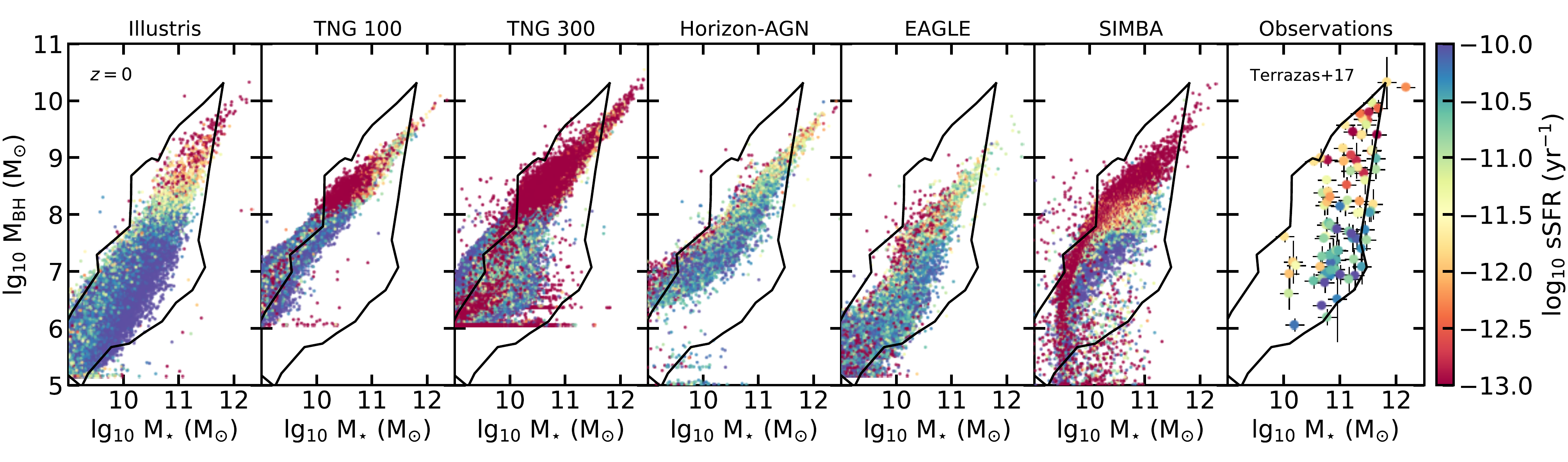}
\caption{$M_{\rm BH}-M_{\star}$ diagram at $z=0$ for all the simulations. We color-code the simulated BHs with the specific star formation rate (sSFR) of their host galaxies. We set $\rm sSFR=10^{-13}\, yr^{-1}$ for galaxies with lower sSFR. The limit $\rm sSFR=10^{-11}\, \rm yr^{-1}$ is often use to define star-forming galaxies (with higher sSFR) or quiescent galaxies (with lower sSFR). In the figure, galaxies with blue colors are forming stars, and yellow to red galaxies are quiescent. 
We show in the last right panel the observed star-forming and quiescent galaxies of \citet{2017ApJ...844..170T} for references, color-coded by their sSFR as well. To guide the eye, we again reproduce the black region representing the observational sample of \citet{2015arXiv150806274R} in each panel. A similar figure (Fig.~\ref{fig:scaling_sf_q_bis}) shows the behavior at higher redshifts.}
\label{fig:scaling_sf_q}
\end{figure*}

In Fig.~\ref{fig:scaling_sf_q}, we reproduce again the $M_{\rm BH}-M_{\star}$ diagram at $z=0$ for the different simulations, but this time we color-code the BHs by the specific star formation rate (sSFR) of their host galaxies. In the literature, we often use the convenient limit $\rm sSFR=10^{-11}\, \rm yr^{-1}$ to separate star-forming galaxies from galaxies with low star formation rate or quenched galaxies (i.e., with low SFR sustained in time). In Fig.~\ref{fig:scaling_sf_q}, star-forming galaxies appear in bluish colors and quiescent galaxies in reddish ones. On the last panel, we show the sample of \citet{2017ApJ...844..170T}\footnote{We already compared this sample to the TNG simulations in \citet{2019arXiv190602747T}.} that only include BHs with dynamical mass measurements, and galaxies with SFR estimated from far-infrared fluxes (from IRAS).

We find a good qualitative agreement between the observations and the simulations. 
In \citet{2017ApJ...844..170T}, observed galaxies of $M_{\star}=10^{10}-10^{12}\, \rm M_{\odot}$ hosting the most massive BHs ($M_{\rm BH}\geqslant 10^{8.5}\, \rm M_{\odot}$) have lower sSFR ($\rm sSFR\leqslant 10^{-11}\, yr^{-1}$). Similarly, in \citet{2015arXiv150806274R} these galaxies are ellipticals, and galaxies with classical bulges (mostly early-type galaxies).
In the simulations, these galaxies also tend to have low specific star formation rates, and to be quenched. 
We note that some of the simulations have a large fraction of quenched galaxies with $\rm sSFR\leqslant 10^{-13}\, \rm yr^{-1}$ (Illustris, TNG, SIMBA), while the other simulations (Horizon-AGN, EAGLE) have quenched galaxies with a broader range of sSFR ($\rm 10^{-13}\leqslant sSFR\leqslant 10^{-11}\, \rm yr^{-1}$). While all simulations rely on AGN feedback to solve the overcooling process in massive galaxies and produce a galaxy stellar mass function in good agreement with observations, they all use different feedback modelings. Consequently and as seen here,  these modelings produce different behaviors and level of quenching in the simulations.

In the sample of \citet{2017ApJ...844..170T}, most of the galaxies with lower-mass BHs ($M_{\rm BH}\leqslant 10^{8}\, \rm M_{\odot}$) are forming stars more efficiently and have higher sSFR of $\rm sSFR> 10^{-11}\, yr^{-1}$. This is true for galaxies with $M_{\star}\geqslant 10^{10.5}\, \rm M_{\odot}$, and we note that, in this sample, galaxies with stellar masses of $M_{\star}\sim 10^{10}\, \rm M_{\odot}$ have lower sSFR. In the sample of \citet{2015arXiv150806274R} these galaxies are also forming stars as they are mostly spiral galaxies and pseudo-bulges galaxies (found in late-type galaxies). 

As we discussed earlier in this paper, the region of the $M_{\rm BH}-M_{\star}$ diagram corresponding to the low-mass BHs in the observational samples of \citet{2015arXiv150806274R} and \citet{2017ApJ...844..170T} does not overlap very well with the low-mass BHs formed in the simulations. 
In the observations BHs of $M_{\rm BH}=10^{6-8}\, \rm M_{\odot}$ are found in galaxies of $M_{\star}=10^{10.5-11.5}\, \rm M_{\odot}$, while in the simulations these BHs are generally found in less massive galaxies. If we compare the sSFR of these observed galaxies with BHs of $M_{\rm BH}=10^{6-8}\, \rm M_{\odot}$ with the sSFR of lower stellar mass galaxies with the same-mass BHs, we find a good agreement. Moreover, we note that some of the same-mass BHs in even lower-mass galaxies also have lower sSFR in the simulations, as found in the observations of \citet{2017ApJ...844..170T}.\\

To summarize, we do find a good qualitative agreement between the sSFR properties of the simulated and observed galaxies hosting the most massive BHs, and we find the same trend for lower-mass BH host galaxies (higher sSFR for BHs in more massive galaxies) but the observed and simulated galaxies populate different regions of the $M_{\rm BH}-M_{\star}$ diagram.

We do not aim at comparing in detail the samples of \citet{2017ApJ...844..170T} and \citet{2015arXiv150806274R}, but we note some correlations in the following.
The quenched galaxies of \citet{2017ApJ...844..170T} overlap with the inactive galaxies of \citet{2015arXiv150806274R} (elliptical and spiral/S0 galaxies), whose BHs are also inactive and have dynamical mass measurement.
BHs with lower masses in \citet{2015arXiv150806274R} are mostly AGN, whose masses are estimated from broad-line emission, and BHs found in spiral and pseudo-bulge galaxies; they cover the same region as the star-forming galaxies of \citet{2017ApJ...844..170T}.
From the comparison of these two samples, we note a correlation between quenched galaxies and inactive massive BHs, and similarly between lower-mass BHs detected as AGN and star-forming galaxies.  
Such correlations need to be taken with a grain of salt, as distinct regions of the $M_{\rm BH}-M_{\star}$ diagram may be populated by different selections and methods of mass measurements/estimates. Mass estimates for the AGN can be based on empirical relations derived from specific galaxy populations (e.g., star-forming galaxies). While BHs with mass dynamical measurements can also be biased to galaxies with e.g., higher velocity dispersion \citep[][and references therein]{Bernardi2007,2016MNRAS.460.3119S}.

We now describe what we find in the simulations at $z=0$ (in the last row of Fig.~\ref{fig:scaling_relation_Lbol}). In the TNG and EAGLE simulations, the massive BHs in massive galaxies ($M_{\rm BH}\geqslant 10^{8.5}\, \rm M_{\odot}$, $M_{\star}\geqslant 10^{10.5}\, $) tend to have low accretion rates. In Illustris, Horizon-AGN, we find both non-active and active BHs among the massive BHs. There is also a significant high fraction of massive BHs with high accretion rates in SIMBA.
Comparing Fig.~\ref{fig:scaling_relation_Lbol} and Fig.~\ref{fig:scaling_sf_q}, we find a clear correlation between quenched galaxies and non-active massive BHs in Illustris, TNG, as in some of the observations mentioned above. The correlation is less obvious in Horizon-AGN and EAGLE. The massive galaxies at $z=0$ in SIMBA can both feed the massive BHs and have low sSFR.
The correlation stands for lower-mass BHs in the TNG simulations, in Horizon-AGN, in SIMBA, i.e. the host galaxies of lower-mass BHs form stars more efficiently. The picture seems harder to establish for the low-mass galaxies in the Illustris and EAGLE simulations. 
Here we only look at the instantaneous accretion rates onto the BHs and luminosities, but the AGN luminosities can vary substantially within short amounts of time \citep[e.g., Fig. 1 of][]{2019MNRAS.483.2712R}. Correlations between star formation rates and average AGN luminosities over a few hundreds Myr could be stronger.
In a forthcoming paper of our series, we will study the correlation between BH activity and star formation of the host galaxies in a more quantitative way. In the following section, we discuss in more details the time evolution, normalization, scatter, and shape of the $M_{\rm BH}-M_{\star}$ relation in the simulations. 

\begin{figure*}
\centering
\includegraphics[scale=0.528]{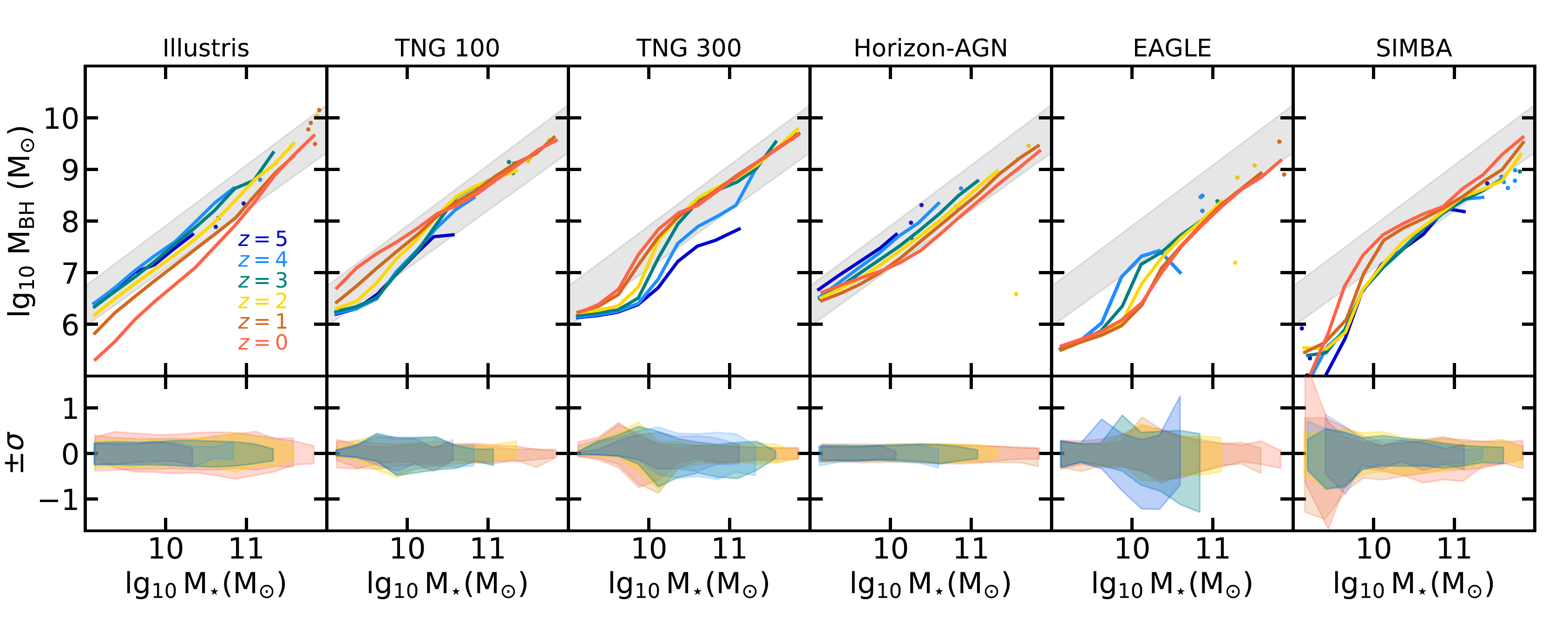}
\caption{{\it Top panels}: Time evolution of the median of the $M_{\rm BH}-M_{\star}$ relation for all the simulations (the mean relations are almost identical to the median relations). The medians are shown in stellar mass bins if they include more than 5 galaxies, otherwise we simply show the points. The grey shaded area in each panels encloses the empirical scaling relations derived at $z=0$ in \citet{2013ARA&A..51..511K} (higher normalization of the scaling relation), \citet{2013ApJ...764..184M}, and \citet{Haring2004} (lower normalization of the relation).
{\it Bottom panels}: 15-85th percentile of the distributions. 
While in general the median of the $M_{\rm BH}-M_{\star}$ relation does not evolve much with time, we do note some differences between the simulations. These differences can also be appreciated in the previous figure. Some simulations increase their scatter toward lower mass BHs with time (at fixed stellar mass), i.e., the median relation moves down in the $M_{\rm BH}-M_{\star}$ diagram for lower redshift. This is true for Illustris, Horizon-AGN, and EAGLE. However, the TNG and the SIMBA simulations evolve upward in the diagram.}
\label{fig:scaling_relation_mean}
\end{figure*}

In Fig.~\ref{fig:scaling_sf_q}, all the simulations seem to favor a linear $M_{\rm BH}-M_{\star}$ relation for galaxies of $M_{\star}\geqslant 10^{11}\, \rm M_{\odot}$, a mass range dominated by quiescent galaxies (in red). However, for lower-mass galaxies which are dominated by star-forming galaxies, a linear relation is not supported by all the simulations. Observationally, this correlation between the quiescence of the host galaxies and the shape of the $M_{\rm BH}-M_{\star}$ relation has been discussed \citep[e.g.,][]{2015ApJ...798...54G}. Different slopes of the scaling relation have been found for samples of early-type/core-Sersic and late-type/Sersic galaxies in observations \citep[][]{2018ApJ...869..113D,2019ApJ...887...10S}. In the next section, we investigate the time evolution, scatter, shape and normalization of the $M_{\rm BH}-M_{\star}$ relation.

\section{Scaling relations and evolution with time}
We now turn to study the evolution of the $M_{\rm BH}-M_{\star}$ relation and its scatter with time. Our goal is to identify the different features seen in the simulations; the physical interpretations of these features is given in the next section. 
When possible, we show observational constraints to guide the eye, but we do not aim at building apple-to-apple comparisons between the observations and simulations.

\subsection{Time evolution of the $M_{\rm BH}-M_{\star}$ relation}
We show in the top panels of Fig.~\ref{fig:scaling_relation_mean} the {\it median} of the $M_{\rm BH}-M_{\star}$ relation for several redshifts. We obtain almost identical relations for the {\it mean} relations. 
For reference, we show a grey shaded area in each panel enclosing the local $M_{\rm BH}-M_{\rm bulge}$ empirical relations of \citet{2013ARA&A..51..511K,2013ApJ...764..184M,Haring2004}, which were used to calibrate the simulations (with different assumptions for $M_{\rm bulge}$). In Fig.~\ref{fig:scaling_relation_mean} we show the total stellar mass of the simulated galaxies $M_{\star}$, which should be comparable to the bulge mass for the highest-mass galaxies.
In the bottom panels, we show the 15th-85th percentiles of the $M_{\rm BH}$ distributions in stellar mass bins.
Please note that in this section we use the same terminology {\it scaling relation} for the mean/median of the $M_{\rm BH}-M_{\star}$ population in the simulations and for the empirical scaling relations (which are not simply defined by the mean/median of the observational samples). 

We find that the scaling relations of all the simulations evolve with time. However, the evolution of the overall normalization is small, particularly in the redshift range that can currently be probed by observations $0\leqslant z \leqslant 2$ (red, yellow, and brown lines in Fig.~\ref{fig:scaling_relation_mean}). In this redshift range, the time evolution is smaller than one order of magnitude in BH mass. 
The strongest evolution is found for the Illustris simulation for these redshifts and for $M_{\star}\leqslant 10^{11}\, \rm M_{\odot}$.

When looking at the time evolution of the overall normalization, two trends emerge. Half of the simulations have a lower overall median $M_{\rm BH}-M_{\star}$ relation with decreasing redshift; this is the case for the Illustris, Horizon-AGN, and EAGLE simulations. In the EAGLE simulation, the relation only evolves with time for intermediate-mass galaxies. The other TNG and SIMBA simulations have scaling relations with higher overall normalization.

The time evolution of the scaling relation is likely the result of several physical processes affecting the growth of both the BHs and their host galaxies. Among these processes, the ability of BHs to accrete gas, the feedback of the AGN and of SNe, the growth of the host galaxies, can play a crucial role. 
There are different hypotheses to explain the two trends in the $M_{\rm BH}-M_{\star}$ diagram.
A higher overall normalization at higher redshift (Illustris, Horizon-AGN, EAGLE) can be the signature of a more rapid growth of the BHs at higher redshifts. Theoretically, we expect galaxies to be gas-rich and more compact at higher redshift, which favors efficient BH accretion.
While at lower redshift (e.g., $z\leqslant 1$), BH growth could be less efficient.
The second hypothesis is that BH growth has a constant efficiency with time, but that galaxies have a relative growth faster at low redshift than at high redshift compared to their BHs. It would lead to a shift of the $M_{\rm BH}-M_{\star}$ relation towards more massive galaxies at lower redshifts.
An increase of the overall scaling relations with time (TNG, SIMBA) could be explained by a relative more efficient growth of BHs at low redshifts with respect to their galaxies, or by a smaller growth of the galaxies with respect to their BHs.
We investigate how these hypotheses could drive the two evolution patterns found here in more detail in the following section of the paper.

For reference, most of the observational studies have concluded that the time evolution of the scaling relation between $z\sim 2$ and $z\sim 0$ was consistent with no evolution \citep{2003ApJ...583..124S,2009ApJ...706L.215J,2011ApJ...741L..11C,2013ApJ...767...13S,2013ApJ...764...80S,2015ApJ...802...14S,2019arXiv191202824S}. A few studies have shown that the ratio $M_{\rm BH}/M_{\star}$ could be larger at higher redshifts \citep[e.g.,][]{2006MNRAS.368.1395M,2019arXiv191011875D}. For example, \citet{2019arXiv191011875D} find that the small evolution in their sample (32 X-ray selected broad-line AGN, $1.2<z<1.7$) goes into this direction, i.e., higher BH masses or lower stellar masses at higher redshift.

Finally, studying the time evolution of the scaling relation for the most massive galaxies of $M_{\star}\geqslant 10^{11}\, \rm M_{\odot}$ is problematic with the current large-scale simulations of side length $\sim 100\,\rm cMpc$, as they suffer from low number statistics. 
The large volume of TNG300 captures the evolution of a higher number of massive galaxies, especially at higher redshift. Indeed, there are only 270 galaxies in TNG100 and 4378 in TNG300 with $10\leqslant \log_{10}\, M_{\star}/\rm M_{\odot}\leqslant 11$ at $z=4$. Only two galaxies with $\log_{10}\, M_{\star}/\rm M_{\odot}\geqslant 11$ in TNG100, and 66 in TNG300. We find that there is a rapid evolution of the massive end ($\log_{10}\, M_{\star}/\rm M_{\odot}\geqslant 10.5$) of the relation for $3\leqslant z\leqslant 5$ in TNG300, a regime that is not accurately covered by the other simulations\footnote{SIMBA (147 cMpc box length) also produces a high number of massive galaxies, e.g. 30 galaxies of  $\log_{10}\, M_{\star}/\rm M_{\odot}\geqslant 11$ at $z=4$.}. At $z\leqslant 3$, we note the absence of an evolution of the relation with time. The BHs embedded in these massive galaxies are regulated by the efficient kinetic feedback. Therefore, their growth is likely driven by mergers only (and not by gas accretion); the growth of their host galaxies is likely also mostly driven by mergers. From the central-limit theorem we expect the $M_{\rm BH}-M_{\star}$ relation to be mostly independent of redshift for the most massive galaxies, and the distributions to have a smaller scatter. 
This is true for the TNG simulations \citep{2017MNRAS.465.3291W,2018MNRAS.479.4056W}, and can also be seen in the other simulations with an efficient AGN feedback even if lacking statistics for massive galaxies. 
In SIMBA, we find an evolution of the relation for these massive galaxies of $M_{\star}\geqslant 10^{11}\, \rm M_{\odot}$: BH growth (through the Bondi accretion channel) exceeds the relative galaxy stellar mass growth by mergers (Cui et al. in prep).

\subsection{Evolution of the scatter of the scaling relation}
We now turn to study the evolution of the scatter of the $M_{\rm BH}-M_{\star}$ relation with both with stellar mass and redshift. We define the scatter as the 15th-85th percentiles of the BH mass distributions in stellar mass bins. We provide the stellar mass bins (bin size is 0.4 dex), redshifts, mean and median BH mass in the bins, and the 15th-85th percentiles in Table~\ref{table:tab_scatter_1} and Table~\ref{table:tab_scatter_2}. We split the tables in two: simulations with an increasing $M_{\rm BH}-M_{\star}$ overall normalization with decreasing redshift (Illustris, Horizon-AGN, EAGLE), and those with a decreasing overall normalization (TNG100, SIMBA).
We find some differences between all of these simulations. For most of the simulations, the scatter of the scaling relation is below 1 dex in $\log_{10} M_{\rm BH}/\rm M_{\odot}$, except for EAGLE which has the largest scatter (e.g., $> 1 \, \rm dex$ for $M_{\star}=10^{10}, 10^{10.5}, 10^{11}\, \rm M_{\odot}$ for $z\leqslant 3$). Horizon-AGN has the smallest scatter, below half a dex in $\log_{10} M_{\rm BH}/\rm M_{\odot}$ (for all redshifts and $M_{\star}$ bins).

\subsubsection{Evolution of the scatter with time} 
For galaxies of $M_{\star}\leqslant 10^{11}\, \rm M_{\odot}$, the scatter in Illustris and TNG100 generally increases with decreasing redshift to $z=0$. The scatter decreases in EAGLE with decreasing redshift, slightly oscillates in SIMBA for different redshifts, and does not evolve with redshift in Horizon-AGN. For all the simulations, the evolution of the scatter with redshift is smaller than 1 dex in BH mass.
We define the time variation of the scatter by the difference of the 15th-85th percentiles between two given redshifts, i.e. $\Delta \sigma=|\sigma_{\rm i}-\sigma_{\rm j}|$. 
Horizon-AGN has the smallest scatter evolution with $\Delta \sigma \leqslant 0.1\, \rm dex$, while EAGLE has a variation of almost one order of magnitude in BH mass (i.e., $\Delta \sigma=0.84 \,\rm dex$), the largest among all the simulations. 
According to the central-limit theorem (i.e., assuming that growth is driven only by mergers) any distribution with a large scatter at high redshift would have a smaller scatter with time. The mild evolution of the scatter that we find here for $M_{\star}\leqslant 10^{11}\, \rm M_{\odot}$ suggests that BH growth by gas accretion, in addition to mergers, play a role in the build-up of the scaling relation.

Comparing the time evolution of the scatter in simulations with observations is challenging. The scatter in observations is often defined by the $1\sigma$ confidence interval of the empirical scaling relations (about half a dex in $\log_{10} M_{\rm BH}/\rm M_{\odot}$), which relies more on the uncertainties of the mass estimates\footnote{Uncertainties on BH mass estimates are of about $\sim 0.5\, \rm dex$ \citep{2015arXiv150806274R,2006ApJ...641..689V}. The uncertainties depend on the method used to measure/estimate BH masses. Uncertainties are smaller for masers, reverberation mapping, dynamical measurements, and larger for single epoch measurements.} rather than the intrinsic scatter of the BH mass distribution in stellar mass bins, as derived here for the simulations. Given that it is harder to estimate BH masses accurately at higher redshift, the scatter in observations is expected to increase with increasing redshifts. Recently, the intrinsic scatter of the $M_{\rm BH}-M_{\star}$ relation was found to be similar at $z\sim 1.5$ and $z=0$ \citep[e.g.,][]{2019arXiv191011875D}. 

\subsubsection{Dependence of the scatter with stellar mass}
Variations of the scatter as a function of the stellar mass (for $M_{\star}\leqslant 10^{11}\, \rm M_{\odot}$) are small in the simulations, on average.
The amplitude of those would probably be lower than the various uncertainties in estimating BH masses and host stellar masses in observations.

Interestingly, the simulations produce different evolutions of the scatter with $M_{\star}$.
At fixed redshift, the scatter decreases with larger $M_{\star}$ in SIMBA, and slightly increases for Illustris and Horizon-AGN.
There is a large variation of the scatter in TNG100, TNG300, and EAGLE: the scatter is smaller in low-mass galaxies of $M_{\star}\leqslant 10^{9.5-10}\, \rm M_{\odot}$ (depending on the redshift), is larger for galaxies of $M_{\star}\sim 10^{10}\, \rm M_{\odot}$, and decreases for more massive galaxies. The dependence with stellar masses in these simulations correlates with the efficiency of BH growth (see next section with the shape of the $M_{\rm BH}-M_{\star}$ relation). 
In the low-mass galaxies, BH growth is not efficient and consequently a small scatter is found. However, when BHs start growing efficiently in galaxies of $M_{\star}\sim 10^{10}\, \rm M_{\odot}$ the scatter is more pronounced \citep[e.g., see][for EAGLE]{2018MNRAS.481.3118M}. Finally, when the growth of the BHs is regulated by AGN feedback in more massive galaxies the scatter decreases.
With the larger volume of TNG300 we find that the scatter for even more massive galaxies of $M_{\star}\geqslant 10^{11}\, \rm M_{\odot}$ in the redshift range $0\leqslant z\leqslant 3$ (for which we have more statistics) is even smaller.

For reference, we compute the scatter (15th-85th percentiles) of the two observational samples of  \citet{2015arXiv150806274R,2019MNRAS.487.3404B} for $M_{\star}\leqslant 10^{11}\, \rm M_{\odot}$ and $z\sim 0$. Our simple method does not reflect the scatter found in observations, and we do not add any correction (e.g., completeness of the sample, volume, low statistics). 
The scatter found in observations also probably does not represent accurately the intrinsic scatter of the entire BH population in the Universe. This is even more true at the BH low-mass end due to very low number of detections. We discuss this further in the discussion section.
We find that the percentile 15th-85th varies in the range $\sigma=0.9-1.9$ in the stellar mass range $M_{\star}=10^{9.5}-10^{11}\, \rm M_{\odot}$ for the sample of \citet{2015arXiv150806274R}, and within $\sigma=1.2,1.6$ for \citet{2019MNRAS.487.3404B}. The scatter found in the simulations is smaller than for these two observational samples with on average more than a dex in $\log_{10} M_{\rm BH}$ of scatter. The scatter increases with increasing $M_{\star}$ for \citet{2015arXiv150806274R}, from $\sigma=1.1$ ($M_{\star}=10^{9.5}\, \rm M_{\odot}$) to $\sigma=1.9$ ($M_{\star}=10^{11}\, \rm M_{\odot}$), but the scatter decreases from $\sigma=1.6$ ($M_{\star}=10^{9.5}\, \rm M_{\odot}$) to $\sigma=1.2$ ($M_{\star}=10^{11}\, \rm M_{\odot}$) for \citet{2019MNRAS.487.3404B}.

Some essential physical processes are not consistently modeled in large-scale simulations but could modulate the accretion, growth, and feedback of BHs, possibly impacting the scaling relation and its scatter.
Here we only discuss the impact of BH spin, and mention more processes in the discussion. 
BH spin is closely tied to both accretion and the energy that can be released by AGN feedback. 
BHs with $M_{\rm BH}<10^{8}\, \rm M_{\odot}$ build their mass by coherent accretion of gas and a few mergers: they have high spins \citep{2014MNRAS.440.1590D}.
More massive BHs of $M_{\rm BH}>10^{8}\, \rm M_{\odot}$ experience more mergers and less coherent accretion: they have more moderate spins \citep[see also][]{2020MNRAS.495.4681I}.
BHs with high spins will release more specific energy than non-rotating BHs, which alters as well the amount of gas that is accreted by the BHs, and consequently, their spins. 
Adding the spin evolution in simulations increases the scatter, especially in the massive end of the scaling relation, where the relation is very tight. In the TNG model, the scatter is mostly increased for stellar masses of $M_{\star}\sim 10^{10}\, \rm M_{\odot}$ and higher \citep{2019MNRAS.490.4133B}.

\subsection{Shape of the median $M_{\rm BH}-M_{\star}$ relation and presence of a characteristic mass for efficient BH growth}

The $M_{\rm BH}-M_{\star}$ relations presented in Fig.~\ref{fig:scaling_relation_mean} have different shapes, and we investigate {\it qualitatively} the presence of a change of slope in the relations for the six simulations. 

The median $M_{\rm BH}-M_{\star}$ relation of Illustris is linear at all redshifts and for all stellar masses.
The linear relationship between BH mass and galaxy total stellar mass indicates that, on average, the BHs and their host galaxies grow at similar rates. When looking at the population of BHs {\it statistically}, there is no galaxy mass regime in Illustris at which BH growth is strongly regulated with respect to the growth of the galaxies and vice-versa.
We also cannot identify any change of slope of the scaling relation of Horizon-AGN for $z\geqslant 2$. However, we do see a transition 
around $M_{\star}\sim 10^{10}\, \rm M_{\odot}$ at lower redshift. This is a signature of BH seeding: the formation of BHs stops at $z=1.5$ in Horizon-AGN, and there are no more newly formed BHs to bring down the scaling relation at lower redshift.

All the other simulations have a linear evolution in massive galaxies ($M_{\star}\geqslant 10^{10.5}\, \rm M_{\odot}$), but present a change of slope in the regime of the lower-mass galaxies ($M_{\star}\leqslant 10^{10.5}\, \rm M_{\odot}$). 
This change of slope reflects a transition between two phases: a first phase in which BHs have a hard time growing in low-mass galaxies, and a second phase for which BH growth is more efficient in more massive galaxies. At the transition between these two regimes, we can define a {\it characteristic galaxy stellar mass}, from which BHs can start growing efficiently. 
In TNG100, we note a change of slope at $z\geqslant 2$ only, while in TNG300 the change of slope is present at all redshifts. The transition between inefficient BH growth and efficient BH growth changes with redshift in the range $M_{\star}\sim 10^{9.5}-10^{10}\, \rm M_{\odot}$: the transition takes place in lower-mass galaxies with time. The inefficient BH growth phase is longer at earlier times than at later times in TNG. 
The EAGLE simulation also presents different phases of BH-galaxy growth. There is a first phase with a lower relative growth of BHs compared to their host galaxies in low-mass galaxies of $M_{\star}\leqslant 10^{9.5-10}\, \rm M_{\odot}$. Galaxies of $M_{\star}\sim 10^{10}\, \rm M_{\odot}$ experience a phase of efficient BH growth. BHs hosted by more massive galaxies of $M_{\star}\geqslant 10^{10-10.5}\, \rm M_{\odot}$ have again a less efficient growth than their galaxies.
As in TNG, the characteristic galaxy mass between the first phase and the second phase depends on redshift. However, in EAGLE the inefficient BH growth phase is shorter at earlier times than at later times.
We do not discuss the shape in SIMBA, since BH seeding takes place in galaxies of $M_{\star}=10^{9.5}\,\rm M_{\odot}$, i.e. where we could expect a change of slope. However, we note that the $M_{\rm BH}-M_{\star}$ mean relation transitions to a steeper slope at $M_{\star}\sim 10^{10.8}\, \rm M_{\odot}$ for $z\leqslant 1$.

The regime of galaxies with $M_{\star}\leqslant 10^{10}\, \rm M_{\odot}$ is important to understand the build-up of the BH population and the $M_{\rm BH}-M_{\star}$ scaling relation. 
As shown below, this regime is subject to resolution effects for the range of resolutions covered in the simulations presented here. 
In the following sections, we investigate {\it qualitatively} the physical processes responsible for the shape of the $M_{\rm BH}-M_{\star}$ relation and its evolution with time.
Before going into details, we mention here that the $M_{\rm BH}-M_{\star}$ relation can show some imprints of the strength and modeling of stellar feedback. In \citet{2017MNRAS.468.3935H}, we discussed the build-up of the BH population in the $M_{\rm BH}-M_{\star}$ diagram in large-scale cosmological hydrodynamical  simulations with a physical model of BH formation, and highlighted the ability of SN feedback to stunt BH growth in low-mass galaxies of $M_{\star}\leqslant 10^{9}\, \rm M_{\odot}$ \citep[see also,][]{2015MNRAS.452.1502D,2017MNRAS.465...32B,2017MNRAS.472L.109A,2018MNRAS.481.3118M,2019arXiv190408431D}. The role of SN feedback was also pointed out in the semi-analytical galaxy formation model {\sc MORGANA} \citep{2015MNRAS.453.4112F}.
It has been shown in several simulations that below the characteristic mass, SN-driven winds are fast enough to overcome the escape velocity of the gravitational potential of the galaxies \citep{2015MNRAS.452.1502D}, therefore depleting the center of galaxies from their cold gas reservoir (which could have fed the BHs). BHs are only able to grow when their host galaxies are massive enough to overcome the effect of SN feedback.

In all these simulations BHs are re-positioned to the potential minimum (of the whole galaxies or within smaller regions) every timestep and thus do not move around in their galaxies, except in Horizon-AGN \citep[see also][]{2017MNRAS.468.3935H}. Doing so favors the accretion onto BHs. Off-center BHs could also stunt BH growth, as shown in the cosmological simulation NewHorizon \citep{2019arXiv190408431D}, and the zoom-in MARVEL-ous dwarf simulations \citep{2019MNRAS.482.2913B} and the zoom-in FIRE simulations \citep{2017MNRAS.472L.109A,2020arXiv200712185C}. Off-center (or wandering) BHs have been found in recent high-resolution radio observations \citep{2020ApJ...888...36R}.

\subsubsection{Impact of the resolution with TNG100/TNG300}
TNG100 and TNG300 use the same galaxy and BH models but have different resolutions (baryonic resolution of $1.4\times 10^{6}\,\rm M_{\odot}$ and $1.1\times 10^{7}\,\rm M_{\odot}$, and spatial resolution of $0.74\, \rm kpc$ and $1.48\, \rm kpc$, for TNG100 and TNG300), and volumes ($100^{3},300^{3}\, \rm cMpc^{3}$). 
The parameters of the sub-grid models do not depend on mass or spatial resolution.

The better resolution of TNG100 allows to more accurately resolve the BH surrounding gas density, and thus their gas accretion \citep{2018MNRAS.479.4056W}. As a result, the BHs and galaxies tend to have slightly higher masses in TNG100 at fixed halo mass.
We find that the shape of the scaling relation at the low-mass end ($\log_{10}\, M_{\star}/\rm M_{\odot}\sim 9$) is different. While we note the presence of a change of slope at all redshifts in TNG300, this is not the case for TNG100 at $z=1-0$. 
The better resolution of TNG100 leads to more efficient growth of the BHs embedded in galaxies of $M_{\star}\leqslant 10^{10}\, \rm M_{\odot}$. Indeed, the simulation is more resolved and the surroundings of BHs have higher gas densities, which leads to increased accretion rates onto the BHs. 
The normalization of the scaling relation of TNG100 should, therefore, be higher, particularly in the regime where gas accretion is more important than BH-BH mergers for BH growth ($M_{\star}\leqslant 10^{10.5}\, \rm M_{\odot}$). 
The resolution also affects the strength of SN winds \citep{2017arXiv170302970P}; the overall impact of the SN feedback is weaker in the more resolved simulation TNG100.
The consequence of the more efficient SN feedback in low-mass galaxies of the TNG model compared to the Illustris model is not as strong in the more resolved simulation TNG100.
No other simulation project produced lower resolution volume, thus we cannot assess their level of resolution convergence in a similar manner as the TNG model.

\section{Understanding the evolution of the mean $M_{\rm BH}-M_{\star}$ relation}
Illustris and TNG100 share the same initial conditions, and therefore one can follow the evolution of individual matched galaxies in the two simulations, and understand how the galaxy/BH sub-grid models affect the build-up of the scaling relation with time.
We use the catalog of \citet{2014MNRAS.439..300L,2018MNRAS.481.1950L} that provides the IDs of the matched galaxies.
We use the Illustris and TNG100 examples to interpret the evolution of the Horizon-AGN, EAGLE, and SIMBA $M_{\rm BH}-M_{\star}$ relations.

\subsection{Final BH mass for the sample of matched BHs}
\begin{figure}
\centering
\includegraphics[scale=0.47]{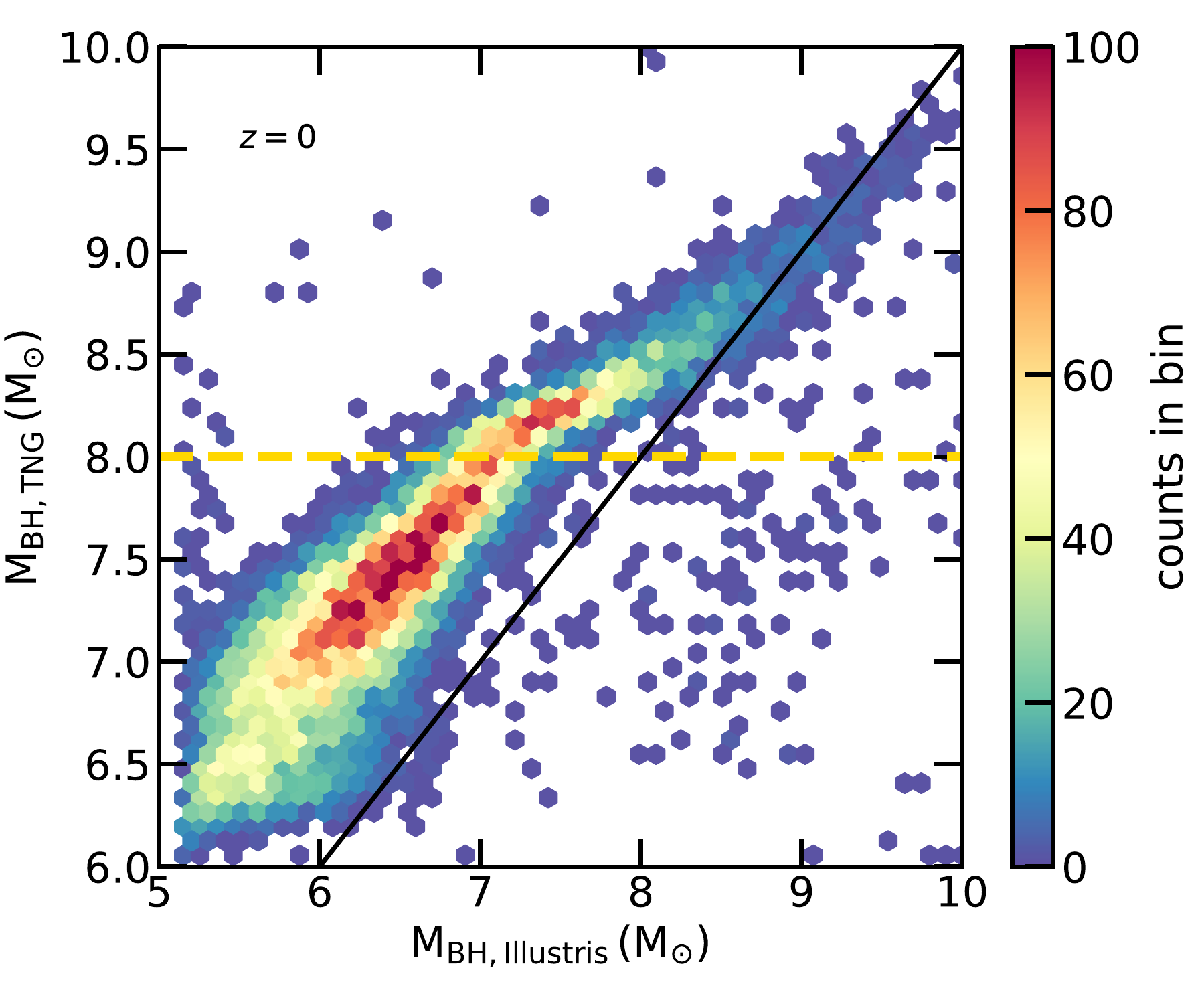}
\caption{Comparison of the BH final mass at $z=0$ in matched Illustris and TNG100 galaxies (with $M_{\star}\geqslant 10^{9}\, \rm M_{\odot}$).
For reference, the BH seeding mass is $M_{\rm BH}\sim 10^{5}\, \rm M_{\odot}$ in Illustris, and $M_{\rm BH}\sim 10^{6}\, \rm M_{\odot}$ in TNG100.
The black line indicates hypothetic identical mass BH pairs. 
In low-mass galaxies (i.e., with $M_{\rm BH}\leqslant 10^{8}\, \rm M_{\odot}$) BHs are more massive in TNG100. However, when the TNG100 BHs transition to the kinetic AGN feedback mode (upper side of the yellow dashed line), their growth as well as their host galaxies growth is strongly regulated, while the same BHs in Illustris are still able to grow. Consequently, the Illustris BHs catch up their TNG counterparts, and in the most massive matched galaxies TNG100 and Illustris BHs eventually end up with similar masses.}
\label{fig:matched_gal_z0}
\end{figure}

We compare the final BH mass of all the Illustris and TNG100 matched galaxies in Fig.~\ref{fig:matched_gal_z0}. Our sample includes about 18000 matched galaxies at $z=0$.
We find that TNG100 BHs are more massive than their matched BHs in Illustris in low-mass galaxies, i.e. for BHs of $M_{\rm BH}\leqslant 10^{8}\, \rm M_{\odot}$. In TNG100, $M_{\rm BH, \, TNG}\sim10^{8}\, \rm M_{\odot}$ is the characteristic mass (yellow dashed line in Fig.~\ref{fig:matched_gal_z0}) for which many BHs transition from the thermal high accretion mode to the kinetic low accretion mode of AGN feedback. The kinetic mode is responsible for efficiently quenching star formation in the BH host galaxies, but also for stunting BH gas accretion.
Consequently, it leads to a change of slope\footnote{The transition to the AGN feedback kinetic  mode implies a change of slope for other quantities/relations, e.g., BH mass scaling relations with $\sigma$ and the Sersic index \citep{2019arXiv191000017L}.} in Fig.~\ref{fig:matched_gal_z0} at this characteristic BH mass of $M_{\rm BH,\, TNG}\sim 10^{8}\, \rm M_{\odot}$. 
BHs with larger masses have a hard time growing in TNG100, but BHs in the matched Illustris galaxies are still able to do so because of a less effective low accretion mode of AGN feedback. The ability of the Illustris massive BHs to grow when the TNG100 BHs are stunted compensates the lower initial mass of the Illustris BHs. 
As a result, we find that the most massive matched BHs in Illustris and TNG100 with $M_{\rm BH}\geqslant 10^{8.5}\, \rm M_{\odot}$ have similar masses in Fig.~\ref{fig:matched_gal_z0}.

\begin{figure*}
\centering
\includegraphics[scale=0.57]{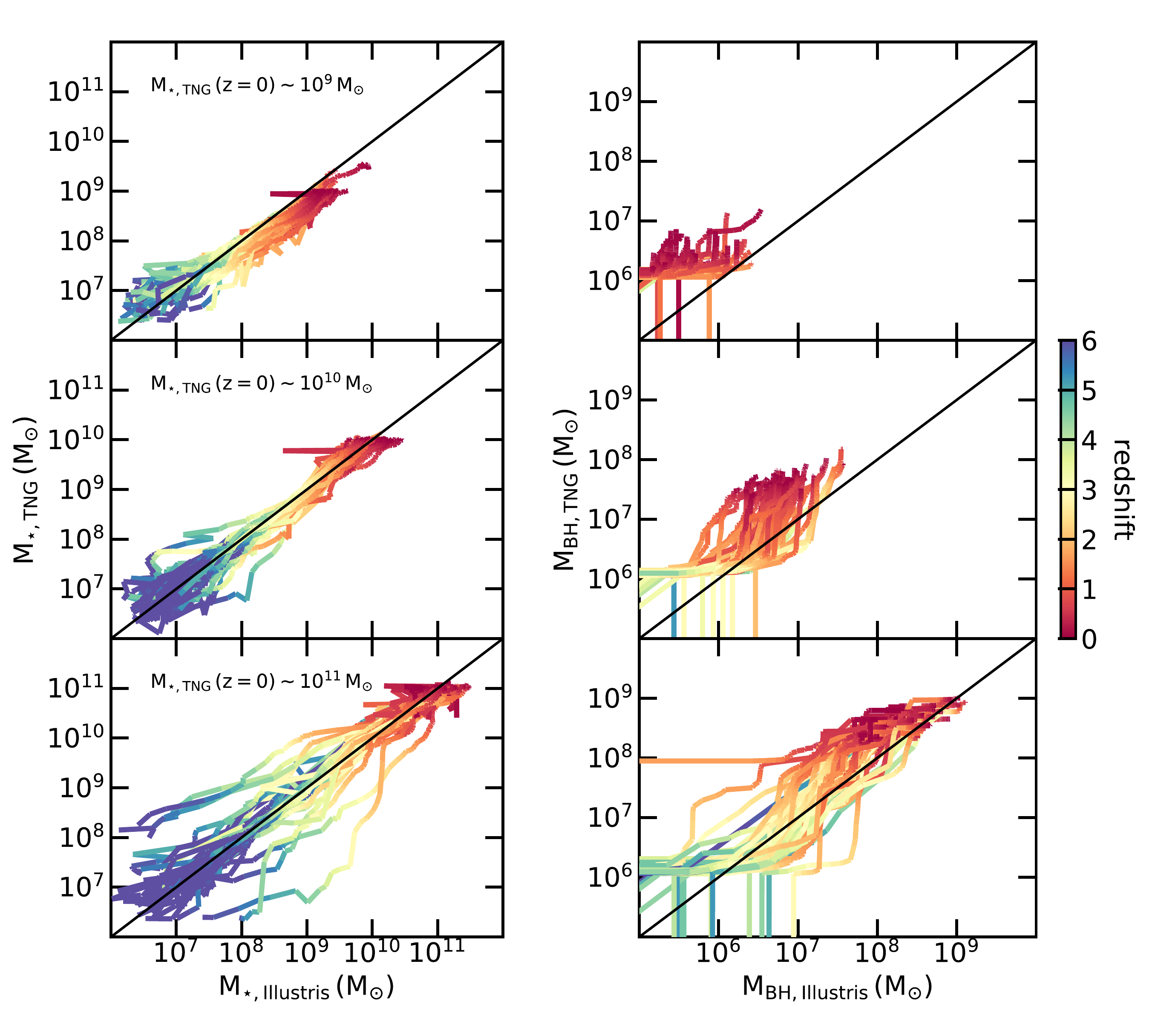}
\caption{Comparison of the stellar mass and BH mass in TNG100 (y-axes) and Illustris (x-axes) matched galaxies. Black lines indicate hypothetic identical host stellar and BH masses. Matched galaxies have very similar time evolution of their stellar mass (left panels).
The average evolution of the galaxy stellar mass is similar in the two simulations, with the exception that the low-mass galaxies have lower stellar mass in TNG100 (top panels, $M_{\star,\, {\rm TNG},\, z=0}\sim 10^{9}\, \rm M_{\odot}$). BHs in these galaxies are more massive in TNG100 because of the larger seeding mass (right top panel). The initial growth of BHs takes place later in TNG100 than in Illustris. However, when they start growing, they grow more efficiently than in Illustris (middle right panel).
However, when the TNG BHs transition to the kinetic AGN feedback mode (bottom right panel), their growth as well as their host galaxies growth is strongly regulated, while the same BHs in Illustris are still able to grow. Consequently, in the most massive matched galaxies ($M_{\star}\geqslant 10^{10.5}\, \rm M_{\odot}$) BHs have similar masses. }
\label{fig:matched_gal_evol}
\end{figure*}

\begin{figure*}
\centering
\includegraphics[scale=0.57]{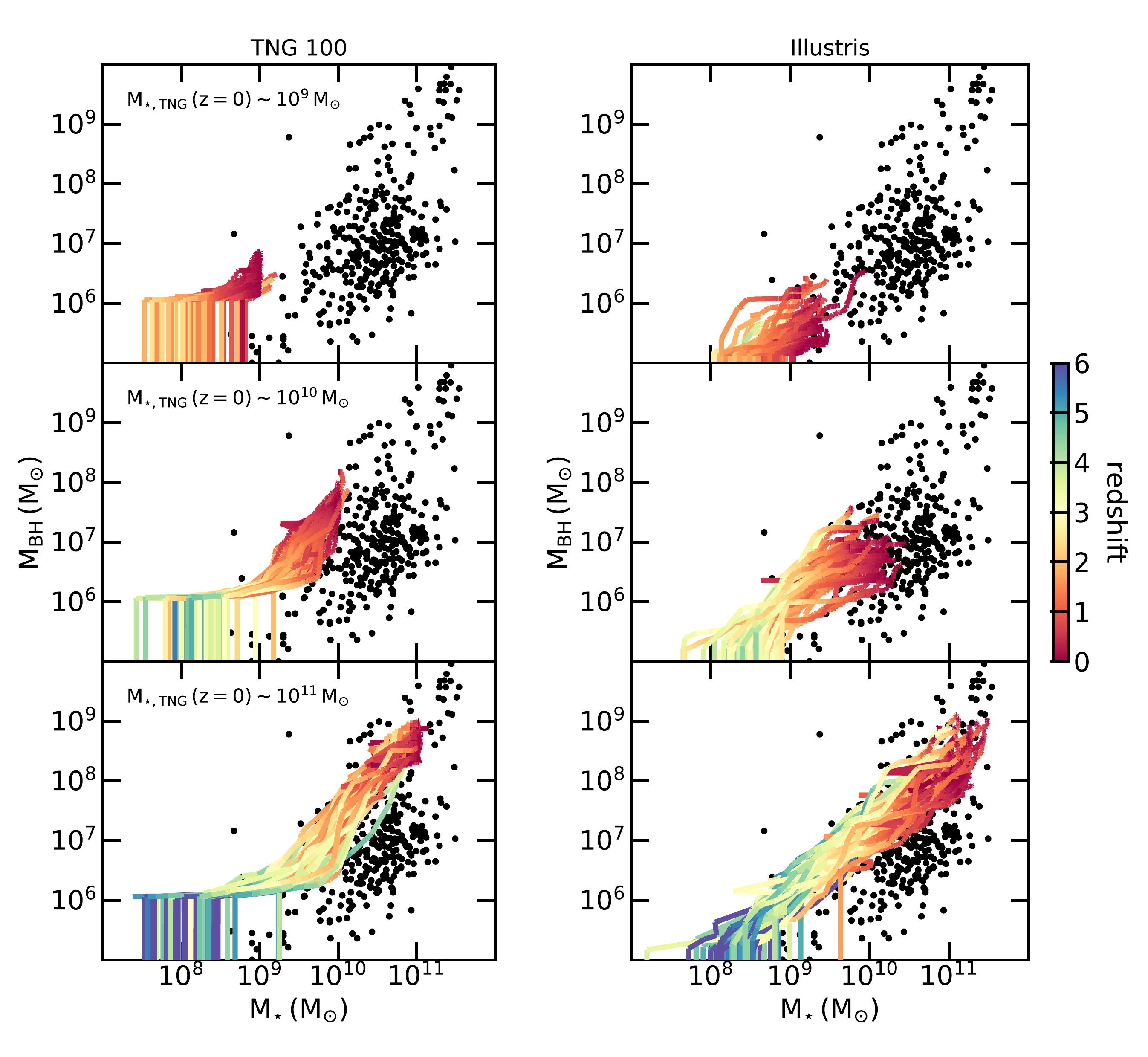}
\caption{$\rm M_{BH}-M_{\star}$ diagram for different final TNG stellar mass ranges: $M_{\star,\, {\rm TNG},\, z=0}\sim 10^{9}\, \rm M_{\odot}$ (top panels), $M_{\star,\, {\rm TNG},\, z=0}\sim 10^{10}\, \rm M_{\odot}$ (middle panels), and $M_{\star,\, {\rm TNG},\, z=0}\sim 10^{11}\, \rm M_{\odot}$ (bottom panels). Results from TNG100 are shown on the left panels, and Illustris on the right. Black dots show the observational sample of \citet{2015arXiv150806274R} ($z\sim 0$). The initial growth of the TNG BHs is delayed because of the stronger SN feedback in low-mass galaxies. However, when they start growing, they grow very efficiently. The early growth histories of the Illustris BHs are more diverse and their growth is on average less efficient. There is no strong decrease of BH growth in the massive galaxies of $M_{\star}\sim 10^{11}\, M_{\odot}$ in Illustris, whereas their growth is stunted in TNG because of the efficient kinetic AGN feedback.}
\label{fig:matched_gal_evol_scaling}
\end{figure*}

\subsection{Time evolution of the matched galaxies and their BHs}
We investigate the time evolution of three different samples of $\sim 50$ matched galaxies each. Given the differences found in Fig.~\ref{fig:matched_gal_z0}, we choose these samples to include galaxies with three key stellar masses at $z=0$: \\
\textbullet\, {\bf Low-mass galaxies}: $M_{\star,\,{\rm TNG},\,z=0}\sim 10^{9}\, \rm M_{\odot}$, corresponding to BHs with $M_{\rm BH, TNG}\sim 10^{6-7}\, \rm M_{\odot}$ at $z=0$, \\
\textbullet\, {\bf Intermediate mass galaxies}: $M_{\star,\, {\rm TNG},\, z=0}\sim 10^{10}\, \rm M_{\odot}$, corresponding to BHs of $M_{\rm BH, TNG}\sim 10^{7-8}\, \rm M_{\odot}$ at $z=0$, \\
\textbullet\, {\bf Massive galaxies}: $M_{\star,\, {\rm TNG},\,z=0}\sim 10^{11}\, \rm M_{\odot}$, corresponding to BHs of $M_{\rm BH, TNG}\sim 10^{8-9}\, \rm M_{\odot}$ at $z=0$. \\

Because the $M_{\rm BH}-M_{\star}$ relation is tight in TNG100, the corresponding TNG BH mass ranges that we quote above are within one order of magnitude in BH mass. The $M_{\rm BH}-M_{\star}$ relation in the Illustris simulation has a larger scatter; the BH mass ranges covered by these galaxy mass samples are larger, between one and two orders of magnitude in BH mass. We first select galaxies that fulfilled our final stellar mass criterion in TNG100, and then look for their matched galaxies in Illustris. 
We then follow back in time all these galaxies using the galaxy merger trees of the two simulations \citep{2015MNRAS.449...49R}, which are based on stellar particles and gas cells. 
We show the time evolution of both their stellar masses (left panels) and their BH masses (right panels) in Fig.~\ref{fig:matched_gal_evol}. 

In the following, we start by analyzing the stellar mass content of galaxies in the three samples.
We find that the relatively low-mass galaxies shown in the top left panel (i.e., $M_{\star,\, {\rm TNG},\,z=0}\sim 10^{9}\, \rm M_{\odot}$) tend to be slightly more massive in Illustris than in TNG100. The TNG100 SN feedback is stronger in low-mass galaxies at all redshifts (addition of velocity floor) and at lower redshifts (new scaling of the wind velocity with the Hubble constant) than in Illustris (see our section~\ref{sec:SN_feedback}). 
For the intermediate galaxies ($M_{\star, \,{\rm TNG},\,z=0}\sim 10^{10}\, \rm M_{\odot}$, middle left panel), the overall stellar mass content of the matched Illustris and TNG100 galaxies is similar. This is again the case for the evolution of the most massive galaxies ($M_{\star, \,{\rm TNG},\, z=0}\sim 10^{11}\, \rm M_{\odot}$, bottom left panel), but we note that some matched galaxies have very different evolutionary paths in this mass range.

While most of the matched galaxies evolve their mass in a similar way, we find that their BHs have different growth histories (right side panels in Fig.~\ref{fig:matched_gal_evol}). We set a null BH mass for galaxies that have not been seeded with a BH yet, which translates into the vertical lines in all panels of Fig.~\ref{fig:matched_gal_evol} for TNG BHs. In the following, we investigate how the Illustris and TNG100 sub-grid models affect the different phases of BH evolution (see the items below) with the right panels of Fig.~\ref{fig:matched_gal_evol}, and what the consequences on the BHs/galaxies co-evolution are with Fig.~\ref{fig:matched_gal_evol_scaling}. In Fig.~\ref{fig:matched_gal_evol_scaling} we show the time evolution of the matched BHs and their host galaxies in the $M_{\rm BH}-M_{\star}$ diagram, dividing in the same three stellar mass bins. 

\begin{itemize}
\item {\bf Initial mass of BHs:}
the initial mass of the TNG100 BHs is higher (by a factor of 10) than the seeding mass of the matched Illustris BHs. As a result, recently formed low-mass galaxies have more massive BHs in TNG100 than in Illustris, as shown in the top right panel of Fig.~\ref{fig:matched_gal_evol}.\\

\item{\bf First phase of BH growth:}
the first phase of BH growth is delayed in TNG100 and takes place earlier in Illustris (middle and bottom right panels of Fig.~\ref{fig:matched_gal_evol}). The stronger feedback from SNe in TNG100 is responsible for delaying the first episode of gas accretion, and not the differences in the seeding or accretion models.
Indeed, both Illustris and TNG100 are seeded in dark matter halos of same mass and thus at about the same cosmic time; the Illustris BHs are not seeded before the TNG100 BHs. At birth, the accretion rates onto the Illustris and TNG100 BHs, given by the Bondi model $\dot{M}_{\rm BH}\propto \alpha M^{2}_{\rm BH}$, should be similar\footnote{Here, we make the assumption that the gas reservoir is the same in the two simulations in these relatively early stages of galaxy formation (i.e., galaxies of $M_{\star}\sim 10^{9}\, \rm M_{\odot}$). }: the seeding mass ($M_{\rm seed}\sim 10^{6}, 10^{5}\, \rm M_{\odot}$ for TNG100 and Illustris, respectively) and the boost factor parameters ($\alpha=1, 100$) compensate one another. 

The stronger SN feedback is responsible for the change of slope in the $M_{\rm BH}-M_{\star}$ diagram of TNG100 and the presence of a characteristic mass in the stellar mass range $M_{\star}=10^{9}-10^{10}\, \rm M_{\odot}$. With the left panels of Fig.~\ref{fig:matched_gal_evol_scaling}, we find that the TNG100 characteristic galaxy stellar mass is moving toward less massive stellar mass with time, and is absent for $z\leqslant 1$. Indeed, the first quiescent phase of BH growth is very short in the top panel of Fig.~\ref{fig:matched_gal_evol_scaling} (orange to red colored lines) for galaxies of $M_{\star}\sim 10^{8.5}-10^{9}\, \rm M_{\odot}$, and longer for the galaxies of same mass but at high redshift (bottom left panel, $z=5-3$, green to yellow colored lines). We find here that the modeling of SN feedback in TNG100 impacts the growth of BHs more substantially at high redshift.

The stronger SN feedback in TNG100 is also responsible for the slightly lower normalization of the $M_{\rm BH}-M_{\star}$ mean relation in TNG100 compared to Illustris, at the low-mass end ($M_{\star}\leqslant 10^{10}\, \rm M_{\odot}$) at high redshift (see Fig.~\ref{fig:scaling_relation_mean}). 
At high redshift ($z\sim 5$) and because of the less effective SN feedback, the Illustris BHs can grow compared to the TNG matched BHs. Therefore, the Illustris BHs embedded in galaxies of $M_{\star}\sim 10^{9}\, \rm M_{\odot}$ (e.g., bottom right panel of Fig.\ref{fig:matched_gal_evol_scaling}) are on average slightly more massive, even if the seeding mass of Illustris is smaller than in TNG100.\\

\item{\bf Later phases of BH growth:}
Once the TNG100 BHs start growing, they grow more efficiently than the Illustris BHs, as demonstrated in the middle and bottom right panels of Fig.~\ref{fig:matched_gal_evol}. The difference can also be appreciated in Fig.~\ref{fig:matched_gal_evol_scaling} (e.g., middle panels). While most of the TNG100 BHs grow efficiently, there is a flattening in the $M_{\rm BH}-M_{\star}$ diagram of Illustris: BH growth is there less efficient than their host galaxy growth, on average.

In addition to the potential different gas content in the surroundings of the BHs (see Fig.~\ref{fig:trends}, top panels), there are two important differences in the accretion schemes of the two simulations. Illustris uses the gas content of the BH parent cell to compute the accretion rates, while TNG100 uses a kernel. The accretion rates onto the TNG100 BHs could also be boosted by the presence of the magnetic fields in the simulation \citep[see Fig. 8 of][where a higher normalization of $M_{\rm BH}-M_{\rm h}$ is found with magnetic field at $z=0$, a higher (or smaller) gas fraction, and smaller (or larger) galaxy size for lower-mass galaxies of a few $\leqslant 10^{10}\, \rm M_{\odot}$ (or for more massive galaxies)]{2017arXiv170302970P}. 

All the TNG100 BHs grow efficiently, which leads to little diversity of growth histories in the $M_{\rm BH}-M_{\star}$ plane in Fig.~\ref{fig:matched_gal_evol_scaling}. 
Instead, the Illustris BHs alternate between episodes of high accretion rates and more quiescent periods with lower accretion rates. BH growth is more diverse in Illustris, at all galaxy mass. Consequently, the scatter of the Illustris $M_{\rm BH}-M_{\star}$ relation is almost independent of stellar mass (see Fig.~\ref{fig:matched_gal_evol_scaling}), and also, on average, larger than in TNG100, except for the phase of efficient TNG100 BH growth that we are discussing here. 
The scatter of TNG100 depends on the stellar mass: smaller scatter in the first inefficient BH growth in low-mass galaxies, smaller in the phase where BH growth is regulated by AGN feedback, and larger in the phase of efficient BH growth. \\

\item{\bf Towards BH quiescence:}
The impact of the strong kinetic low accretion mode AGN feedback in TNG100 can be seen in the bottom right panel of Fig.~\ref{fig:matched_gal_evol}. 
While the growth of the TNG100 BHs is completely stunted because of AGN feedback for BHs of $M_{\rm BH}\sim 10^{8}\, \rm M_{\odot}$ (horizontal lines in the panel), the Illustris BHs are less regulated by the Illustris AGN feedback (injection of thermal energy in bubbles displaced from the galaxies) and are, therefore, still able to grow.
As noticed in Fig.~\ref{fig:matched_gal_z0}, the BHs of these massive galaxies reach similar masses by $z=0$.

\end{itemize}
\begin{figure}
\centering
\includegraphics[scale=0.44]{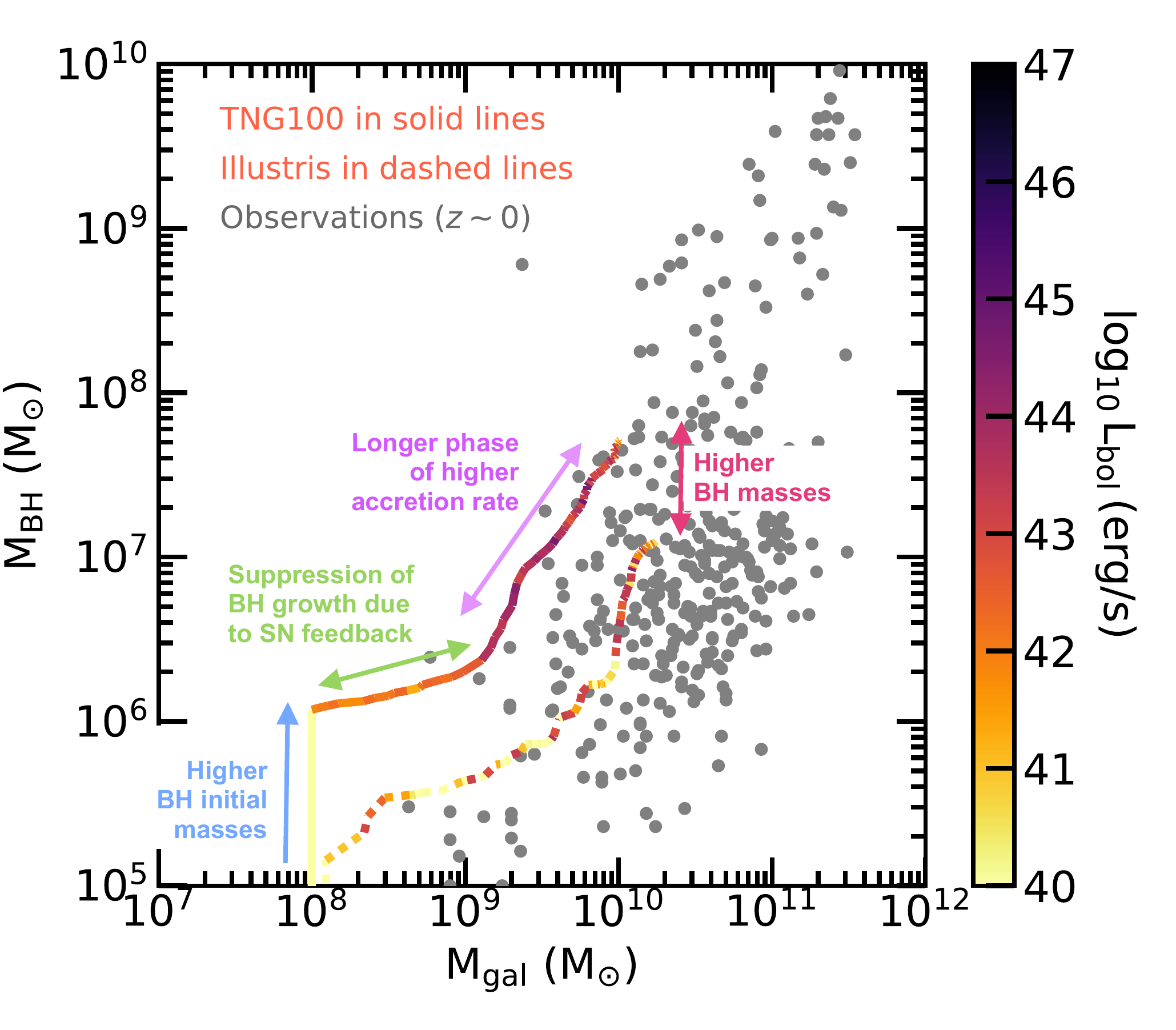}
\caption{Illustration of one set of matched galaxies in TNG (solid line) and Illustris (dashed line) with total stellar mass of $M_{\star}=10^{10}\, \rm M_{\odot}$ at $z=0$. Not all BHs in the same galaxy mass range follow the same track, but at least a large fraction does so. The BHs in TNG start with a higher initial mass, are stunted by an efficient SN feedback, experience longer phases of higher accretion rate than the BHs in Illustris. Consequently, the BHs in TNG reach larger masses by $z=0$ in galaxies with $M_{\rm \star,\, z=0}=10^{10}\, \rm M_{\odot}$. A few examples for galaxies of $M_{\rm \star,\, z=0}=10^{10}\, \rm M_{\odot}$ and $M_{\rm \star,\, z=0}=10^{11}\, \rm M_{\odot}$ are given in Appendix~\ref{sec:illustration}. }
\label{fig:4gal_illustration}
\end{figure}

In Appendix~\ref{sec:illustration}, we analyze in detail eight pairs of matched Illustris and TNG100 galaxies, with final galaxy mass of $M_{\star}=10^{10}\, \rm M_{\odot}$ and $M_{\star}=10^{11}\, \rm M_{\odot}$ at $z=0$. The phases of BH growth and feedback described above are illustrated for individual galaxies. The properties of BHs and their galaxies are followed in time, and used to understand the build-up of the $M_{\rm BH}-M_{\star}$ relation with time in Illustris and TNG100, as described in the next section.
We provide in Fig.~\ref{fig:4gal_illustration} an illustration of the different BH growth histories that two TNG100 and Illustris matched BHs can have.

\begin{figure}
\centering
\includegraphics[scale=0.54]{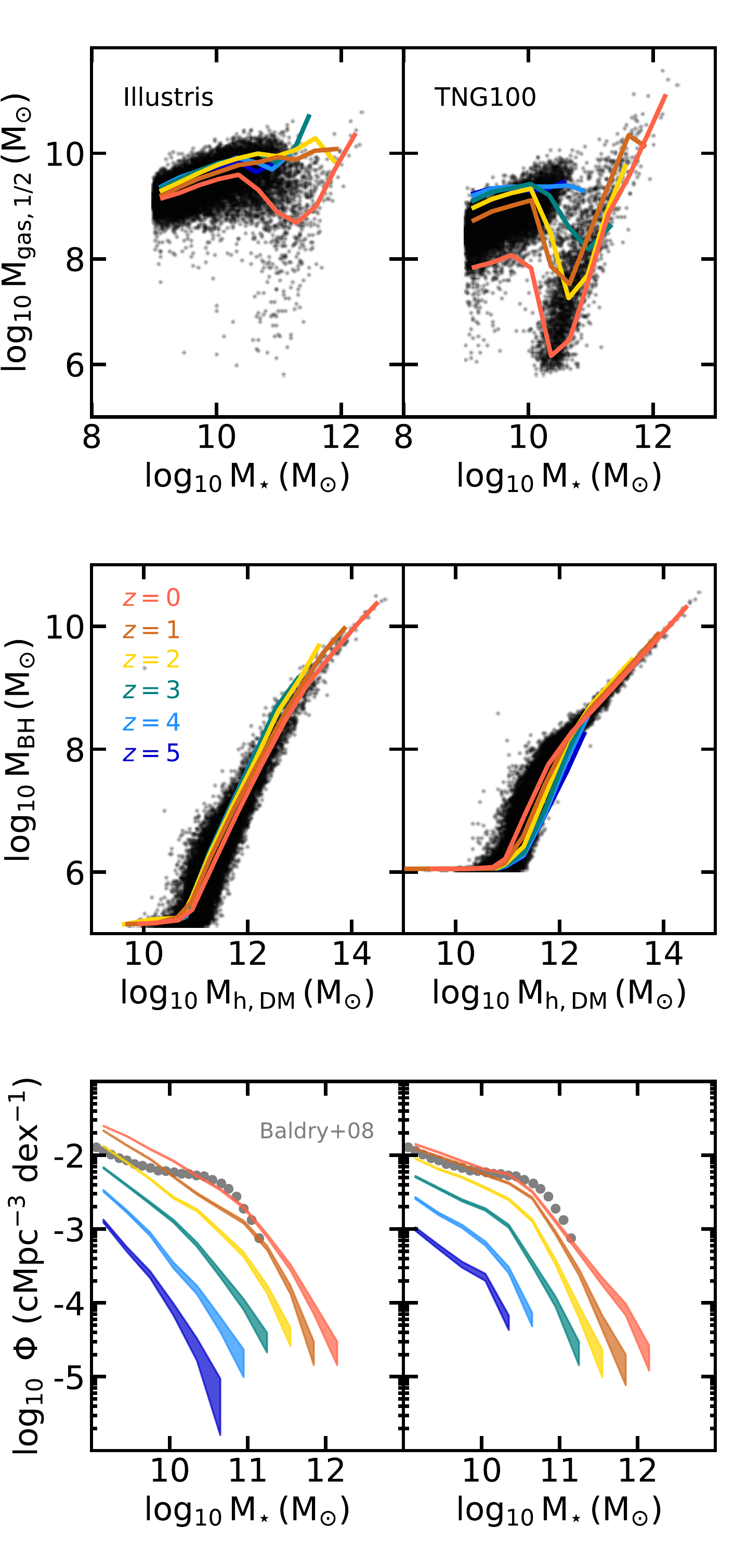}
\caption{Different diagnostics to understand the evolution of the overall normalization of the $M_{\rm BH}-M_{\star}$ mean relation with time. 
{\it Top panels:} Mean mass of the gas contained in the half-mass radius of the galaxies; the entire population at $z=0$ is shown in black. 
The gas content in the inner region of the galaxies diminishes with time.
{\it Middle panels:} Mean BH mass as a function of the dark matter mass of their host halos. We show in black all the halos with BHs at $z=0$. 
At fixed dark matter halo mass, we find a larger evolution in TNG100, and almost no evolution in Illustris. 
{\it Bottom panels:} Galaxy stellar mass functions. Grey symbols show the observational constraints of \citet{2008MNRAS.388..945B} ($z\sim 0$). The low-mass end of the galaxy mass function is overestimated in Illustris ($M_{\star}\leqslant 10^{10.5}\, \rm M_{\odot}$); a better agreement is found for TNG100.}
\label{fig:trends}
\end{figure}

\subsection{Build-up of the scaling relation in Illustris and TNG100}
In Fig.~\ref{fig:summary_simu}, we present a summary of the features discussed in this section with the scaling relation of the simulations at $z=3$ and $z=0$.

\subsubsection{Illustris}
In Illustris, the overall normalization of the $M_{\rm BH}-M_{\star}$ mean relation decreases with time from $z=3$ to $z=0$, at all galaxy stellar masses.
This can be due to two effects: a less efficient growth of BHs with time and/or a more efficient growth of the host galaxies than their BHs with time. In this section, we use the conclusions from Fig.~\ref{fig:matched_gal_evol}, Fig.~\ref{fig:matched_gal_evol_scaling}, and Fig.~\ref{fig:trends} which shows the gas content of the Illustris and TNG galaxies, the mass of the BHs as a function of their halo mass, and the stellar mass functions of the simulations.

We find that the average mass of gas in the half-mass radius of the galaxies $M_{\rm gas,\, 1/2}$ decreases with time. This could favor the first option above (less efficient BH growth at later times), however one has to consider the relative growth efficiency of BHs and galaxies.
We note here that the average gas content decreases with a smaller amplitude in Illustris than in TNG100: statistically the Illustris galaxies have more gas than the TNG100 galaxies (for any galaxy stellar mass). 
Having a decreasing average gas mass does not ensure that the gas content decreases in time for a given galaxy. In fact, when analyzing the evolution of individual galaxies (Appendix B, Fig.~\ref{fig:4gal_1e10}) we find that the gas mass of the galaxies does not evolve significantly with time for $z\leqslant1$ for many galaxies of $M_{\star, z=0}\sim 10^{10}\, \rm M_{\odot}$. However, the inner gas content of these galaxies diminishes with time with respect to their total stellar mass and their SFR is constant in the range $z\leqslant1$: galaxies still form stars with the same efficiency and grow their total stellar content more efficiently with respect to their inner gas content. Consequently, galaxy growth becomes more efficient than BH feeding (which scales with the inner gas content, see Fig.~\ref{fig:4gal_1e10}) with time.

The $M_{\rm BH}-M_{\rm h,\, DM}$ diagram in Fig.~\ref{fig:trends} shows very little evolution between the average BH mass and the dark matter mass of their host halos, indicating that the efficiency of BH growth does not change much with time, on average.
The more efficient growth at low redshifts of the galaxies compared to their BHs is also corroborated by the evolution of the galaxy stellar mass function, shown in Fig.\ref{fig:trends} (bottom panels). In Illustris, we find that the galaxy mass function keeps evolving in the redshift range $z=1-0$. The mean $M_{\rm BH}-M_{\star}$ is shifted toward more massive galaxies with time.

\subsubsection{TNG100}
TNG100 shows the opposite trend: the overall normalization of the mean $M_{\rm BH}-M_{\star}$ relation increases with time in the redshift range $4\geqslant z \geqslant 0$ and for $M_{\star}\leqslant 10^{10.5}\, \rm M_{\odot}$. Our results for the larger simulation TNG300 indicates that the evolution at high redshift $z\geqslant 3$ could extend to larger stellar mass galaxies.

A more efficient growth of BHs at low redshift, or a slower galaxy growth compared to BH growth in galaxies of $M_{\star}\leqslant 10^{10.5}\, \rm M_{\odot}$, could explain the trend identified in TNG100.
For TNG100, there is no evolution of the galaxy stellar mass function for $M_{\star}\leqslant 10^{10.5}\, \rm M_{\odot}$ and $z\leqslant 1$. Moreover, we find a time evolution of the $M_{\rm BH}-M_{\rm h, \, DM}$ mean relation, suggesting that the time evolution in the $M_{\rm BH}-M_{\star}$ diagram is driven by changes in the efficiency of BH growth, and that these changes do not affect the host galaxies as much as the BHs. 
As explained previously, the stronger modeling of the SN feedback is responsible for stunting BH growth at high redshift in TNG100 (see Fig.~\ref{fig:summary_simu}). 
We find that the impact of SN feedback on BH growth in low-mass galaxies is weaker at low redshift $z=1-0$ than at higher redshifts.
As a result, the overall normalization of the mean $M_{\rm BH}-M_{\star}$ relation increases with time.

In more massive galaxies of $M_{\star}\geqslant 10^{10.5}\, \rm M_{\odot}$ and low redshift, the efficient kinetic AGN feedback takes over and stops BH growth via gas accretion. It is responsible for the absence of time evolution in massive galaxies at low redshift.

\subsection{Build-up of the scaling relation in Horizon-AGN, EAGLE, and SIMBA}

\subsubsection{Horizon-AGN}
The overall normalization of the $M_{
\rm BH}-M_{\star}$ relation decreases with time in Horizon-AGN \citep[Fig.~\ref{fig:scaling_relation_mean}, and][]{2016MNRAS.460.2979V}, for all stellar masses, except at $z=0$ in low-mass galaxies with $M_{\star}\leqslant 10^{10}\, \rm M_{\odot}$ whose normalization is higher than at higher redshift.

As in Illustris and TNG100, the amount of gas in the Horizon-AGN galaxies diminishes with time \citep{2017MNRAS.472..949B}. As shown in the previous section with the TNG100 simulation counterexample, this does not explain on its own the decrease of the normalization with time. One has to consider both the relative growth of the BHs and the growth of their galaxies. To build a complete picture, we analyze the stellar mass function of Horizon-AGN with redshift and compare it to the observational constraints of \citet{2008MNRAS.388..945B} at $z=0$ (not shown)\footnote{While here we discuss the galaxy mass function, the galaxy luminosity function of Horizon-AGN can be found in \citet{2017MNRAS.tmp..224K}.}.

The stellar mass function of Horizon-AGN cleary overpredicts the observational constraints at $z=0$.
The simulation produces a higher number of galaxies below the knee of the function (i.e., $M_{\star}\leqslant 10^{11}\, \rm M_{\odot}$) at $z=0$ and at higher redshifs ($z\leqslant 3$). 
It also overpredicts the number of more massive galaxies of $M_{\star}>10^{11}\, \rm M_{\odot}$ at $z=0$.
The stellar mass function evolves towards more massive galaxies from $z=4$ to $z=0$. We also note that the stellar mass function at $z=0$ for lower-mass galaxies of $M_{\star}\geqslant 10^{10.3}\, \rm M_{\odot}$ is slightly smaller than at higher redshifts.

We now compare the time evolution of the stellar mass function for given stellar mass ranges, to the evolution of BH mass function for the corresponding BH mass ranges.
From Fig.~\ref{fig:scaling_relation_Lbol} we know that in Horizon-AGN galaxies of $M_{\star}\sim 10^{10.5}-10^{11}\, \rm M_{\odot}$ host BHs of $M_{\rm BH}\sim 10^{8}\, \rm M_{ \odot}$ at $z=0$, and BHs of $M_{\rm BH}\sim 10^{7.5}-10^{8.5}\, \rm M_{\odot}$ at $z=2$. 
Here we anticipate on our following section where we discuss the BH mass function and its evolution with time. The corresponding BH mass function (shown in Fig.~\ref{fig:mass_function_2}) does not evolve in the redshift range $z=2-0$ for these BHs of $M_{\rm BH}=10^{7.5}-10^{8}\, \rm M_{\odot}$. Statistically, the galaxies in this stellar mass range grow more than their BHs.
This effect can also be seen by comparing Fig.~\ref{fig:scaling_relation_Lbol} and Fig.~\ref{fig:scaling_sf_q} at $z=0$. We find a larger fraction of quite faint AGN with $L_{\rm bol}\leqslant 10^{42} \, \rm M_{\odot}$ in galaxies of $M_{\star}\sim 10^{10.5}-10^{11}\, \rm M_{\odot}$ (Fig.~\ref{fig:scaling_relation_Lbol}, bottom panel), while a large fraction of these galaxies still form stars with $\rm sSFR\geqslant 10^{-10}\, \rm yr^{-1}$.

Lower-mass galaxies with $M_{\star}\sim 10^{10}\, \rm M_{\odot}$ host BHs of $M_{\rm BH}\sim 10^{6.5}-10^{7.5}\, \rm M_{\odot}$ at $z=0-2$, and BHs of $M_{\rm BH}\sim 10^{7}-10^{7.5}\, \rm M_{\odot}$ at $z=4$. 
The BH mass function for $M_{\rm BH}\sim 10^{7}\, \rm M_{\odot}$ decreases from $z=1$ to $z=0$. There are less low-mass BHs of $\leqslant 10^{7}\, \rm M_{\odot}$ in the simulation at $z=0$ than at $z=1$. This is due to the non-replenishment of low-mass BHs in the simulation after $z=1.5$, when BH formation ceases.

To conclude, we find that the overall decrease of the $M_{\rm BH}-M_{\star}$ normalization in Horizon-AGN is due to a more important growth of the galaxies than the BHs (suffering from reduced amount of gas in their surroundings), as shown by the galaxy stellar mass function and the BH mass function. 
As in Illustris, the $M_{\rm BH}-M_{\star}$ relation of Horizon-AGN is shifted towards more massive galaxies with time. 
The higher $M_{\rm BH}-M_{\star}$ normalization for $M_{\star}\leqslant 10^{10}\, \rm M_{\odot}$ at $z<1$ is explained by the suppression of BH sink particle formation at $z=1.5$ in the simulation (see Fig.~\ref{fig:summary_simu}). At fixed galaxy stellar mass, the tail of the BH distribution present at high redshift and composed of newly formed BHs does not exist anymore at $z=0$, and leads to a higher mean BH mass value.

\subsubsection{EAGLE}
The overall normalization of the EAGLE simulation decreases with time for galaxies with $M_{\star}\leqslant 10^{10.7}\, \rm M_{\odot}$, and does not evolve with time for more massive galaxies. The relation stops evolving at $z\leqslant 1$ for all galaxy stellar masses, and is unchanged to $z=0$.
The evolution with time of the $M_{\rm BH}-M_{\star}$ mean relation has been analyzed in detail in \citet{2016MNRAS.462..190R,2018MNRAS.481.3118M,2017MNRAS.465...32B}, we briefly summarize their findings.

As in the TNG simulations, the non-linear shape of the $M_{\rm BH}-M_{\star}$ relation in low-mass galaxies arises from the low efficiency of BH growth, because of the strength of SN feedback (see Fig.~\ref{fig:summary_simu}). 
When a weaker SN feedback is employed, the relation becomes linear \citep[][their Fig. 10]{2015MNRAS.450.1937C}. An even stronger feedback than the fiducial EAGLE model stunts BH growth even more and the simulation does not form any massive BHs of $M_{\rm BH}>10^{7}\, \rm M_{\odot}$.
As shown in our Fig.~\ref{fig:trends} with TNG100, the relation between BH mass and dark matter halo mass changes as well with time, demonstrating that the change in the $M_{\rm BH}-M_{\star}$ normalization is mainly due to a variation in the efficiency of BH growth and not galaxy growth \citep[][their Fig.6]{2016MNRAS.462..190R}.
The change of overall normalization is due to the prevention of BH growth because of SN feedback. 

In EAGLE, the characteristic mass at which BHs start growing efficiently, i.e. $M_{\star}\sim 10^{10}\, \rm M_{\odot}$, depends on redshift (Fig.~\ref{fig:scaling_relation_mean}). At high redshift, BHs start growing in less massive galaxies. This is interesting because while the TNG simulations have a similar shape because of SN feedback, the evolution with redshift of the characteristic mass follows the opposite trend.
\citet{2018MNRAS.481.3118M} show that the transition to the BH efficient growth phase in EAGLE is set by the development of the a hot halo, which traps the SN-driven winds. The transition takes place at fixed virial temperature of $T_{\rm vir}\sim 10^{5.6}\, \rm K$, independently of redshift (their Fig. 5). 
The mass of the galaxies reaching this fixed temperature changes with redshift, which cause the evolution of the characteristic mass with redshift.

We also note here that at some level the transition to the rapid BH growth phase also depends on the modeling of BH accretion. In EAGLE, a modified version of the Bondi accretion model is employed, and takes into account the angular momentum of the gas to be accreted by the BHs. Instead of directly falling onto the BHs, the gas first settles into an accretion disk. 
When a higher viscosity is assumed (i.e., lower $C_{\rm visc}$ parameter in Eq. 12), the phase of rapid BH growth takes place in less massive galaxies. The phase starts in more massive galaxies for a smaller viscosity parameter \citep{2015MNRAS.450.1937C}. This shows that the characteristic mass at which BHs start growing efficiently can also depend on the modeling of accretion rate onto the BHs.

\subsubsection{SIMBA}
\begin{figure*}
\centering
\includegraphics[scale=0.95]{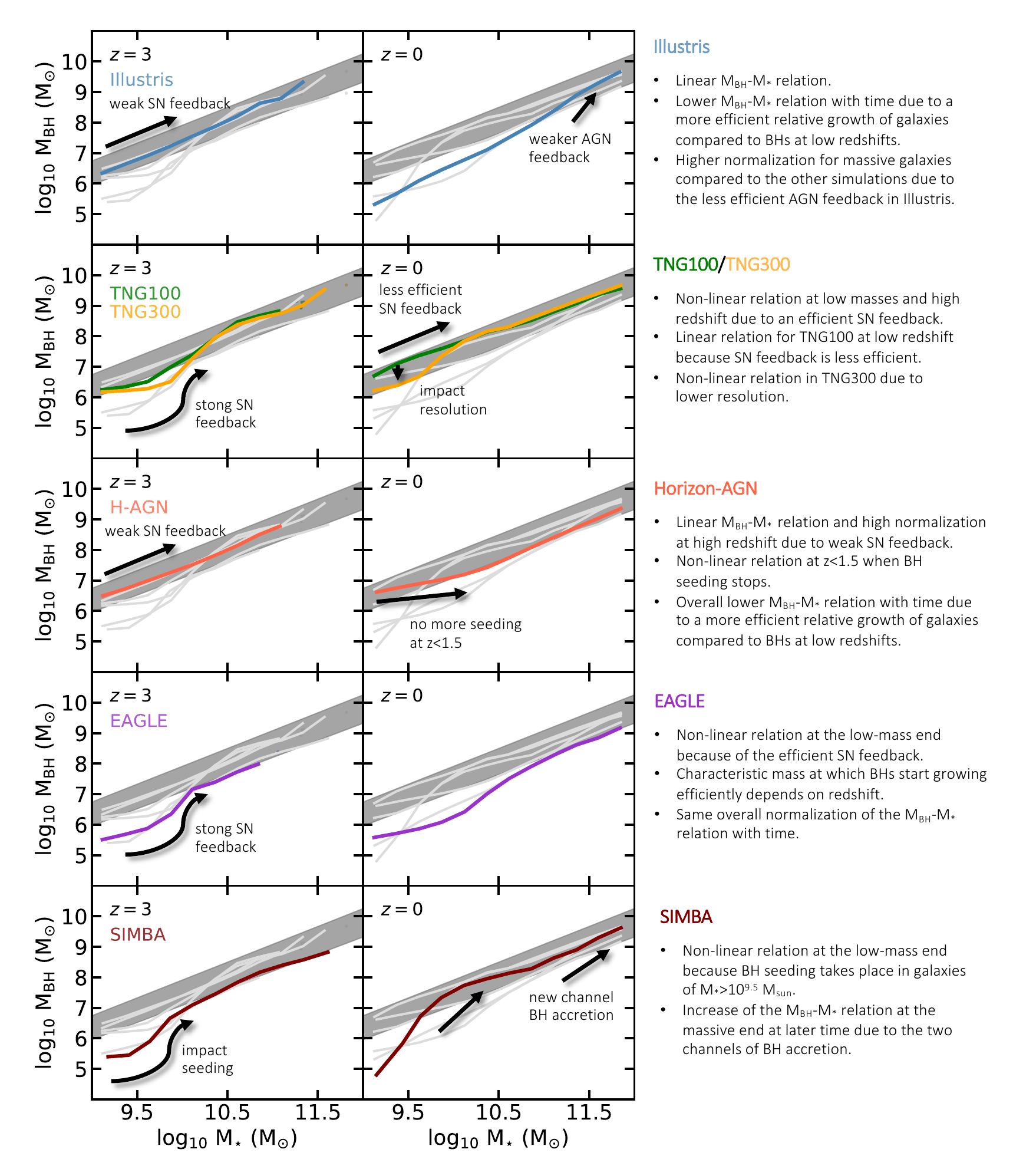}
\caption{Impact of the different subgrid models of the Illustris, TNG100, TNG300, Horizon-AGN, EAGLE, and SIMBA simulations on their mean/median $M_{\rm BH}-M_{\star}$ relation at $z=3$ (left panels) and $z=0$ (right panels). The dark grey shaded area in each panel represents the empirical scaling relations derived at $z=0$ in \citet{2013ARA&A..51..511K,2013ApJ...764..184M,Haring2004}. In each panel we also show the $M_{\rm BH}-M_{\star}$ relations of the other simulations in light grey. The detailed descriptions of the processes are included in the sections 5.3.1 (Illustris), 5.3.2 (TNG100/TNG300), 5.4.1 (Horizon-AGN), 5.4.2 (EAGLE), 5.4.3 (SIMBA).
}
\label{fig:summary_simu}
\end{figure*}

The mean $M_{\rm BH}-M_{\star}$ relation does not evolve at $z\geqslant 2$, but its normalization increases with decreasing redshift for $z=2-0$ and all galaxy stellar masses \citep{2019MNRAS.487.5764T}. The shape of the relation at high redshift ressembles the shape identified in TNG. However, since in SIMBA the seeding takes place in galaxies of $M_{\star}\geqslant 10^{9.5}\, \rm M_{\odot}$, the non-linear shape may be due to the seeding rather than SN feedback (see Fig.~\ref{fig:summary_simu}). We note here also that the SN feedback in SIMBA is less efficient (lower wind velocities) than in TNG.

While all the simulations studied here use a version of the Bondi accretion model, SIMBA uses a two-mode model, i.e., a combination of the torque model (for the cold gas) and the Bondi model (for the hot gas). In pratice, the torque model dominates at early times (i.e., for most BHs, except the most massive). 
In that case, the accretion rates onto the BHs is mostly a function of the gas and stellar content in the inner galactic disk, i.e. within a sphere of radius $2 h^{-1}\, \rm ckpc$ around the BHs, rather than the BH mass (only proportional to $M_{\rm BH}^{1/6}$ in the torque model, against $M_{\rm BH}^{2}$ in the Bondi model). 
The growth of BHs being tied to the growth of the inner galactic disks, there is no evolution with redshift.

More massive BHs also accrete hot gas via the Bondi model. These BHs have, on average, lower accretion rates and Eddington ratios. For these BHs, the impact of AGN feedback increases as $f_{\rm Edd}$ decreases (AGN wind velocity increases with decreasing $f_{\rm Edd}$) and contributes to increase the amount of hot gas in the galaxies. A runaway process takes place: increase of BH mass, more important AGN feedback, enhancement of the hot environment, and in return increase of BH mass again via the Bondi accretion channel. 
In the second paper of this series we studied in detail the $f_{\rm Edd}$ distributions of all the simulations, but we give here a few key features. 
At $z=2$ a large fraction of the BHs in the mass range $M_{\rm BH}=10^{8}-10^{9}\, \rm M_{\odot}$ enter the regime of low accretion rates of $\log_{10}\, f_{\rm Edd}\leqslant -2$, while the distribution of lower-mass BHs still peaks at $\log_{10}\, f_{\rm Edd}\leqslant -1$. By $z=0$ the majority of the BHs (independently of their mass) have low Eddington ratios ($\log_{10}\, f_{\rm Edd}\leqslant -2$). Therefore, between $z=2$ and $z=0$ more and more BHs enter a regime in which they quench their host galaxies, but also increase the hot environment of their close surrounding (gas ejected in jets is heated to the virial temperature of the halos, assuming that they thermalise their energy into the surrounding hot gas) and favor their growth through the Bondi channel. The less efficient growth of the galaxies and the additional growth channel of BH growth both contribute to increase the overall normalization of the mean $M_{\rm BH}-M_{\star}$ relation in the redshift range $z=2-0$.

In SIMBA, we also note a redshift evolution of the massive end of the $M_{\rm BH}-M_{\star}$ relation, for $M_{\star}\geqslant 10^{11}\, \rm M_{\odot}$. The Illustris and Horizon-AGN also present an evolution with redshift; there is an overall decrease of the $M_{\rm BH}-M_{\star}$ relation with time. 
As we explained already, in these simulations the galaxies keep growing in mass while BH growth is slower, the mean relations actually move towards more massive galaxies. The more efficient AGN feedback of the TNG and EAGLE simulations suppresses both BH gas accretion and galaxy growth, and the growth of both systems is dominated by mergers in this regime, preventing any time evolution of the $M_{\rm BH}-M_{\star}$ relation. 
In SIMBA, there is an overall increase of the relation with time. This is due to a more important BH growth via gas accretion with respect to galaxy growth in this regime of massive galaxies (Cui et al., in prep).  

\begin{figure*}
\centering
\includegraphics[scale=0.5]{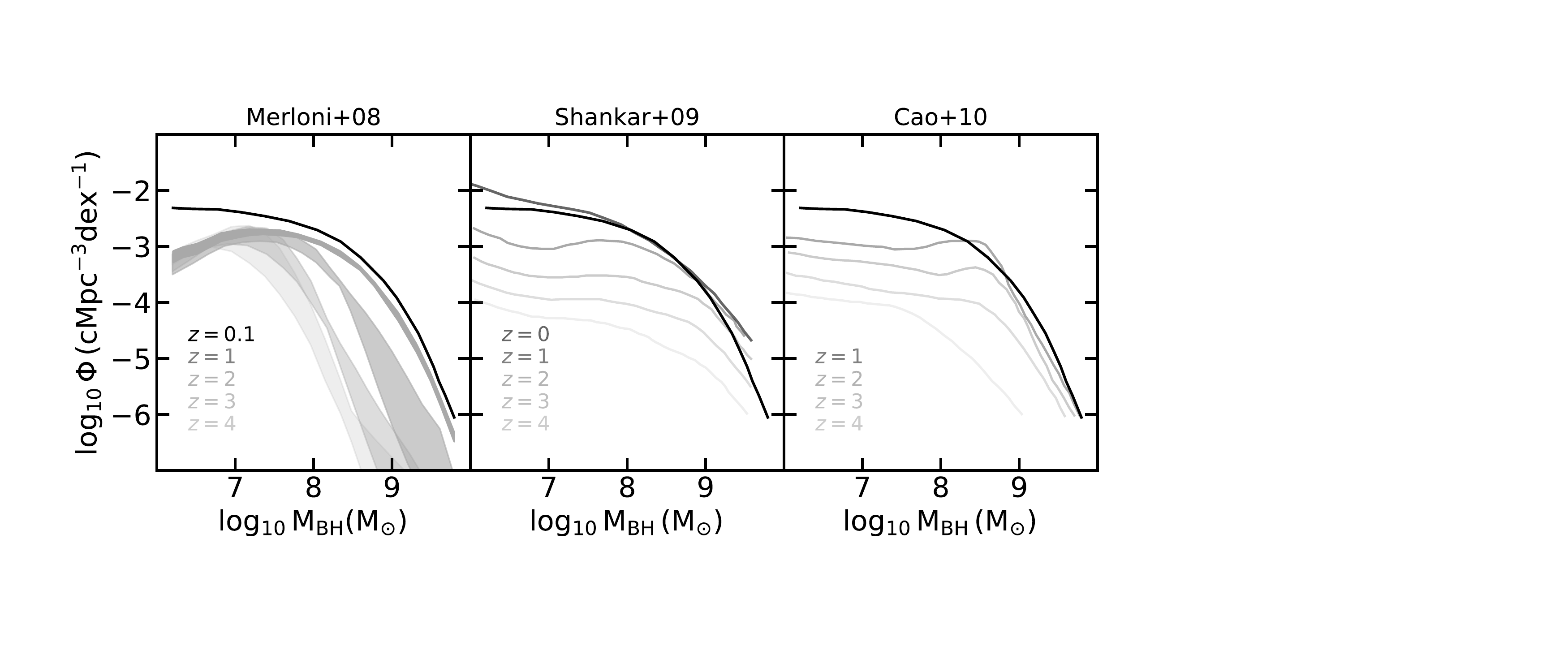}
\caption{Observational constraints of the BH mass function of \citet{Merloni08}, \citet{2009ApJ...690...20S}, and \citet{2010ApJ...725..388C}. We reproduce the constraint of \citet{Merloni08} for $z=0.1$ in all panels (in black). The overall normalization of the BH mass functions increases with time. However, there is no consensus among these constraints on the low-mass end of the BH mass function, or the details of its time evolution.}
\label{fig:bh_mass_fct_obs}
\end{figure*}

\begin{figure*}
\centering
\includegraphics[scale=0.6]{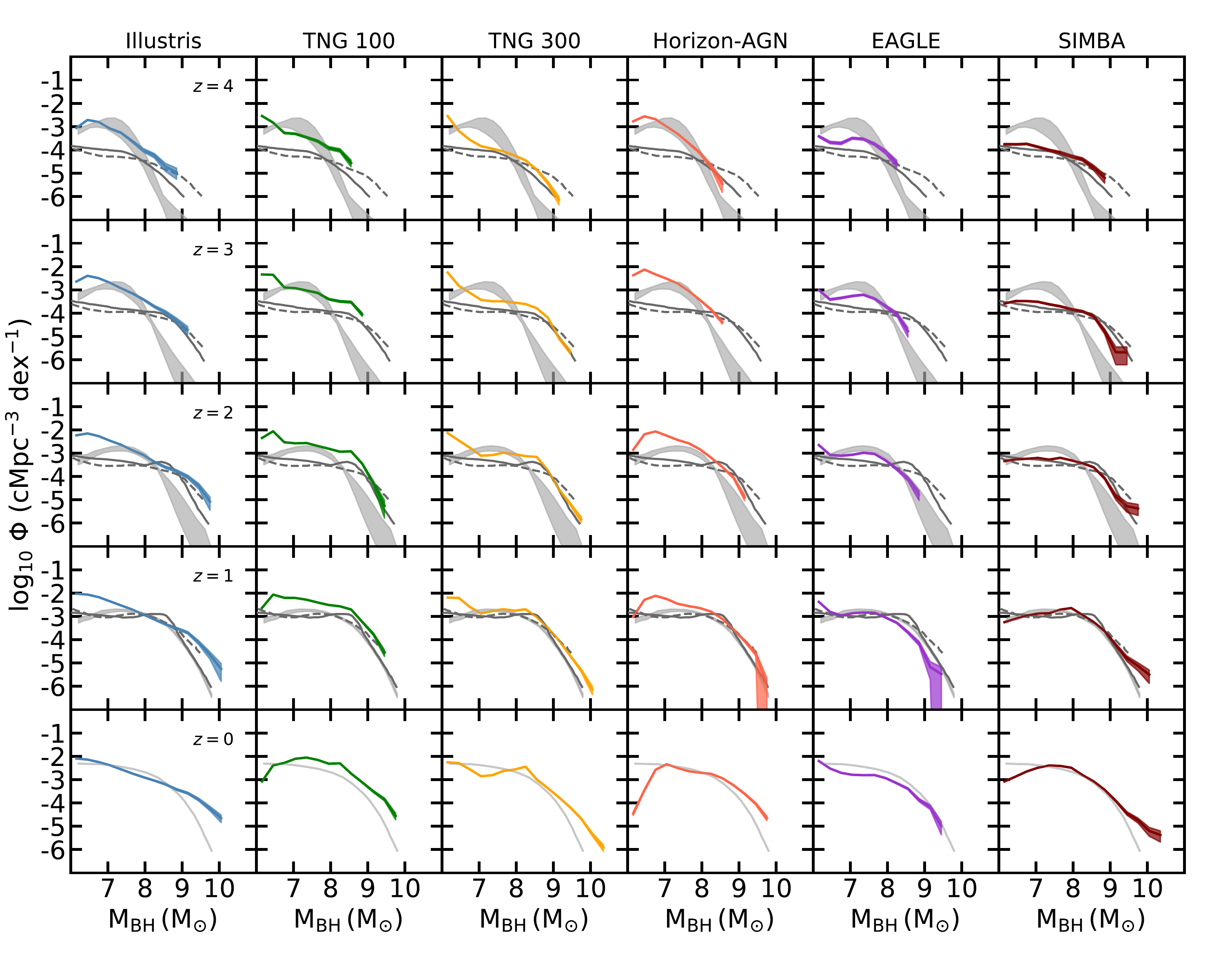}
\caption{Redshift evolution of the BH mass function. Grey shaded areas represent the observational constraints of \citet{Merloni08} for the same redshifts as for the simulations, with the exception of $z=0$ for which we show the observations at $z=0.1$. We also show the observational constraints of \citet{2010ApJ...725..388C} (solid grey lines) and \citet{2009ApJ...690...20S} (dashed grey lines). The global trend seen in the observational constraints is reproduced, but some discrepancies are noticeable. At the low-mass end, some simulations produce too many $M_{\rm BH}\sim 10^{7}\, \rm M_{\odot}$ BHs at some redshift (Illustris, TNG, Horizon-AGN), or not enough (EAGLE), compared to the constraints of \citet{Merloni08}. All the simulations (except EAGLE) overproduce BHs of $M_{\rm BH}\sim 10^{10}\, \rm M_{\odot}$ at $z=0$.}
\label{fig:mass_function}
\end{figure*}

\begin{figure*}
\centering
\includegraphics[scale=0.516]{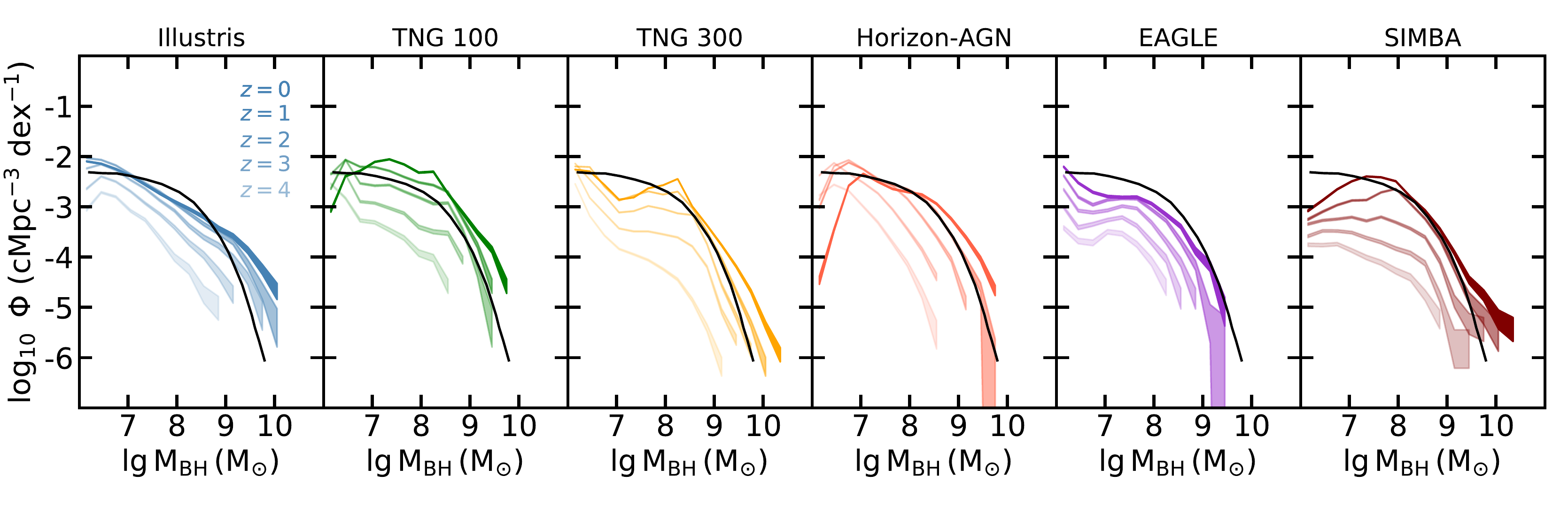}
\caption{Redshift evolution of the BH mass function from all the simulations. Black lines show the BH mass function constraints at $z=0.1$ \citep{Merloni08}. We note for some of the simulations an excess of the most massive BHs. While some simulations do not show a strong change of their normalization between $z=2$ and $z=1$, this is not true for all of them.}
\label{fig:mass_function_2}
\end{figure*}

\begin{figure*}
\centering
\includegraphics[scale=0.516]{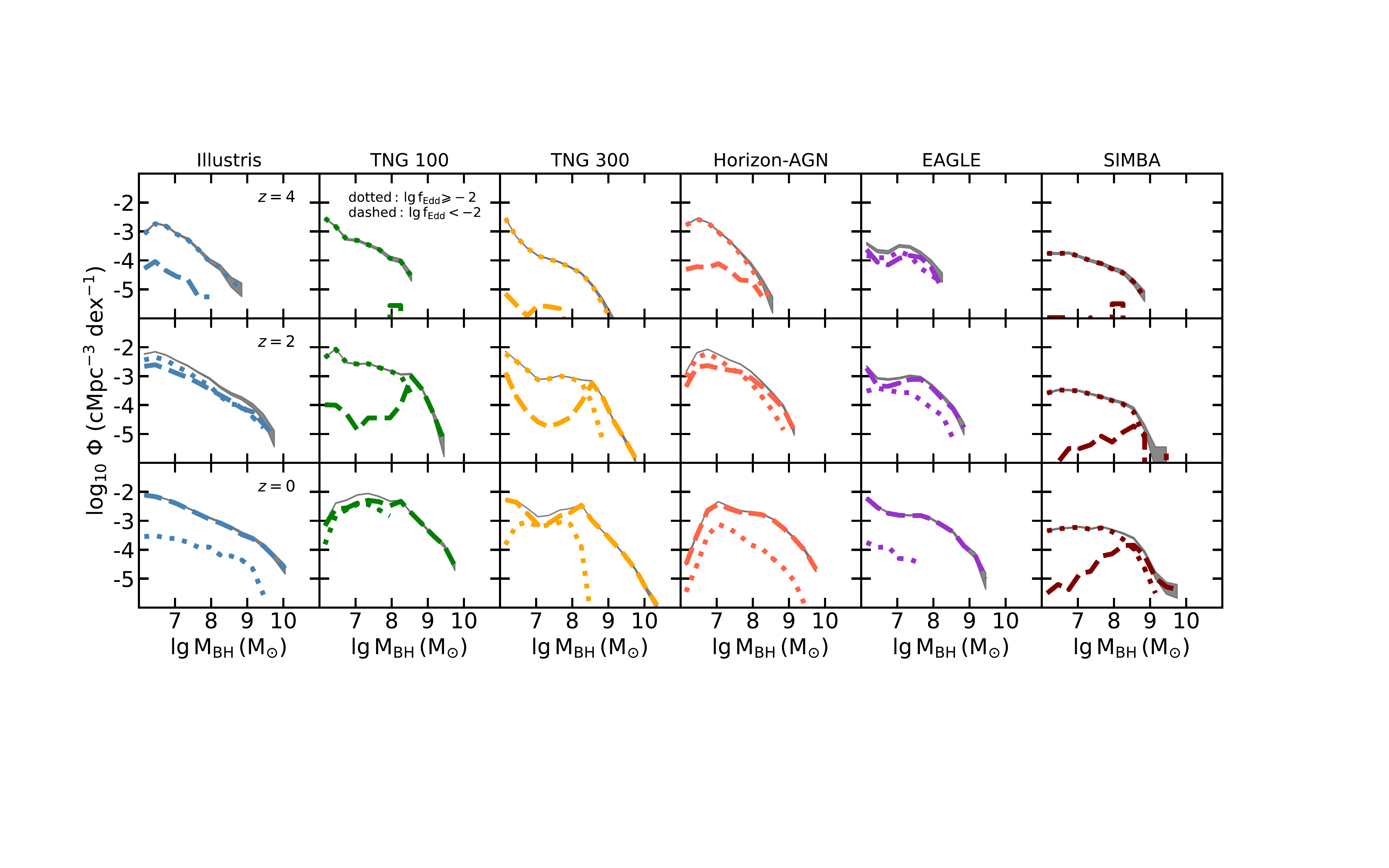}
\caption{BH mass function with time, divided by accreting and quiescent AGN. We show active AGN with $\log_{10} f_{\rm Edd}\geqslant -2$ with the dotted lines and  inactive AGN with $\log_{10} f_{\rm Edd}<-2$ with dashed lines. The grey lines show the global BH mass function. In all simulations, active AGN dominate the BH mass function at high redshift. Inactive BHs dominate at low redshift. We note that in most simulations the inactive BHs start to dominate the high-mass end of the BH mass function first before dominating over all BH masses, with the exception of the simulation SIMBA in which active BHs still dominate the low-mass end of the BH mass function at $z=0$.} 
\label{fig:mass_function_3}
\end{figure*}

\section{BH mass function}
\subsection{Observational constraints}
In this section, we present the observational constraints of \citet{Merloni08}, \citet{2010ApJ...725..388C}, and \citet{2009ApJ...690...20S}. 
BH mass functions are often derived from a given relation between BH mass and a given property of their host galaxies in the local Universe, which can be their total stellar masses, or more likely their stellar velocity dispersion or the luminosity of their bulges. In this case, deriving estimations of the BH mass function at higher redshifts means assuming an evolution of the given scaling relation with time (which is not well constrained at the moment). Other methods combine the estimate of the BH mass function at $z=0$ and either an AGN lightcurve \citep[e.g.,][]{2009ApJ...690...20S, 2010ApJ...725..388C} or the BH mass function of AGN \citep[e.g.,][]{Merloni08}. For both, the underlying goal is to estimate how much the BHs accrete as a function of BH mass and time (i.e., the average growth rate).
In the first case, the methods need to state/assume for how long BHs are efficiently accreting, i.e., AGN lifetime/duty cycle, radiative efficiency, which can be a function of BH mass, redshift, BH populations, obscured/non-obscured AGN, etc. Different lightcurves have been employed. For example, \citet{2010ApJ...725..388C} use a lightcurve with a power-law decay mimicking the decrease of BH gas accretion due to AGN feedback.
In the second case, the methods rely again on scaling relations to derive the distribution of BH accretion rates empirically (instead of assuming a given AGN lightcurve). The review of \citet{2012AdAst2012E...7K} provides a full description of all the methods that have been used in the literature and the pros and cons for each of them. 

The three constraints that we show in Fig.\ref{fig:bh_mass_fct_obs} do not reach a consensus on either the normalization of the BH mass function, the low-mass end, and do present some differences in their time evolutions. We summarize this below.
\begin{itemize}
\item {\bf Normalization of the BH mass function}. All the constraints have an increased overall normalization with time (i.e., decreasing redshift). The normalization of the functions increases at all BH masses in \citet{2010ApJ...725..388C} and \citet{2009ApJ...690...20S}, while the normalization of the low-mass end ($M_{\rm BH}\leqslant 10^{7}\, \rm M_{\odot}$) of the \citet{Merloni08} constraints does not evolve with time.
The shapes of the BH mass functions are also different, and this, to some degree, depends on the Eddington ratio distribution employed in the models \citep{2012AdAst2012E...7K,2013MNRAS.428..421S}.
\item {\bf Low-mass end of the BH mass function}. The constraint of \citet{Merloni08} presents a turn over at the low-mass end for $z\geqslant 1$. 
Instead, the BH mass functions of \citet{2010ApJ...725..388C} and \citet{2009ApJ...690...20S} more or less reach a plateau at the low-mass end.
In these models, these differences are the by-product of the assumptions made when imposing the boundary conditions at the low-mass end \citep{2009ApJ...690...20S,2010ApJ...725..388C}.

\item {\bf Time evolution}. While all the overall constraints evolve upward with time, we note some differences in the time evolutions. The BH mass function of \citet{2009ApJ...690...20S} evolves gradually in intervals of $\Delta z=1$, at all BH masses. The massive end of the BH mass function evolves slightly more rapidly in \citet{2009ApJ...690...20S}, and even more so in \citet{2010ApJ...725..388C}. In \citet{Merloni08}, the BH mass function does not evolve much at high redshift, between $z=4$ and $z=3$ independently of BH mass. However, a stronger evolution is observed for $z\leqslant 3$ (i.e., more or less at the peak of cosmic AGN activity) for $M_{\rm BH}\geqslant 10^{7.5}\, \rm M_{\odot}$. 
The peak of the BH mass functions evolves towards higher mass BHs in the constraints of \citet{Merloni08} and \citet{2010ApJ...725..388C}; the identification of a peak in \citet{2009ApJ...690...20S} is harder.
In \citet{Merloni08}, the low-mass end does not evolve in the redshift range $z=4-1$, while the high-mass end does. 
The evolution can, therefore, be seen as {\it anti-hierarchical}: the massive end of the BH mass function assembles first, before the full low-mass BH population does (at $z\sim 0$). 
The build-up of the BH mass growth is different in the constraints of \citet{2009ApJ...690...20S} and \citet{2010ApJ...725..388C}. The BH mass functions, instead, build up uniformly for all BH masses up to $z\sim2$. At $z\sim2$ while the massive end is already built and does not evolve much anymore, the low-mass end keeps building up with time up to $z=0$.
\end{itemize}

Given these discrepancies between the observational constraints, one needs to be careful when addressing the differences with large-scale cosmological simulations.
Moreover, to derive their BH mass functions, the studies described here assume a radiative efficiency, which is different (lower) from what is assumed in the large-scale simulations. For example, \citet{Merloni08} assume a radiative $\epsilon_{\rm r}=0.07$. Since $\epsilon_{\rm r}$ is largely degenerate with other parameters in the simulations and used more or less as a free parameter to match a given scaling relation, it will not affect strongly our qualitative comparison in the following.

\subsection{Time evolution of the BH mass function}

In Fig.~\ref{fig:mass_function}, we show the simulation BH mass functions. We select all galaxies above the resolution limit $M_{\star}\geqslant 5\times 10^{8}\, \rm M_{\odot}$. We compare the simulation results to the observational constraints presented above: \citet{Merloni08}, \citet{2010ApJ...725..388C} (solid grey lines) and \citet{2009ApJ...690...20S} (dashed grey lines).

The global trends of the observations are reproduced, i.e., that the normalization of the BH mass function increases with time, and less massive BHs are more abundant than their most massive counterpart.
While these global trends are in agreement with observations, we do find some noticeable differences with observations. 
In the following, we start by presenting the results for $z>0$ and then we address the comparison between the constraints and the simulations at $z=0$. We also discuss the time evolution of the BH mass function obtained in the simulations.\\

\noindent {\bf Low-mass end of the BH mass function:}
\noindent In \citet{Merloni08}, the BH mass function peaks at about $M_{\rm BH}\sim 10^{7-8}\, \rm M_{\odot}$. We use this peak as a reference and call the low-mass end the part of the function below that peak.  
We find that the Illustris, TNG100, and Horizon-AGN simulations, over-predict the number of lower-mass BHs ($M_{\rm BH}\leqslant 10^{7-8}\, \rm M_{\odot}$). TNG300 over-predicts the number of BHs only for the less massive BHs of $M_{\rm BH}\sim 10^{6.5}\, \rm M_{\odot}$. The EAGLE and SIMBA simulations have a lower normalization of the low-mass end. The number of BHs produced is lower than the constraints of \citet{Merloni08} at $z\geqslant 3$. EAGLE has a good agreement with the constraints of \citet{Merloni08} at lower redshifts, while we find a good agreement between the simulation SIMBA and the constraints of \citet{2009ApJ...690...20S} and \citet{2010ApJ...725..388C} at all redshift.\\

\noindent{\bf High-mass end of the BH mass function:}
\noindent 
For more massive BHs of $M_{\rm BH}\geqslant 10^{8}\, \rm M_{\odot}$, Horizon-AGN and EAGLE provide a good match to the constraints of \citet{Merloni08}.
Illustris, TNG, and SIMBA seem to form too many massive BHs when compared to \citet{Merloni08}, but in good agreement with \citet{2010ApJ...725..388C} and \citet{2009ApJ...690...20S} for $z\geqslant 1$.
Interestingly, the simulation SIMBA produces a very good agreement with the BH mass function derived in \citet{2010ApJ...725..388C} (for $z\geqslant 1$) and \citet{2009ApJ...690...20S} for the entire BH mass range and all redshifts. Illustris and TNG produce a slightly higher number of massive BHs with respect to the constraints presented here (at $z=1,2$ for Illustris, and at $z\geqslant 2$ for TNG). \\
Finally, all the simulations except EAGLE over-predict the number of massive BHs with $10^{9}-10^{10}\, \rm M_{\odot}$ compared to the constraints of \citet{Merloni08}, for all redshifts.\\

\noindent{\bf BH mass function at $z=0$:}
By $z=0$, we find that the overall normalizations of the BH mass functions in the simulations are in good agreement with the constraints of \citet{Merloni08}. At higher redshift, \citet{Merloni08} predict a BH mass distribution that peaks at $M_{\rm BH}\geqslant 10^{7-8}\, \rm M_{\odot}$, with fewer lower-mass BHs. At $z=0$ the distribution flattens at the BH low-mass end (presence of more low-mass BHs of $M_{\rm BH}\leqslant 10^{7}\, \rm M_{\odot}$ at $z\sim 0$ than at $z\geqslant 1$). 
We see the same behavior in Illustris. Most of the other simulations (TNG100, HorizonAGN, and SIMBA) predict a drop of the BH mass function for the less massive BHs, with the exception of the EAGLE simulation, whose shape is very similar to the shape found in \citet{2009ApJ...690...20S}.
The low-mass end of Horizon-AGN is the lowest at $z=0$; the drop increases between $z=1$ and $z=0$. The number of BHs with $M_{\rm BH}\sim 10^{6}\, \rm M_{\odot}$ is reduced by one order of magnitude. This is due to the modeling of BH formation in the simulation; BHs cannot form after $z<1.5$, and therefore there are no newly formed BHs to fill the gap at the low-mass end for $z<1$.

At $z=0$, the constraint of \citet{Merloni08} predicts that one BH of $M_{\rm BH}\sim 10^{10}\, \rm M_{\odot}$ should be found in a volume of $(100 \, \rm cMpc)^{3}$ (grey line in the bottom panels of Fig.~\ref{fig:mass_function}). Except for EAGLE, all the simulations produce more than a single BH at this mass. This can also clearly be seen in the bottom panels of Fig.~\ref{fig:scaling_relation}, where the number of these massive BHs (located at the top right side of the black line region) is higher in the simulations than in observations.\\

\noindent{\bf Time evolution of the BH mass function:}
In Fig.~\ref{fig:mass_function_2}, we show the same BH mass functions for the same redshift but on a single panel for each simulation, where it is easier to see the build-up of the BH mass function with time. Black lines indicates the observational constraint of \citet{Merloni08} for $z=0.1$. 
In all the simulations, the normalization of the BH mass function evolves over one order of magnitude. 
In most of them, the normalization increases for all BH masses with time: 
quite strongly between $z=4$ to $z=1$, and then slows down to $z=0$. 
Some of the $M_{\rm BH}\sim10^{7-9}\rm M_{\odot}$ populations are almost already in place at $z=1$.

The homogeneous increase of the normalization in the simulations is more similar to the constraints of \citet{2009ApJ...690...20S,2010ApJ...725..388C} than \citet{Merloni08}. However, at $z\leqslant 2-1$ the massive end of their BH mass functions ($M_{\rm BH}\geqslant 10^{9}\, \rm M_{\odot}$) is already built and do not evolve anymore ({\it anti-hierarchical} build-up), only the low-mass end still assembles.
In the simulations, the massive end of the BH mass function ($M_{\rm BH}\geqslant 10^{9}\, \rm M_{\odot}$) keeps evolving after $z\sim 2-1$.
However, it seems that the simulations also present an excess of the most massive BHs compared to the observational constraints. If the growth of these massive BHs was more regulated, it could lead to a more {\it anti-hierarchical} build-up of the BH mass function.

\subsection{Contribution from active and inactive BHs}

\begin{figure}
\centering
\includegraphics[scale=0.6]{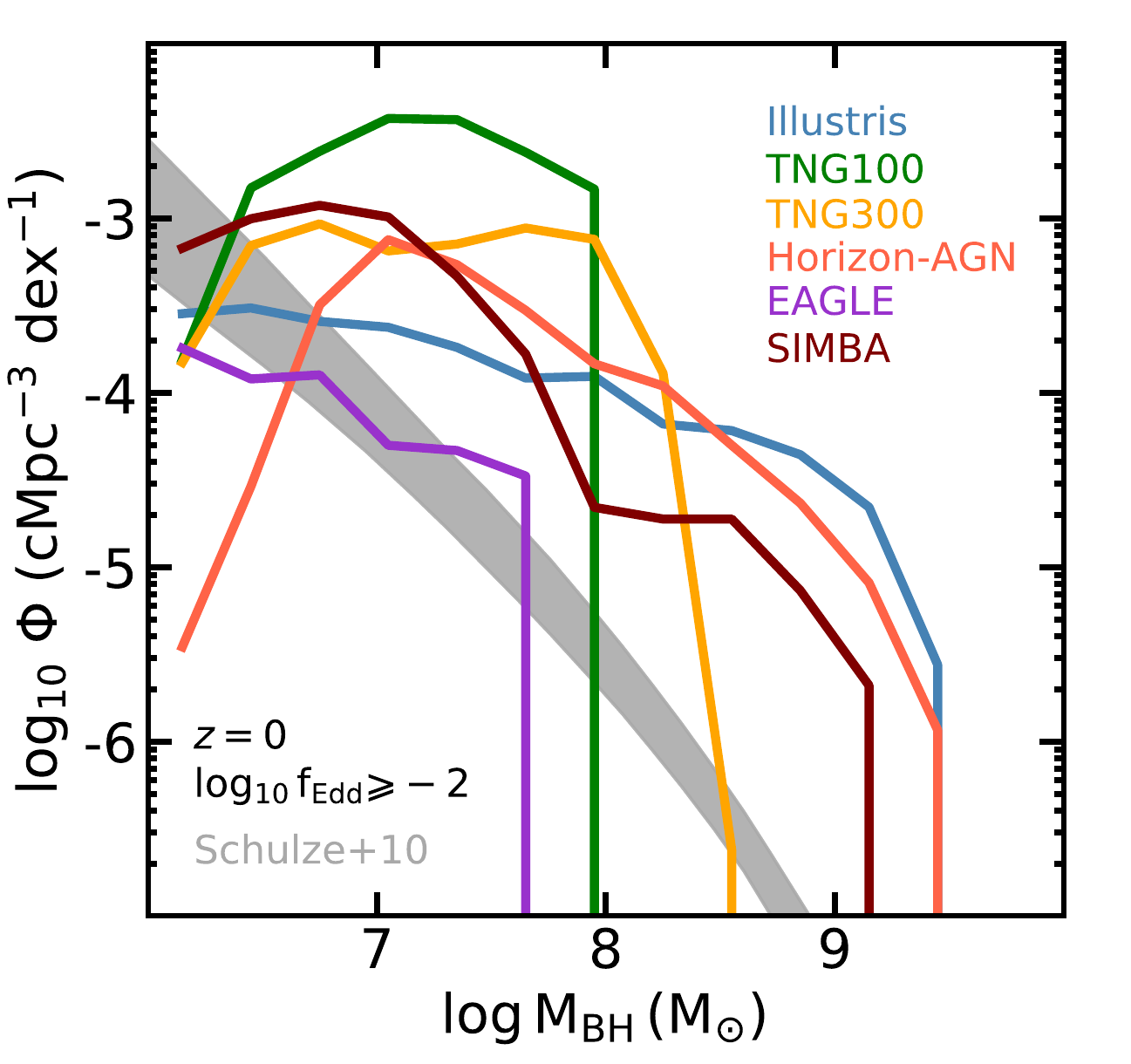}
\caption{BH mass function of the rapidly growing BHs, defined as $\log_{10} \, f_{\rm Edd}\geqslant -2$. We add the constraints of \citet{2010A&A...516A..87S} assuming the same definition $\log_{10} \, f_{\rm Edd}\geqslant -2$. }
\label{fig:mass_function_active}
\end{figure}

To understand the build-up of the BH mass function, we divide in Fig.~\ref{fig:mass_function_active} the function by efficiently accreting BHs (also called active BHs and defined as $\log_{10} \, f_{\rm Edd}\geqslant -2$, dotted lines) and BHs with lower Eddington ratios ($\log_{10} \,f_{\rm Edd}<-2$, dashed lines). The global BH mass functions are shown with grey lines.
At early times ($z\geqslant 4$) the BH mass function is driven by BHs with high accretion rates in all the simulations. In the EAGLE simulation, the massive end of the function is already dominated by inactive BHs by that time.
For the other simulations the contribution of inactive BHs becomes predominant around $z=2$ for the massive BHs, and then gradually covers the lower mass end of the BH mass function. The effect is sometimes called {\it downsizing} or {\it anti-hierarchical growth} (while using a different definition than in the previous section): at high redshift the inactive BHs are massive, and at lower redshift inactive BHs encompass more lower-mass BHs. In other words, massive BHs at $z=0$ are less active than the same mass BHs at earlier times \citep[e.g.][]{Merloni2004,2007ApJ...667..131G,Vestergaard2008,Vestergaard2009,Hirschmann2012}.
At $z=0$, we still note some differences among the simulations. While in Illustris, Horizon-AGN, and EAGLE, the contribution from active BHs is low (at all BH masses), which means that the BH mass function is the result of BHs with $\log_{10} \,f_{\rm Edd}<-2$, the contribution of inactive and active BHs are more or less the same for the low-mass end of TNG100. In SIMBA, the massive end is dominated by inactive BHs, but the active BHs still completely take over for the low-mass of the BH mass function (for $M_{\rm BH}\leqslant10^{8.5}\, \rm M_{\odot}$). 
The large fraction of active BHs, i.e., BHs with luminosity of $L_{\rm bol}\geqslant 10^{42}\, \rm erg/s$, for this mass range at $z=0$ in SIMBA can be seen in Fig.~\ref{fig:scaling_relation_Lbol}. 
While comparing the results for Illustris and TNG, we see a very strong decrease in the number of active BHs (dotted lines) at the characteristic mass of $M_{\rm BH}\sim 10^{8}\, \rm M_{\odot}$ in TNG compared to Illustris. This is a signature of the kinetic AGN feedback mode, which is able to suppress BH accretion in these massive BHs significantly.

The picture that emerges from our analysis is the following. In simulations, BHs grow primarily as active BHs rapidly accreting (i.e., $\log_{10} \, f_{\rm Edd}\geqslant -2$). Indeed, at high redshift we see that the contribution of non-active BHs (i.e., $\log_{10} \, f_{\rm Edd}< -2$) is modest. 
The massive end of the BH mass function is built first with active BHs, and is rapidly dominated by non-active BHs as soon as $z=2$.
The above picture is more or less the same for all the simulations. Still, we have noted some differences among the simulations (as stated above), which indicate that the simulations can have different growth histories for their BHs.
The final contribution of active BHs and inactive BHs to the BH mass function at $z=0$ is also different for the different simulations. To stress the differences at $z=0$, we show the active BH mass function of the simulations in Fig.~\ref{fig:mass_function_active}.

A few papers investigate the empirical active BH mass function \citep{Marconi2004,2007ApJ...667..131G,Vestergaard2008,Vestergaard2009,2010ApJ...719.1315K,2010A&A...516A..87S}, by estimation from broad-line AGN. At $z=0$, it appears that the number of broad-line AGN is much lower than the empirical estimates of the full BH mass function, at all masses. At higher redshift (e.g., $z=2$), AGN still seem to represent a small fraction of the BHs, except for the most massive BHs. \citet{2010ApJ...719.1315K} find that the number of broad-line AGN with $M_{\rm BH}\geqslant 10^{9}\, \rm M_{\odot}$ ($z=2$) could be high and within the scatter of all the empirical estimates of the BH mass function. 
In Fig.~\ref{fig:mass_function_active}, we also show the observational constraints of \citet{2010A&A...516A..87S} which use the same definition for active BHs as the one that we employ for the simulations, i.e. $\log_{10}\, f_{\rm Edd}\geqslant -2$\footnote{The definition of active BHs is important, and impacts the normalization but also the shape of the active BH mass function, both in simulations and observations. However, our main conclusions on the evolution and $z=0$ contributions of active and inactive BHs to the BH mass function are the same with a different definition such as $\log_{10}\, f_{\rm Edd}\geqslant -1$.}. 
The active BH mass function in the constraints is reconstructed: it takes into account the fact that BHs in the observational sample of \citet{2010A&A...516A..87S} are flux selected (only AGN above the flux limit are detected) and not BH mass selected. These selection effects primarily impact the low-mass end of the active BH mass function; the uncorrected data of \citet{2010A&A...516A..87S} present a decrease of the function towards the low-mass end \citep[which is also present in][for the same reason\footnote{Possible explanations for the discrepancies among the estimates of the active BH mass function can be found in \citet{2010A&A...516A..87S}. The authors particularly discuss the discrepancies of several orders of magnitude ($\Phi=10^{-8}-10^{-3}\, \rm cMpc^{-3} \,dex^{-1} $) for BHs of $M_{\rm BH}\leqslant 10^{7.5}\, \rm M_{\odot}$ with respect to the studies of \citet{2007ApJ...667..131G,2009ApJ...704.1743G,Vestergaard2008,Vestergaard2009}.}]{2007ApJ...667..131G}.
We find that all the simulations (except EAGLE) have a higher normalization of their low-mass end of the active BH mass function ($M_{\rm BH}\leqslant 10^{8}\, M_{\odot}$). As previously said, we note the complete absence of active BHs among the $M_{\rm BH}\geqslant 10^{8}\, M_{\odot}$ BHs in some of the simulations (EAGLE, TNG). Horizon-AGN, SIMBA, and Illustris have higher normalization at the high-mass end than the constraints.

\section{Discussion}
\subsection{The $M_{\rm BH}-M_{\star}$ diagram and relation in simulations and observations}

While all the simulations follow, by design, one of the empirical $M_{\rm BH}-M_{\star}$ scaling relations, they struggle to produce the diversity of BHs observed in galaxies in the local Universe \citep[e.g.,][]{2015arXiv150806274R,2019MNRAS.487.3404B}.\\

\noindent {\bf BHs in low-mass galaxies of $M_{\star}\sim 10^{9}\, \rm M_{\odot}$.}
Systematic searches for low-mass BHs in dwarf galaxies have led to the discovery of BHs (mostly AGN) of $M_{\rm BH}\leqslant 10^{6}\, \rm M_{\odot}$ in galaxies of $M_{\star}\sim 10^{9}\, \rm M_{\odot}$ \citep{2012NatCo...3E1304G,2013ApJ...775..116R,2015ApJ...809L..14B}. The seeding mass in the TNG model is $M_{\rm BH}\sim 10^{6}\, \rm M_{\odot}$, and SIMBA starts seeding galaxies of $M_{\star}\geqslant 10^{9.5}\, \rm M_{\odot}$ (Fig.~\ref{fig:scaling_relation}). By their design, these simulations do not cover the regime of low-mass BHs ($M_{\rm BH}\leqslant 10^{6}\, \rm M_{\odot}$) in low-mass galaxies. The predictions from all the simulations on the low-mass galaxy regime are uncertain due to limited resolution, and seeding prescriptions.
While massive BHs of $M_{\rm BH}\geqslant 10^{7}\, \rm M_{\odot}$ have not been found yet in observations of low-mass galaxies, the TNG, Horizon-AGN and SIMBA (a few only) predict some. While these BHs are inactive in SIMBA, some of them can be qualified as AGN in TNG and Horizon-AGN (Fig.~\ref{fig:scaling_relation_Lbol}). If existing in the local Universe, these massive AGN should have been easier to identify in low-mass galaxies than the lower-mass AGN that have been already detected. \\

\noindent {\bf BHs in intermediate mass galaxies of $M_{\star}\sim 10^{10-11}\, \rm M_{\odot}$.}
Some BHs of $M_{\rm BH}\geqslant 10^{8}\, \rm M_{\odot}$ in galaxies of $M_{\star}\sim 10^{10}\, \rm M_{\odot}$ have been observed, but are often lacking in the simulations. In EAGLE, all these BHs were found in very compact satellite galaxies that experienced a combination of early stellar assembly and tidal stripping of their stellar component \citep{2016MNRAS.460.1147B}. 
In general, the simulations do not produce the lowest-mass and highest-mass BHs at fixed galaxy stellar mass found in observations in the range $M_{\star}\sim 10^{10-11.5}\, \rm M_{\odot}$.
Particularly, the low-mass AGN \citep[broad-line AGN in][]{2015arXiv150806274R} of $M_{\rm BH}\sim 10^{6}-10^{7.5}\, \rm M_{\odot}$ observed in galaxies of $M_{\star}\sim10^{10}-10^{11.5}\, \rm M_{\odot}$ are not formed in many simulations.\\

\noindent {\bf BHs in massive galaxies of $M_{\star}\geqslant 10^{12}\, \rm M_{\odot}$.}
We demonstrated with the BH mass function that there is probably an excess of the most massive BHs ($M_{\rm BH}\sim 10^{9}-10^{10}\, \rm M_{\odot}$) in simulations, compared to the constraints of \citet{Merloni08}. 
The most massive of these simulated BHs ($M_{\rm BH}\sim 10^{10}\, \rm M_{\odot}$) are embedded in massive galaxies of $M_{\star}\sim 10^{12}\, \rm M_{\odot}$, a region not covered by \citet{2015arXiv150806274R}.
The luminosity function of such observed galaxies with $M_{\star}\sim 10^{12}\, \rm M_{\odot}$ at $z\sim 0$ is within the range $\Phi \sim 5\times 10^{-6}-10^{-5.5}\, \rm  cMpc^{-3} \, dex^{-1}$ \citep{2008MNRAS.388..945B,2013ApJ...777...18M,2013MNRAS.436..697B,2015MNRAS.454.4027D}.
All the simulations tend to have extended and tight $M_{\rm BH}-M_{\star}$ relation after $M_{\star}\geqslant 10^{11.5}\, \rm M_{\odot}$. 
BHs of $M_{\rm BH}\leqslant 10^{10.5}$ have been found in BCGs (brightest cluster galaxies) with BH dynamical mass measurements \citep{2013ARA&A..51..511K,2013ApJ...764..184M}, and up to $M_{\rm BH}\leqslant 10^{12}\, \rm M_{\odot}$ for BH mass estimates (e.g., mass inferred from the fundamental plane, BCG K-band luminosities, X-ray and radio luminosities). The most massive BHs reside in BCGs with bulge mass up to $M_{\rm bulge}<10^{12.5} \, \rm M_{\odot}$. 
A non-negligeable fraction of the BHs residing in BGGs and BCGs are overmassive with respect to the bulge stellar mass or the stellar velocity dispersion of their host galaxies \citep{2018ApJ...852..131B,Phipps_2019}, compared to the scaling relation of \citet{2013ApJ...764..184M} for example. 
In the comparison of \citet{2018ApJ...852..131B} with the Horizon-AGN simulation, a very good agreement is found in the plane $M_{\rm BH}-M_{\rm 500}$ ($M_{\rm 500}$ the mass enclosed in a sphere of mean mass density 500 times the Universe critical density). While a relatively good agreement is also found for the $M_{\rm BH}-M_{\rm bulge}$ relation (tighter relation in Horizon-AGN).
This tends to show that cosmological simulations produce tighter relations even for the most massive systems when we compare to actual samples of BHs in BCGs/BGGs.
Interestingly, a few simulated galaxies were identified in Horizon-AGN with $M_{\rm bulge}\geqslant 10^{12.5}\, \rm M_{\odot}$, i.e., more massive than the observed ones \citep{2018ApJ...852..131B,Phipps_2019}, and overmassive with respect to their $M_{\rm 500}$. These simulated galaxies could be too massive while providing a good agreement for the mass of their BHs. Further comparisons with simulations are needed to understand this regime of very massive systems, and whether the growth of their BHs is triggered in a different way (enhanced number of galaxy mergers, BH mergers, dry mergers, but also low-angular momentum and cold gas flows) than in more normal galaxies.\\

Two scaling relations are favored in the local sample of \citet{2015arXiv150806274R}:
one for inactive BHs in elliptical and bulge galaxies with dynamical BH mass measurements, and one with a lower normalization for active BHs (i.e., broad-line AGN with masses estimated via reverberation mapping or single-epoch virial theorem). 
While the two subsamples seem distinct at first glance, \citet{2015arXiv150806274R} discuss how the quiescent BH population also likely overlaps with the lower AGN relation, but for the most part, they are not detected. The galaxies with pseudo-bulges and dynamical BH masses also overlap with the broad-line AGN.
The large sample of \citet{2019MNRAS.487.3404B} with BH mass estimates from narrow-line emissions includes a larger number of more massive BHs ($M_{\rm BH}\sim 10^{7.5}-10^{9}\, \rm M_{\odot}$) in galaxies of $M_{\star}\sim 10^{10}\, \rm M_{\odot}$. It shows that AGN also populate the region between the two scaling relations for inactive and active BHs of \citet{2015arXiv150806274R}, i.e., BHs of $M_{\rm BH}\sim 10^{7}-10^{8.5}\, \rm M_{\odot}$ in galaxies of $M_{\star}\sim 10^{11}-10^{11.8}\, \rm M_{\odot}$. This suggests again that the two previous samples of inactive and active BHs may just be sub-groups of a more global population of BHs. The sample of \citet{2019MNRAS.487.3404B} also reinforces our finding that the population of BHs in simulations may not successfully reproduce the diversity of observed BHs (smaller scatter in simulations than observations at fixed $M_{\star}$).
Given the low number of objects in the observational samples, incompleteness, selection biases, samples based on AGN (which are rare objects), the uncertainties on BH mass estimates, the scatter of the $M_{\rm BH}-M_{\star}$ relations probably does not reflect accurately the intrinsic scatter of the BH population in the Universe. In the simulations we consider all the BHs, AGN or inactive, and we do not take into account the volume of the samples and simulations that we compare. 
Finally, here the galaxy total stellar mass is used for both simulations and observations, but these computations may not be the same and affect our comparison.
For all these reasons, our comparison with observations need to be taken with a grain of salt. 
As observational samples become larger, we must include volume-weighted comparisons with simulations, and not only compare which regions of the $M_{\rm BH}-M_{\star}$ observational plane are produced by simulations but also whether the corresponding galaxies have the same types/morphologies as in the observations. Here, we started this process by comparing the star forming properties of the observed and simulated galaxies in the $M_{\rm BH}-M_{\star}$ plane.

\subsection{Moving forward: Improving the diversity of BHs in simulations}
The lower scatter of the $M_{\rm BH}-M_{\star}$ relation in the simulations may be due to the lack of stochasticity in the galaxy/BH sub-grid models. They all use free parameters for the efficiency of physical processes (e.g., radiative efficiency, BH accretion, SN feedback, AGN feedback). These parameters are often set the same for all galaxies, independently of e.g., their mass, evolution, environment, redshift. Instead, these parameters could be more physically motivated, and/or randomly drawn from distributions rather than being fixed values. The latter solution would allow us to investigate the likely larger scatter of the scaling relations.

Instead of seeding all galaxies/halos reaching a given mass, the seeding could be based on the local properties of the gas such as density \citep{2014MNRAS.444.1453D} and gas metallicity \citep{Bellovary2011,2017MNRAS.468.3935H,2017MNRAS.470.1121T}. The least massive supermassive BH observed in the local Universe has a mass of $M_{\rm BH}\sim 5\times 10^{4}\, \rm M_{\odot}$ and is located in a $M_{\star}\sim 2.5 \times 10^{9}\, \rm M_{\odot}$ galaxy \citep{2015ApJ...809L..14B}. While large-scale simulations of $\sim 100\,\rm cMpc$ do not have the resolution to implement completely physical BH formation mechanisms, they could use seeding mass allowing for the formation of the low-mass BHs observed locally (i.e., $M_{\rm BH}=10^{4}\, \rm M_{\odot}$). 
Decreasing the initial mass of BHs is also challenging numerically. To follow the dynamics of lower-mass BHs and their growth, the seed mass of the BH particles must be sufficiently higher than the mass of the other particles. This allows to avoid scattering of the BHs with the other particles, i.e., stellar, dark matter, and gas particles \citep{2015MNRAS.451.1868T,2019MNRAS.486..101P}. The lower the BH seed mass, the higher the resolution of the simulation must be. 
To account for the stochasticity of BH formation and unresolved early growth in primordial halos, a distribution of initial BH masses could be employed.

In simulations, BH growth is self-regulated by AGN feedback. A lower AGN feedback efficiency leads to more massive BHs for the same galaxy properties and a higher feedback efficiency to less massive BHs \citep{DiMatteo2005,Springel2005}.
The radiative efficiency of BH accretion, which is one of the factors determining the strength of AGN feedback, is usually set to a fixed value for all BHs ($10\%$ or $20\%$).  The radiative efficiency is closely related to the spin of the BHs, since the spin sets the marginally stable orbit beyond which matter will fall onto the BHs without losing further energy. The BH accretion disk extends closer or further to the BH depending on its spin. Assuming a distribution of radiative efficiencies rather than a fixed value would naturally increase the $M_{\rm BH}-M_{\star}$ scatter. 
BH spins are not followed consistently in the simulations studied here. Doing so could both allow to self-consistently determine the radiative efficiency for every BH, and also increase the scatter especially for the massive end of $M_{\star}\geqslant 10^{10}\, \rm M_{\odot}$ \citep{2014MNRAS.440.1590D,2019MNRAS.490.4133B}. 
Consequently, this would affect AGN feedback, since the amount of energy that is released from AGN feedback is higher for highly spinning BHs, and lower for non-spinning BHs. 
The efficiency of AGN feedback also depends on a variety of other factors that the large-scale cosmological simulations do not resolve, such as the phase structure of the gas into which the energy is deposited. 
The multi-phase interstellar medium (ISM) is not modeled in most of the simulations, and this can significantly impact both the accretion onto the BHs and the coupling efficiency of AGN feedback, and therefore also the scatter in the $M_{\rm BH}-M_{\star}$ plane.
Better simulation resolutions could help capturing greater starburst and accretion rates onto the BHs, potentially increasing the scatter as well.
The modeling of BH accretion itself can affect the scatter, both by the dependence on BH mass (strong dependence in the Bondi model, and weak dependence in the torque model employed in SIMBA), but also by the size of the region considered to compute the accretion rates (parent cell of the BHs, or kernel).

The fact that simulations capture a mean $M_{\rm BH}-M_{\star}$ relation in agreement with empirical scaling relations but not the full scatter at fixed stellar mass (if this is confirmed by new larger observational samples) may not seem like a major problem at first glance, but it may be. The observational samples of \citet{2015arXiv150806274R,2019MNRAS.487.3404B} demonstrate that there is a large number of BHs with $M_{\rm BH}\leqslant 10^{8}\, \rm M_{\odot}$ in galaxies of $M_{\star}= 10^{10.5-11.7}\, \rm M_{\odot}$ in the local Universe. These galaxies would be considered as having under-massive BHs in most of the simulations compared to their mean relation. We need to understand whether these observed galaxies represent simply the tail of the BH mass distribution at fixed stellar mass, or if they represent the mean behavior of the BH population. In the latter case, BHs being less massive than what is found in the simulations would mean that they have experienced less growth and/or mergers and that they have probably injected less energy via AGN feedback in their host galaxies through their history. Would we find the same galaxy properties for these galaxies of $M_{\star}= 10^{10.5-11.7}\, \rm M_{\odot}$ in the simulations if they would host a BH of $M_{\rm BH}\leqslant 10^{6.5-8}\, \rm M_{\odot}$ instead of  $M_{\rm BH}> 10^{8}\, \rm M_{\odot}$? For example, TNG galaxies quench through the kinetic low-accretion mode of AGN feedback, taking place in galaxies with BHs of $M_{\rm BH}\geqslant 10^{8}\, \rm M_{\odot}$. The fraction of quenched massive galaxies may change if a large fraction of these galaxies would host less massive BHs, or we would need to model quenching mechanism differently.

\subsection{Moving forward: Constraining the shape of the $ M_{\rm BH}-M_{\star}$ scaling relation}
A linear $M_{\rm BH}-M_{\star}$ relation favors a coherent growth of the BHs and their host galaxies, while a non-linear relation shows that in some galaxy mass regimes the BHs or their host galaxies grow more efficiently than the other. 
In this paper, we have shown that the BHs formed in simulations with a strong SN feedback, such as EAGLE and TNG (at high redshift), have a reduced growth just after their birth until their host galaxies reach a characteristic mass of $10^{9.5}-10^{10}\, \rm M_{\odot}$; after this mass BHs are free to grow efficiently and to catch up the growth of their host galaxies. 
Only very few AGN candidates have been found in observations at high redshift $z\geqslant 4-5$ \citep{Willott2010,2012A&A...537A..16F,2012ApJ...748...50C,2013ApJ...778..130T,2015A&A...578A..83G,2015MNRAS.448.3167W,2020ApJ...891...69C}. The lack of AGN at high redshift could be due to high level of obscuration, or to reduced growth of BHs at early times (and so lower-mass BHs), in agreement with the shape of the $M_{\rm BH}-M_{\star}$ relation in TNG and EAGLE. 
In the upcoming years, combining X-ray missions (i.e., Athena, LynX, AXIS) to JWST will help us to constraint the existence or absence of AGN at high redshift and the properties of their host galaxies (e.g., stellar masses). 
We discussed the impact of SN feedback on the low-mass end of the $M_{\rm BH}-M_{\star}$ relation, but it is interesting to note that the simulations studied here also have different implementations of the quasar mode of the AGN feedback. It may, to some extend, impact the normalization and shape of the low-mass end of the relation. 
The non-linear relationship in low-mass galaxies is a key point to be explored further in observations in the local Universe (Fig.~\ref{fig:all_scaling_relations}). Interestingly, we note that while BH growth is stunted in the low-mass galaxies at all redshifts in EAGLE, this is not the case in TNG100, for which the reduced growth is only observed for $z\geqslant 2$ and not in the local Universe (Fig.~\ref{fig:scaling_relation_mean}). 
The low-mass regime is also a key point to explore in cosmological simulations with higher resolution \citep{2017MNRAS.468.3935H}. Higher resolution simulations also suggest that resolving the multi-phase ISM is key to capturing the suppression of BH growth by stellar feedback independently of the accretion model \citep{2017MNRAS.472L.109A,2020arXiv200812303A}.

As demonstrated in this paper, the shape and the scatter of the $M_{\rm BH}-M_{\star}$ relation and the BH mass function are all affected by the simulation resolution. 
With a higher resolution, TNG100 resolves better the gas in the BH surroundings and in the whole galaxies, which explains the higher normalization of the low-mass end of the BH mass function in TNG100 (i.e., the excess of BHs with $M_{\rm BH}\leqslant 10^{8.5}\, \rm M_{\odot}$) compared to TNG300. The lower resolution of TNG300 makes the feedback from SNe more effective, and regulates BH growth more efficiently in low-mass galaxies of $M_{\star}\leqslant 10^{10}\, \rm M_{\odot}$ at all redshifts.

The main goal of this paper was to carefully compare the BH populations of the last decade cosmological simulations and understand how the sub-grid physics affect them, and not to make perfect apple-to-apple comparisons with observations (which is particularly challenging for the scaling relation and BH mass function).
We stress here that we have not attempted to understand whether all the observational constraints used in this paper were actually consistent with one another. This is a global concern.  
Complementary observational constraints often use samples that do not overlap, can be biased in different ways, and assume relations established with other empirical constraints. For example, assuming a scaling relation is in most of the cases needed to determine the BH mass function. The scaling relation may be different from the one we found for a given simulation, but we still compare the simulated BH mass function to the one derived from a different scaling relation. A global effort to address more consistent comparisons is needed.

\section{Conclusions}
We have compared six large-scale cosmological hydrodynamical simulations of $\geq100$ cMpc length on a side: Illustris, TNG100, TNG300, Horizon-AGN, EAGLE, and SIMBA. 
We have studied the mass properties of the supermassive BHs of these simulations, more specifically their relations with the stellar mass of their host galaxies, and their mass functions. 
Our main findings are summarized as follows.
\begin{itemize}
\item All the simulations produce a $M_{\rm BH}-M_{\star}$ relation in general agreement with the observations. This is partially by design since they all follow a given empirical scaling relation \citep[e.g.,][]{2013ARA&A..51..511K} to calibrate their sub-grid models.
\item 
Most of the simulations produce a tight $M_{\rm BH}-M_{\star}$ relation, and a smaller diversity of BH masses at fixed stellar masses than in the observational samples of \citet{2015arXiv150806274R} (Fig.~\ref{fig:scaling_relation}) and of \citet{2019MNRAS.487.3404B}.
The most striking difference from observations is that some simulations tend to miss lower-mass BHs ($M_{\rm BH}\leqslant 10^{7.5}\, \rm M_{\odot}$) at fixed stellar mass for $M_{\star}\leqslant 10^{11.5}\, \rm M_{\odot}$.

\item Simulated massive galaxies in the $M_{\rm BH}-M_{\star}$ diagrams (Fig.~\ref{fig:scaling_sf_q}) have sSFR qualifying them as quiescent galaxies (with $\rm sSFR\leqslant 10^{11}\, \rm yr^{-1}$), in agreement with observations of  \citet{2017ApJ...844..170T} and the quiescent elliptical galaxies of \citet{2015arXiv150806274R}. 
Galaxies hosting less massive BHs covering the region $M_{\star}\sim 10^{10-11.5}\, \rm M_{\odot}$ and $M_{\rm BH}\sim 10^{6-8}\, \rm M_{\odot}$ \citep[i.e., covering or close to the observed AGN region of][that consists of a high fraction of spiral galaxies]{2015arXiv150806274R} are generally star-forming galaxies (with $\rm sSFR\leqslant 10^{11}\, \rm yr^{-1}$). For most of the simulations, these galaxies also have luminosities typical of AGN (i.e., $L_{\rm bol}\sim 10^{43}\, \rm erg/s$, Fig.~\ref{fig:scaling_relation}).

\item 
The evolution of the median/mean $M_{\rm BH}-M_{\star}$ relation of the simulations is small in the redshift range $0\leqslant z\leqslant 5$ ($\leq 1$ dex of $\log_{10} \,M_{\rm BH}/\rm M_{\odot}$, Fig.~\ref{fig:scaling_relation_mean}). The normalization of the relation decreases with time in Illustris, Horizon-AGN, and EAGLE, and increases in TNG and SIMBA.

\item We analyzed matched Illustris and TNG100 galaxies and find that the evolution with time of the $M_{\rm BH}-M_{\star}$ mean relation depends on the sub-grid models of galaxy formation and BH physics. 
The overall decrease of the normalization in Illustris can be explained by a more efficient growth of the galaxies with respect to their BHs (particularly for $z\leqslant 2$).
For TNG100, the higher normalization with time for $M_{\star}\leqslant 10^{10}\, \rm M_{\odot}$ is due to reduced BH growth by strong SN feedback at high redshift in low-mass galaxies.

\item 
The overall decrease of the $M_{\rm BH}-M_{\star}$ normalization with time in Horizon-AGN is due to a more efficient relative growth of galaxies compared to their BHs with time, as in Illustris. 
In EAGLE, BH growth is initially reduced by SN feedback (as in TNG) and the rapid growth phase starts in halos of fixed virial temperature (i.e., in less massive galaxies at higher redshift), which explains the overall normalization increase with time \citep{2018MNRAS.481.3118M}.
The normalization of the relation in SIMBA only increases for $z\leqslant2$: the low accretion AGN feedback mode quenches galaxies, but also increases their hot environment, which in return favors an additional growth channel for the BHs (Bondi accretion for the hot gas). The growth of BHs becomes more efficient than for their host galaxies.

\item
We find that the change of slope of the $M_{\rm BH}-M_{\star}$ relation arises from the modeling of SN feedback (e.g., its strength, its dependence with time).
In TNG and EAGLE the modeling of SN feedback is strong enough to stunt BH growth in low-mass galaxies of $M_{\star}\leqslant 10^{9.5}-10^{10}\, \rm M_{\odot}$, creating non-linear relations (Fig.~\ref{fig:scaling_relation_mean}).
Horizon-AGN and Illustris have weaker SN feedback, leading to linear $M_{\rm BH-M_{\star}}$ relations. 
Simulations do not agree on the linearity or non-linearity of the low-mass end of the relation, which is therefore a key point to be explored in observations in the local Universe and at high redshift.

\item The time variation of the $M_{\rm BH}-M_{\star}$ scatter does not exceed one dex in BH mass for $10^{9}\leqslant M_{\star}/\rm M_{\odot}\leqslant 10^{11}$ in all the simulations. In general, the scatter stays the same in Horizon-AGN, increases in Illustris and SIMBA, and decreases in EAGLE and TNG100 with time.

\item 
The scatter of the $M_{\rm BH}-M_{\star}$ relation has different dependence on $M_{\star}$ in different simulations (for $M_{\star}\leqslant 10^{11}\, \rm M_{\odot}$). In general the scatter is constant in Horizon-AGN, slightly increases in Illustris, and decreases in SIMBA.
For TNG and EAGLE, the scatter is small when SN feedback efficiently regulates BH growth, larger when BHs grow efficiently, and decreases when they are regulated by AGN feedback.

\item  
Different simulations show different behaviors at the high-mass end of the $M_{\rm BH}-M_{\star}$ relation ($M_{\star}\geqslant 10^{11}\, \rm M_{\odot}$). In TNG300 (largest simulated volume and more statistics in this regime), the relation does not evolve with time, because BH and galaxy growth are driven by mergers in this regime \citep{2018MNRAS.479.4056W,2017arXiv170703406P}, and gas accretion is stunted by AGN feedback.
The scatter is smaller in this regime and decreases with time, in agreement with predictions from the central-limit theorem. 
However in SIMBA, BH growth exceeds galaxy growth by mergers in this regime; gas accretion plays a more important role (Cui et al., in prep). Consequently, the massive end of the relation evolves with time, and its scatter does not decrease.
\end{itemize}

We analyzed the BH mass functions of the simulations, and summarize below our main results.
\begin{itemize}
\item All the simulations have a BH mass function with an increasing overall normalization with time (Fig.~\ref{fig:mass_function}, Fig.~\ref{fig:mass_function_2}), following the hierarchical build-up of their host galaxies.

\item Most of the simulations, except the EAGLE simulation, overpredict the number of the most massive BHs ($M_{\rm BH}\geqslant 10^{9}\, \rm M_{\odot}$) at $z=0$ compared to the constraint of \citet{Merloni08}. These BHs are embedded in the most massive galaxies of $M_{\star}\sim 10^{12}\, \rm M_{\odot}$ (Fig.~\ref{fig:scaling_relation}).
The TNG100 simulation also predicts an excess of lower-mass BHs with $M_{\rm BH}=10^{7}-10^{9}\, \rm M_{\odot}$ (probably due to the high seeding mass).

\item We find that the most massive BHs are less active at $z=0$ than the same mass BHs at earlier times in all the simulations, and therefore all the simulations show evidence of an {\it anti-hierarchical growth}.

\item We identify some differences in the contribution of rapidly growing BHs (active BHs with $\log_{10} f_{\rm Edd}\geqslant -2$) and inactive BHs to the BH mass function among the simulations. 
At high redshift ($z=4$), the BH mass function is dominated by active BHs for most of the simulations, except EAGLE with already a large contribution from inactive BHs. 
Inactive BHs dominate the massive end of the mass function at $z=2$, and the entire mass function by $z=0$.
However in SIMBA, active BHs still dominate the BH mass function for $M_{\rm BH}\leqslant 10^{8.5}\, \rm M_{\odot}$.

\item At $z=0$, all the simulations (except EAGLE) overpredict the number of active BHs at the low-mass end ($M_{\rm BH}\leqslant 10^{8}\, \rm M_{\odot}$) compared to the constraints of \citet{2010A&A...516A..87S}. At the high-mass end, half of the simulations do not have any active BHs of $M_{\rm BH}\geqslant 10^{8}\, \rm M_{\odot}$, the other half have higher normalization of more than an order of magnitude.

\end{itemize}

\section*{Acknowledgment}
We thank the anonymous referee, Volker Springel, Joop Schaye, and Rainer Weinberger for useful comments on the paper, and Dalya Baron for sharing her recent observational sample.
We acknowledge the support from the The Flatiron Institute, which is supported by the Simons Foundation.
YRG acknowledges the support of the European Research Council through grant number ERCStG/716151, and RB
the support of the Science and Technology Facilities Council [ST/P000541/1].
DAA was supported in part by NSF grant AST-2009687. 

\section*{Data Availability Statement}
The data from the Illustris and the TNG100 simulations can be found on their respective websites: https://www.illustris-project.org, https://www.tng-project.org. The data from the EAGLE simulation can be obtained upon request to the EAGLE team at their website: http://icc.dur.ac.uk/Eagle/. 
The data from the SIMBA simulation can be found on the website: http://simba.roe.ac.uk/.
The Horizon-AGN simulation is not public, but some catalogs are available at: https://www.horizon-simulation.org/data.html.

\appendix
\section{Other scaling relations}
\label{other_scaling_relations}

While the scaling relation $M_{\rm BH}-M_{\star}$ is the most convenient because stellar mass is the easiest quantity to measure both in observations and simulations, it is not the tightest relation that has been found in observations. The relation with bulge mass has a smaller scatter \citep{Magorrian1998,Haring2004,Gultekin2009}, and the relation with stellar velocity dispersion of the host galaxies an even tighter scatter \citep{Ferrarese2000,Gebhardt2000,2013ApJ...764..184M,2013ARA&A..51..511K,2013ApJ...772...49W,2016ApJ...818...47S,2016MNRAS.460.3119S,2016ApJ...818...47S,2019MNRAS.490..600D}. 
Measuring the mass of bulges in observations is challenging and difficult to do in a systematic way for all galaxy morphologies. Bulge+disk decomposition relies on several aspects/assumptions (each study having its own) and can suffer from lack of information on the inclination, surface brightness limits, spatial resolution, and signal-to-noise ratio \citep[see discussions in][]{2011ApJS..196...11S,2012MNRAS.420.2662D,2016MNRAS.460.2979V}.
It has been shown that the $M_{\rm BH}-\sigma$ plane could be more fundamental than those with galaxy stellar luminosity or mass, or effective radius \citep[for a review]{2020FrP.....8...61M}.
At relatively low redshift, there is no observational evidence for a redshift evolution of the $M_{\rm BH}-\sigma$ relation, as demonstrated by \citet{2015ApJ...805...96S} using about 90 SDSS quasars at $z<1$ and by indirect estimates \citep{2009ApJ...694..867S,Zhang2012}.

Regarding cosmological simulations, \citet{2019arXiv191000017L} recently studied the relation $M_{\rm BH}-\sigma$ in Illustris and TNG100.
The $\sigma$ values were computed using rest-frame SDSS-r band luminosity-weighted stellar line-of-sight velocity dispersion measured within a given projected radius, as it is the case for observations.  
They demonstrated some discrepancies between the simulations and the observations, where BHs in Illustris and TNG100 were either too massive at fixed $\sigma$, or conversely $\sigma$ were too small with respect to the BHs \citep{2019arXiv191000017L}.
The same $M_{\rm BH}-\sigma$ scaling relation was also studied for Illustris \citep{2015MNRAS.452..575S}, Horizon-AGN \citep{2016MNRAS.460.2979V}, EAGLE \citep{2019MNRAS.485..396V} and SIMBA \citep{2019MNRAS.487.5764T}, but without an observationally friendly frame.
Interestingly, SIMBA shows a broader scatter in $M_{\rm BH}-\sigma$ than for the other scaling relations \citep{2019MNRAS.487.5764T}.

\section{Time evolution of the specific star formation rate of galaxies in the $M_{\rm BH}-M_{\star}$ diagram}
\label{ssfr_redshift}
\begin{figure*}
\centering
\includegraphics[scale=0.55]{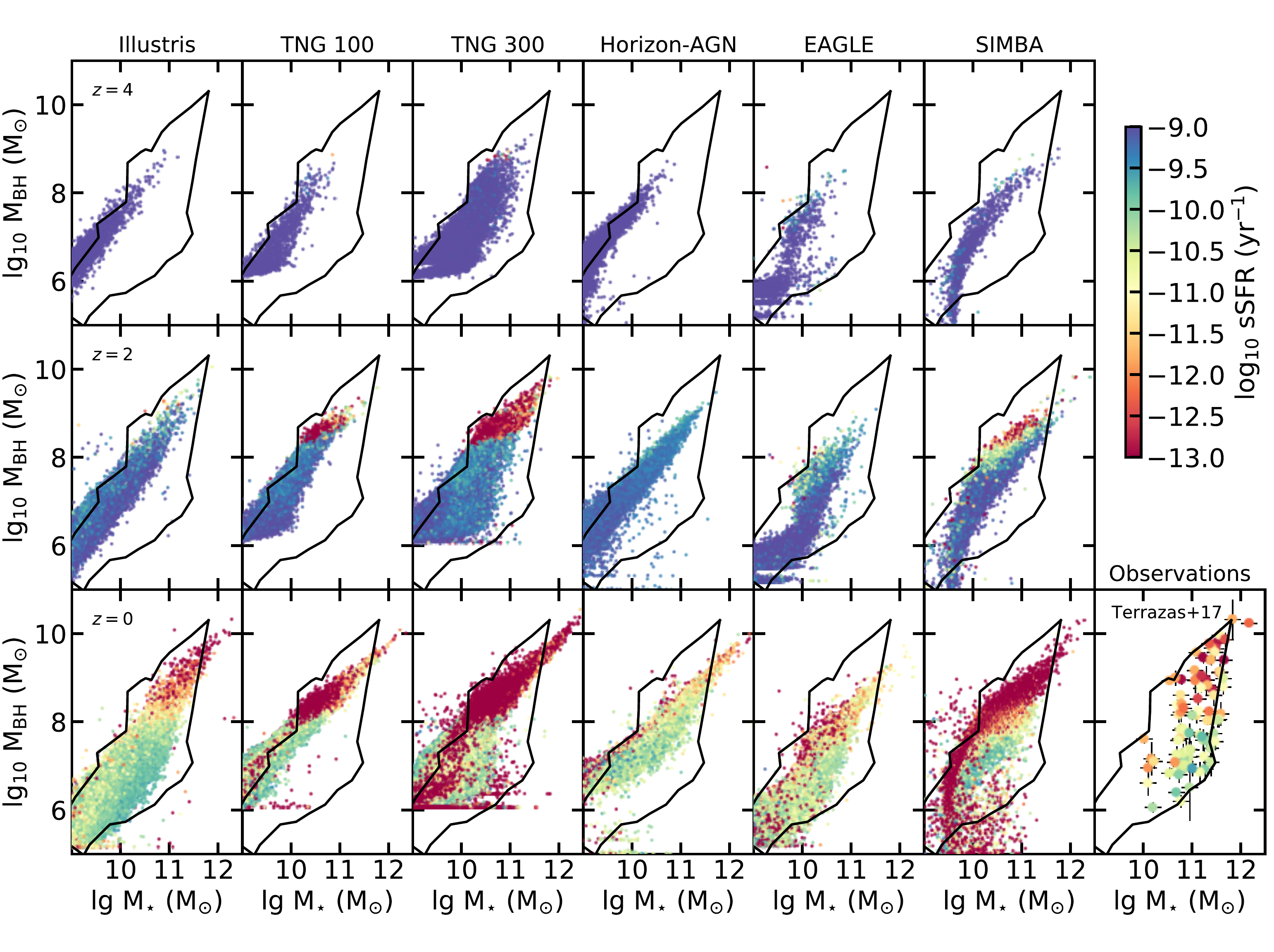}
\caption{$M_{\rm BH}-M_{\star}$ diagram at $z=4,2,0$ for all the simulations. We color-code the simulated BHs with the specific star formation rate (sSFR) of their host galaxies. We set $\rm sSFR=10^{-13}\, yr^{-1}$ for galaxies with lower sSFR. The limit $\rm sSFR=10^{-11}\, \rm yr^{-1}$ is often used to define star-forming galaxies (with higher sSFR) or quiescent galaxies (with lower sSFR). In the figure, galaxies with blue colors are forming stars, and yellow to red galaxies are quiescent. 
We show in the last right panel the observed star-forming and quiescent galaxies of \citet{2017ApJ...844..170T} for references, color-coded by their sSFR as well. To guide the eye, we again reproduce the black region representing the observational sample of \citet{2015arXiv150806274R} at $z=0$ in each panel. }
\label{fig:scaling_sf_q_bis}
\end{figure*}

In Fig.~\ref{fig:scaling_sf_q_bis}, we show the simulation $M_{\rm BH}-M_{\star}$ diagrams color coded by the specific star formation rate of the BH host galaxies. 

\section{Scatter of the $M_{\rm BH}-M_{\star}$ relations in the simulations}
\label{scatter_scaling_relations}
We present the scatter of the $M_{\rm BH}-M_{\star}$ relations for the simulation Illustris, Horizon-AGN, and EAGLE in Table~\ref{table:tab_scatter_1}, for different bins of galaxy total stellar mass and redshift. These simulations have a decreasing overall normalization of the $M_{\rm BH}-M_{\star}$ with time. We show the scatter of the simulations having an increasing overall normalization in Table~\ref{table:tab_scatter_2}, i.e., TNG100 and SIMBA.

\begin{table}
\caption{Time evolution of the scaling relation $M_{\rm BH}-M_{\star}$. We provide the stellar mass bins ($\log_{10} M_{\star}/\rm M_{\odot}$), the redshift in the range $z=3-0$, the corresponding mean ($\langle \log_{10} M_{\rm BH}/\rm M_{\odot}\rangle$) and median ($\log_{10} \tilde{M}_{\rm BH}/\rm M_{\odot}$) of BH mass in the given stellar mass bin, and the scatter (15th-85th percentiles) of the mean scaling relation. For reference, we find that the 15th-85th percentiles of the observations of \citet{2015arXiv150806274R} (at $z\sim 0$) is comprised in the range $\sigma=0.9-1.9$ for $M_{\star}=10^{9.5}-10^{11}\,\rm M_{\odot}$. For the observational sample of \citet{2019MNRAS.487.3404B}, we find $\sigma=1.2-1.6$ in the same $M_{\star}$ range. The simulations produce a smaller scatter than found in the observations.}
\begin{center}
\begin{tabular}{lccccc}
\hline
Simulation & $M_{\star}$ & redshift & $\langle M_{\rm BH}\rangle$  & $\tilde{M}_{\rm BH}$ & $\sigma$ \\
\hline
\hline

Illustris & 9.5 & $z=3$	&  6.76 & 6.77 & 0.54 \\      
	    & 9.5 & $z=2$	&  6.62 & 6.62 & 0.59 \\
	    & 9.5 & $z=1$	&  6.36 & 6.36 & 0.74 \\
	    & 9.5 & $z=0$	&  5.90 & 5.86 & 0.88 \\\cmidrule{2-6}

	    & 10 & $z=3$	&  7.36 & 7.35 & 0.60 \\      
	    & 10 & $z=2$	&  7.20 & 7.20 & 0.58 \\
	    & 10 & $z=1$	&  6.98 & 6.98 & 0.68 \\
	    & 10 & $z=0$	&  6.60 & 6.59 & 0.86 \\\cmidrule{2-6}
	    
	    & 10.5 & $z=3$	&  8.00 & 8.03 & 0.67 \\      
	    & 10.5 & $z=2$	&  7.81 & 7.80 & 0.63 \\
	    & 10.5 & $z=1$	&  7.59 & 7.58 & 0.74 \\
	    & 10.5 & $z=0$	&  7.30 & 7.30 & 0.95 \\\cmidrule{2-6}
	     
	    & 11 & $z=3$	&  8.71 & 8.65 & 0.49 \\      
	    & 11 & $z=2$	&  8.54 & 8.59 & 0.82 \\
	    & 11 & $z=1$	&  8.28 & 8.25 & 0.84 \\
	    & 11 & $z=0$	&  8.05 & 8.08 & 1.04 \\		    

\hline	    
H-AGN & 9.5 & $z=3$	&  6.84 & 6.85 & 0.39\\      
	    & 9.5 & $z=2$	&  6.76 & 6.78 & 0.36 \\
	    & 9.5 & $z=1$	&  6.67 & 6.69 & 0.42 \\
	    & 9.5 & $z=0$	&  6.82 & 6.84 & 0.32 \\\cmidrule{2-6}

	    & 10 & $z=3$	&  7.36 & 7.35 & 0.37 \\      
	    & 10 & $z=2$	&  7.27 & 7.27 & 0.41 \\
	    & 10 & $z=1$	&  7.13 & 7.13 & 0.46 \\
	    & 10 & $z=0$	&  7.10 & 7.10 & 0.35 \\\cmidrule{2-6}
	    
	    & 10.5 & $z=3$	&  7.90 & 7.92 & 0.45 \\      
	    & 10.5 & $z=2$	&  7.82 & 7.82 & 0.43 \\
	    & 10.5 & $z=1$	&  7.75 & 7.75 & 0.48 \\
	    & 10.5 & $z=0$	&  7.59 & 7.59 & 0.51 \\\cmidrule{2-6}
	     
	    & 11 & $z=3$	&  8.54 & 8.52 & 0.42 \\      
	    & 11 & $z=2$	&  8.41 & 8.46 & 0.48 \\
	    & 11 & $z=1$	&  8.34 & 8.35 & 0.43 \\ 
	    & 11 & $z=0$	&  8.22 & 8.23 & 0.47 \\
	    	    
\hline 
EAGLE & 9.5 & $z=3$	&  5.75 & 5.79 & 0.48\\      
	    & 9.5 & $z=2$	&  5.70 & 5.69 & 0.42 \\
	    & 9.5 & $z=1$	&  5.68 & 5.69 & 0.43 \\
	    & 9.5 & $z=0$	&  5.79 & 5.79 & 0.57 \\\cmidrule{2-6}

	    & 10 & $z=3$	&  6.77 & 6.82 & 1.32 \\      
	    & 10 & $z=2$	&  6.48 & 6.30 & 1.31 \\
	    & 10 & $z=1$	&  6.22 & 6.10 & 0.85 \\
	    & 10 & $z=0$	&  6.30 & 6.22 & 0.86 \\\cmidrule{2-6}
	    
	    & 10.5 & $z=3$	&  7.41 & 7.56 & 1.53 \\      
	    & 10.5 & $z=2$	&  7.39 & 7.46 & 1.02 \\
	    & 10.5 & $z=1$	&  7.25 & 7.32 & 1.10 \\
	    & 10.5 & $z=0$	&  7.19 & 7.25 & 1.12 \\\cmidrule{2-6}
	     
	    & 11 & $z=3$	&  7.74 & 8.02 & 1.49 \\      
	    & 11 & $z=2$	&  8.02 & 8.14 & 0.95 \\
	    & 11 & $z=1$	&  8.00 & 8.11 & 0.76 \\
	    & 11 & $z=0$	&  8.03 & 8.07 & 0.65 \\	    
\hline
\end {tabular}
\end{center}
\label{table:tab_scatter_1}
\end{table}

\begin{table}
\caption{Same as Table~\ref{table:tab_scatter_1} for the TNG100 and SIMBA simulations.}
\begin{center}
\begin{tabular}{lccccc}
\hline
Simulation & $M_{\star}$ & redshift & $\langle M_{\rm BH}\rangle$  & $\tilde{M}_{\rm BH}$ & $\sigma$ \\
\hline	    
\hline	
TNG100 & 9.5 & $z=3$	&  6.48 & 6.40 & 0.45\\      
	    & 9.5 & $z=2$	&  6.64 & 6.56 & 0.66 \\
	    & 9.5 & $z=1$	&  6.88 & 6.91 & 0.66 \\
	    & 9.5 & $z=0$	&  7.17 & 7.22 & 0.50 \\\cmidrule{2-6}

	    & 10 & $z=3$	&  7.13 & 7.17 & 0.91 \\      
	    & 10 & $z=2$	&  7.39 & 7.44 & 0.75 \\
	    & 10 & $z=1$	&  7.54 & 7.56 & 0.55 \\
	    & 10 & $z=0$	&  7.70 & 7.70 & 0.40 \\\cmidrule{2-6}
	    
	    & 10.5 & $z=3$	&  8.11 & 8.14 & 0.78 \\      
	    & 10.5 & $z=2$	&  8.26 & 8.30 & 0.62 \\
	    & 10.5 & $z=1$	&  8.22 & 8.25 & 0.52 \\
	    & 10.5 & $z=0$	&  8.20 & 8.21 & 0.37 \\\cmidrule{2-6}
	     
	    & 11 & $z=3$	&  (8.71) & (8.71) & (0.29) \\      
	    & 11 & $z=2$	&  8.74 & 8.73 & 0.28 \\
	    & 11 & $z=1$	&  8.69 & 8.69 & 0.42 \\
	    & 11 & $z=0$	&  8.63 & 8.64 & 0.48 \\

\hline
SIMBA & 9.5 & $z=3$	&  5.65 & 5.74 & 1.24 \\      
	    & 9.5 & $z=2$	&  5.62 & 5.68 & 1.25 \\
	    & 9.5 & $z=1$	&  5.78 & 5.86 & 1.46 \\
	    & 9.5 & $z=0$	&  6.35 & 6.59 & 1.28 \\\cmidrule{2-6}
	    
	    & 10 & $z=3$	&  6.88 & 6.89 & 0.82 \\      
	    & 10 & $z=2$	&  6.94 & 6.94 & 1.02 \\
	    & 10 & $z=1$	&  7.20 & 7.28 & 1.03 \\
	    & 10 & $z=0$	&  7.34 & 7.51 & 0.64 \\\cmidrule{2-6}
	    
	    & 10.5 & $z=3$	&  7.62 & 7.63 & 0.68 \\      
	    & 10.5 & $z=2$	&  7.69 & 7.73 & 0.79 \\
	    & 10.5 & $z=1$	&  7.83 & 7.94 & 0.70 \\
	    & 10.5 & $z=0$	&  7.76 & 8.01 & 0.84 \\\cmidrule{2-6}
	     
	    & 11 & $z=3$	&  8.24 & 7.63 & 0.68 \\      
	    & 11 & $z=2$	&  8.26 & 7.73 & 0.79 \\
	    & 11 & $z=1$	&  8.27 & 7.94 & 0.70 \\
	    & 11 & $z=0$	&  8.17 & 8.01 & 0.84 \\	    
\hline
\end {tabular}
\end{center}
\label{table:tab_scatter_2}
\end{table}

\section{Illustration of the $M_{\rm BH}-M_{\star}$ build-up for individual matched Illustris and TNG100 galaxies}
\label{sec:illustration}
\begin{figure*}
\centering
\includegraphics[scale=0.44]{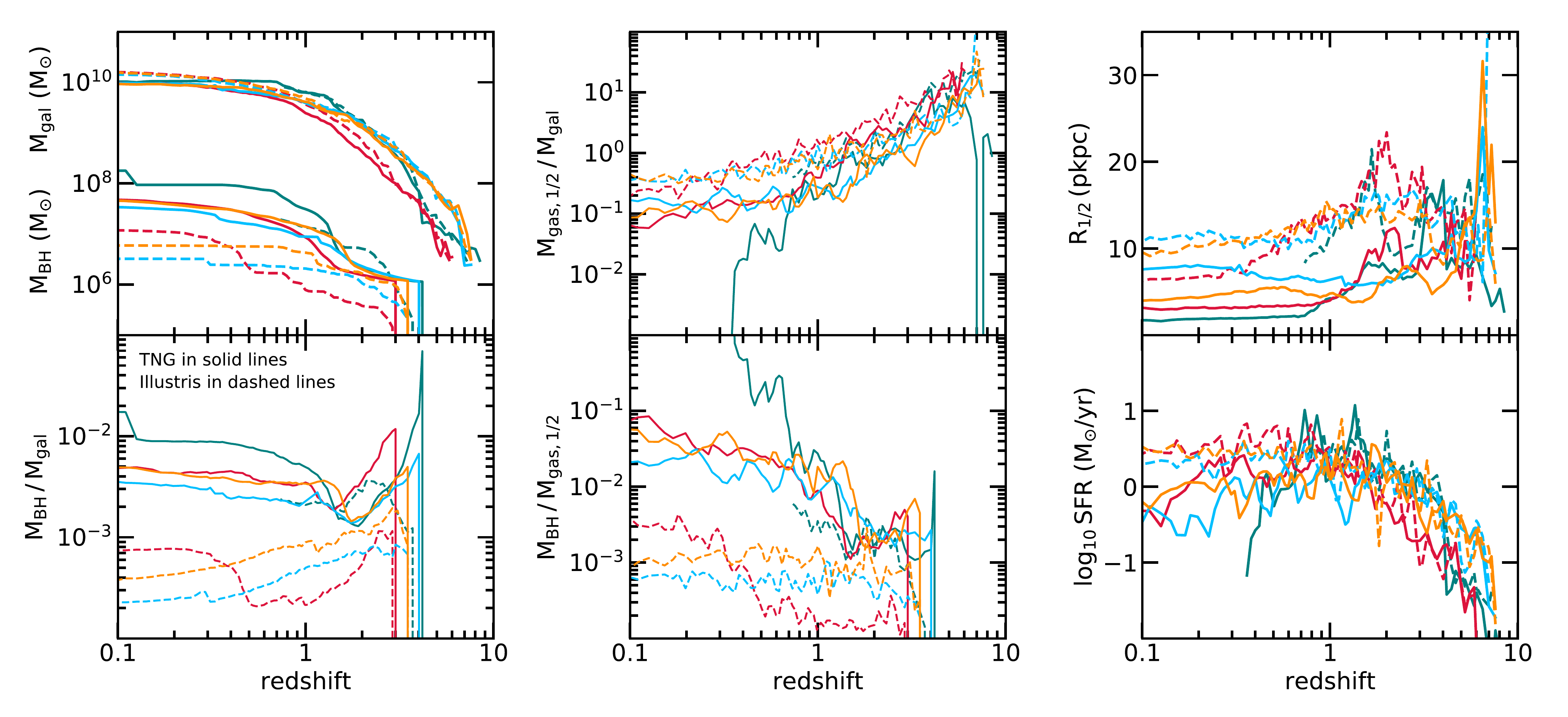}
\includegraphics[scale=0.44]{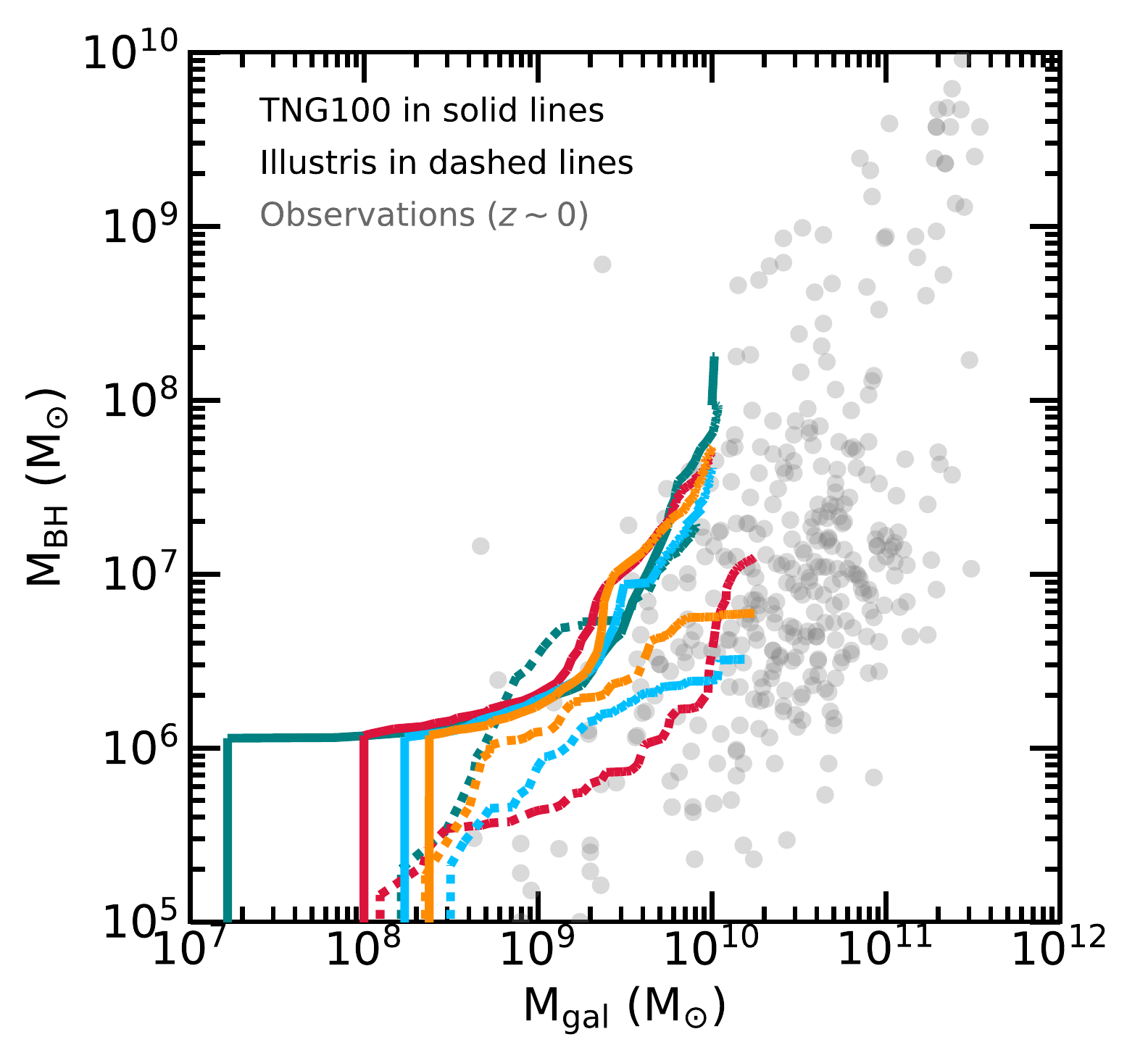}
\caption{Time evolution of four pairs of matched Illustris (dashed lines) and TNG100 (solid lines) galaxies with $M_{{\rm gal}, \, {\rm TNG},\, z=0}=10^{10}\, \rm M_{\odot}$. 
{\it First columns}: BH mass and host galaxy stellar mass evolution (top), and the ratio of these quantities $M_{\rm BH}/M_{\rm gal}$ (bottom). {\it Second columns}: ratios $M_{\rm gas, 1/2}/M_{\rm gal}$ of the gas mass in the half-mass stellar radius to the stellar mass in twice the half-mass radius. {\it Third columns}: galaxy sizes ($R_{\rm 1/2}$), and star formation rates (SFR). {\it Bottom panel}: $M_{\rm BH}-M_{\rm gal}$ diagram. In these galaxies evolving to $M_{\rm gal}=10^{10}\, M_{\odot}$ at $z=0$, we note that a large fraction of their BHs are more massive in TNG100 than in Illustris. This result holds even for the BHs embedded in galaxies with similar stellar mass evolution; the right panels show that these galaxies have however different evolution of their sizes and SFR history with time. The TNG100 galaxies are seeded with massive BHs of $\sim 10^{6}\, \rm M_{\odot}$. The start of the efficient BH accretion phase is delayed. After the low accretion phase, the TNG100 BHs grow more efficiently than their galaxies, as shown by the increasing $M_{\rm BH}/M_{\rm gal}$ ratios with time. In Illustris, we note a large fraction of systems with decreasing $M_{\rm BH}/M_{\rm gal}$ ratios with time, reflecting the more efficient growth of the galaxies compared to their BHs. In this mass range, most of the matched BHs in the two simulations are more massive in TNG100 than in Illustris.}
\label{fig:4gal_1e10}
\end{figure*}

\begin{figure*}
\centering
\includegraphics[scale=0.44]{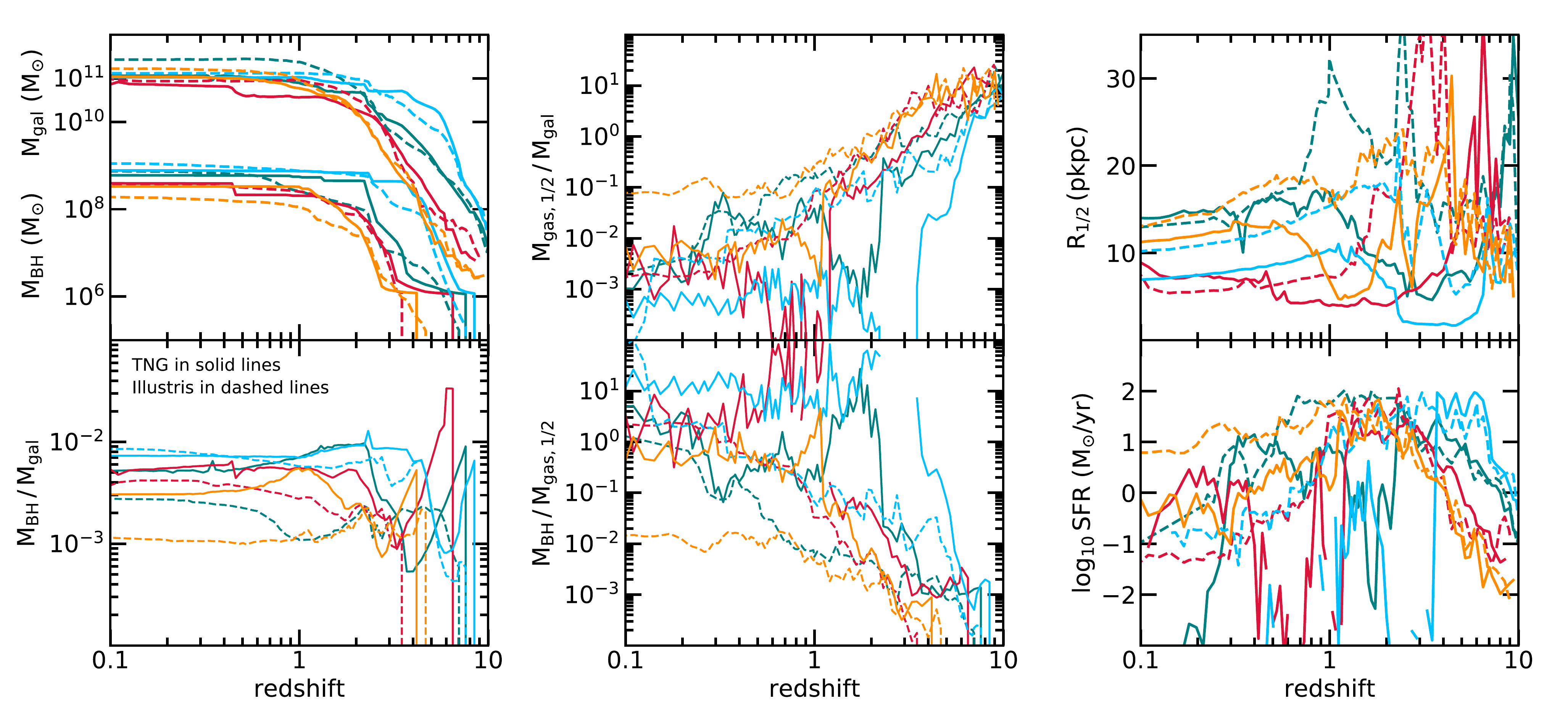}
\includegraphics[scale=0.44]{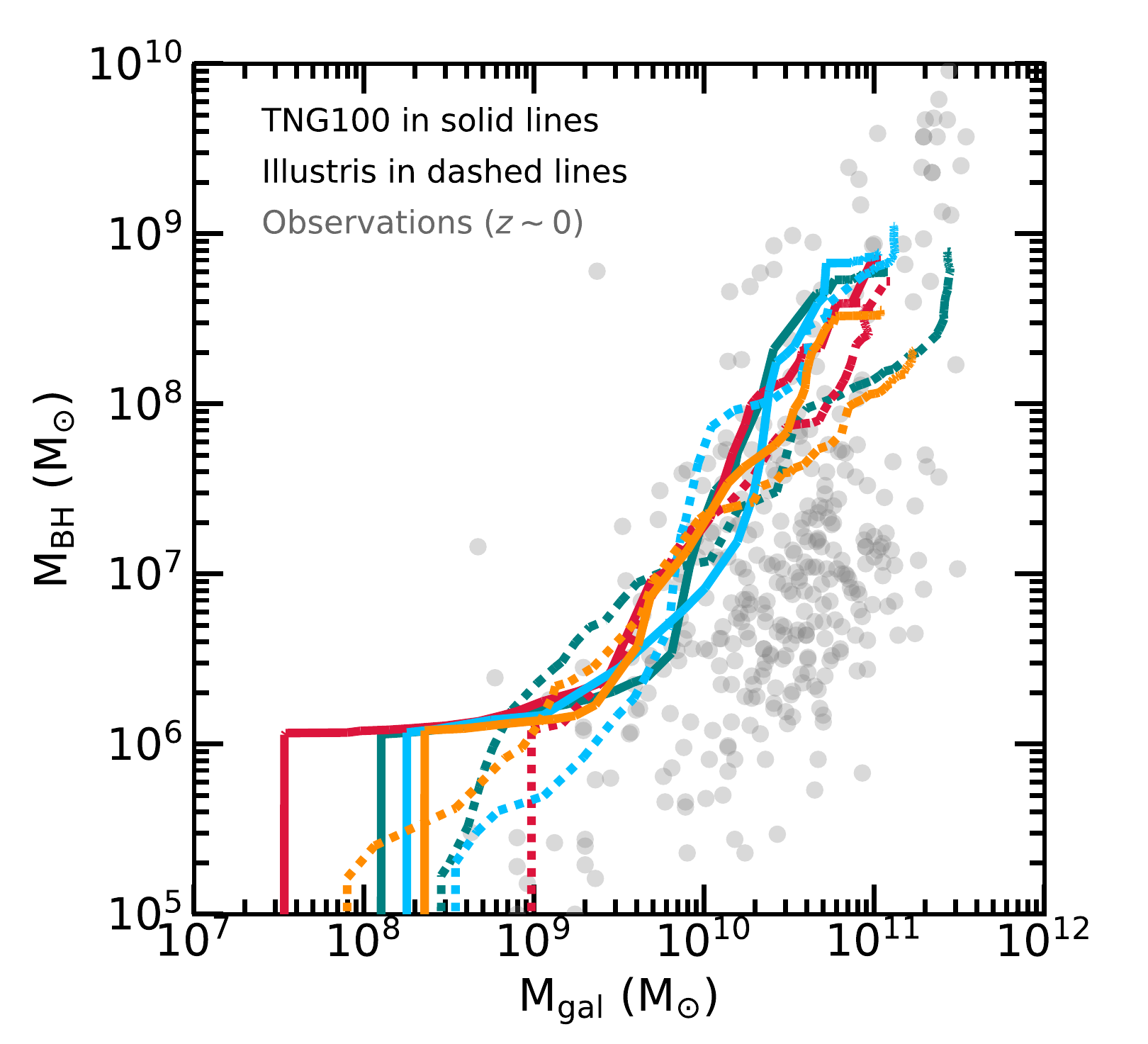}
\caption{Same as Fig.~\ref{fig:4gal_1e10}, but with more massive galaxies of $M_{{\rm gal}, \, {\rm TNG},\, z=0}=10^{11}\, \rm M_{\odot}$. For these more massive galaxies, we note that the difference in BH mass at $z=0$ in the matched TNG100 and Illustris galaxies is not as significant as for the $M_{\rm gal} (z=0)=10^{10}\, M_{\odot}$ matched galaxies. The $M_{\rm BH}/M_{\rm gal}$ ratios of the two simulations differ by less than an order of magnitude.
For reference, the galaxies of $M_{\rm gal} (z=0)=10^{10}\, M_{\odot}$ show stronger differences, with ratios that are about one order of magnitude higher in TNG100 than in Illustris.}
\label{fig:4gal_1e11}
\end{figure*}

To illustrate the results of this paper, we show the time evolution of eight pairs of galaxies from the Illustris and TNG simulations. 
We choose these galaxies as reaching a final stellar mass of $M_{\star}=10^{10}\, \rm M_{\odot}$ (Fig.~\ref{fig:4gal_1e10}) and $M_{\star}=10^{11}\, \rm M_{\odot}$ (Fig.~\ref{fig:4gal_1e11}) at $z=0$. The evolution of the Illustris galaxies is shown with dashed lines, and the TNG100 galaxies with solid lines. We present the quantities: stellar masses in twice the half-mass radius $M_{\star}$, BH masses $M_{\rm BH}$, ratios $M_{\rm gas,\, 1/2}/M_{\star}$  between the gas mass in the half-mass radius and stellar mass in twice the radius, ratios $M_{\rm BH}/M_{\rm gas,\, 1/2}$, half-mass sizes $R_{\rm 1/2}$, star formation rate SFR, and $M_{\rm BH}-M_{\star}$ diagrams.
Some of the matched galaxies have a similar stellar mass evolution with time in Illustris and TNG100 (e.g., in Fig.~\ref{fig:4gal_1e10}), but different BH mass evolution. The different sub-grid physics models also impact galaxy quantities such as the star formation rate or the size of the galaxies (right panels). Galaxies in TNG100 have smaller radii, and more reduced SFR at lower redshift due to the efficient AGN feedback low accretion mode.

In the following, we summarize the main features explaining the differences in the co-evolution of the Illustris and TNG100 BHs and their host galaxies.
On the bottom panel of Fig.~\ref{fig:4gal_1e10}, most of the Illustris BHs are scaled down in the $M_{\rm BH}-M_{\star}$ diagram compared to their matched Illustris BHs. To understand the evolution in this diagram, it is essential to disentangle the growth of the BHs from the growth of their galaxies. At $z=0$, the galaxies are more massive in Illustris than in TNG100 (top left panel) by a factor of about two, which results in a shift toward the right for these galaxies in the $M_{\rm BH}-M_{\star}$ diagram. In some cases (e.g., the blue and orange lines for the Illustris systems), we note that the host Illustris galaxies keep growing for $z\leqslant 2$, while BH growth is less efficient (asymptote in the left top panel). For these BHs, the ratios $M_{\rm BH}/M_{\rm gas,\, 1/2}$ are constant with time, suggesting that the smaller growth of these BHs is due to a diminution of the gas content in the inner regions of their host galaxies (middle panel). The less efficient growth of BHs and the more efficient growth of the galaxies leads to lower overall normalizations of the $M_{\rm BH}-M_{\star}$ mean relation with time in Illustris (see Fig.~\ref{fig:scaling_relation_mean}). We also note a diversity of evolutionary paths in the simulation Illustris, which are responsible for the larger scatter at fixed $M_{\star}$ in the $M_{\rm BH}-M_{\star}$ diagram than in the TNG100 simulation.

The TNG100 galaxies are less massive than the Illustris matched ones, but their BHs are all more massive (left panels) in this galaxy stellar mass range $M_{\star, \,{\rm TNG}, \,z=0}=10^{10}\, \rm M_{\odot}$ (Fig.~\ref{fig:4gal_1e10}).
The evolution of the mass ratio $M_{\rm BH}/M_{\star}$ (left panel) shows us that the growth of the TNG100 BHs is stunted early on (until $z\sim 2$ for $M_{\star, {\rm TNG},\,z=0}=10^{10}\, \rm M_{\odot}$) due to a stronger SN feedback modeling, but that the BHs are then growing more efficiently than their host galaxies. 
This is not due to a higher gas content in the inner regions of the TNG100 galaxies, as shown by the lower $M_{\rm gas,\, 1/2}/M_{\star}$ ratios in TNG100 (middle panel). Moreover, the higher ratios $M_{\rm BH}/M_{\rm gas,\, 1/2}$ in TNG100 demonstrate that BHs are growing more efficiently in TNG100 than in Illustris, even when surrounded by a lower content of gas. This can be due to the use of a kernel to compute the accretion rates onto the BHs (less stochastic than using the parent cell as in Illustris), but also to the addition of the magnetic fields, which boosts the accretion in TNG100. 
We stress here that at fixed environment (galaxy properties, gas content, etc) and without considering the above factors, the accretion rates onto the TNG100 BHs will be more important in TNG100 as soon as the mass of the TNG100 BHs is one order of magnitude larger than their matched Illustris BHs (due to the choice of the parameters in the Bondi modeling).

The largest differences in the evolution of the BH mass (up to one order of magnitude in $M_{\rm BH}$) is found for the stellar mass regime $M_{\rm \star, \, TNG (z=0)}=10^{10}\, \rm M_{\odot}$.
The evolutionary paths of BHs in matched galaxies that reach $M_{\star,\, {\rm TNG},\,z=0}\sim 10^{11}\, \rm M_{\odot}$ at $z=0$ is more diverse than for galaxies with $M_{\star, {\rm TNG},\, z=0}=10^{10}\, \rm M_{\odot}$. We illustrate this with Fig.~\ref{fig:4gal_1e11}. While the BHs tend to be always more massive when reaching $z=0$ in TNG100, we find some periods of time for which the matched Illustris BHs are more massive.

\bibliographystyle{mn2e}
\bibliography{biblio_complete,biblio_complete-2}

\label{lastpage}
\end{document}